\documentclass[12pt]{article}
\usepackage{graphicx} 
\usepackage{amsfonts}
\usepackage{iftex}
\ifPDFTeX
  \usepackage[T1]{fontenc}
  \usepackage[utf8]{inputenc}
  \usepackage{textcomp} 
\else 
  \usepackage{unicode-math}
  \defaultfontfeatures{Scale=MatchLowercase}
  \defaultfontfeatures[\rmfamily]{Ligatures=TeX,Scale=1}
\fi
\usepackage{lmodern}
\usepackage{amsmath,amssymb,color,xcolor}
\usepackage{apacite}
\usepackage{natbib}
\usepackage[ruled]{algorithm2e}
\usepackage{setspace}
\usepackage{subcaption}
\usepackage{booktabs}
\usepackage{mathtools}
\usepackage{comment}
\usepackage{colortbl}
\def\spacingset#1{\renewcommand{\baselinestretch}%
{#1}\small\normalsize} \spacingset{1}

\newcommand{\yc}[1]{{{\small \color{red}  #1}}}
\newcommand{\jk}[1]{{{\small \color{teal}  #1}}}
\newcommand{\sz}[1]{{{\color{blue}  #1}}}

\newcommand{\bfy}{{\mathbf y}}
\newcommand{\bfz}{{\mathbf z}}
\newcommand{\bfx}{{\mathbf x}}
\newcommand{\bfbeta}{{\boldsymbol \beta}}
\newcommand{\bfgamma}{{\boldsymbol \gamma}}
\newcommand{\bftheta}{{\boldsymbol{\theta}}}
\newcommand{\bfzeta}{{\boldsymbol{\zeta}}}
\newcommand{\bfeta}{{\boldsymbol{\eta}}}
\newcommand{\bfxi}{{\boldsymbol{\xi}}}

\newcommand{\bfalpha}{{\boldsymbol{\alpha}}}

\newcommand{\bfkappa}{{\boldsymbol \kappa}}

\usepackage{amsthm}
\usepackage{authblk} 
\theoremstyle{plain}
\newtheorem{theorem}{Theorem}

\newtheorem{remark}{Remark}

\title{\large A Latent Variable Framework for Multiple Imputation with Non-ignorable Missingness: Analyzing Perceptions of Social Justice in Europe}
\author[1]{\vspace*{-0.4cm}\small Siliang Zhang\thanks{\texttt{Corresponding author:slzhang@fem.ecnu.edu.cn}}}
\author[2]{\small Yunxiao Chen}
\author[2]{\small Jouni Kuha}
\affil[1]{\small Key Laboratory of Advanced Theory and Application in Statistics and Data Science-MOE, School of Statistics, East China Normal University}
\affil[2]{\small Department of Statistics, London School of Economics and Political Science}
\date{}

\begin{document}

\spacingset{1.8}

\maketitle
\vspace{-1.2cm}
\begin{abstract}
This paper proposes a general multiple imputation approach for analyzing
large-scale data with missing values. An imputation model is derived
from a joint distribution induced by a latent variable model, which can
flexibly capture associations among variables of mixed types. The model
also allows for missingness which depends on the latent variables and is thus non-ignorable with respect to the observed data. 
We develop a frequentist multiple imputation method for this framework and provide asymptotic theory that establishes valid inference for a broad class of analysis models. Simulation studies confirm the method's theoretical properties and robust practical performance.
The procedure is applied to a cross-national analysis of individuals' perceptions of justice and fairness of income distributions in their societies,   using data from the European Social Survey which has substantial nonresponse. The analysis demonstrates that failing to account for non-ignorable missingness can yield biased conclusions; for instance, complete-case analysis is shown to exaggerate the correlation between personal income and perceived fairness of income distributions in society. Code implementing the proposed methodology is publicly available at https://anonymous.4open.science/r/non-ignorable-missing-data-imputation-E885.

\noindent 
Keywords:\hspace{-0.05cm} Missing not at random; \hspace{-0.05cm}Stochastic approximation; \hspace{-0.05cm}European Social Survey
\end{abstract}


\section{Introduction}\label{sec:intro}

Missing data is a persistent problem in many kinds of research. It can
lead to bias and loss of information in the conclusions drawn from data.
For example, in this paper we examine cross-national comparisons
of people's perceptions about justice and fairness in their societies,
using data from the European Social Survey (ESS). There missingness
arises from nonresponse, when respondents are unable or unwilling to
answer some of the survey questions. 
Cross-national surveys like the ESS are an invaluable resource for understanding how attitudes and behaviors vary across populations, but their analysis is complicated by the nonresponse, especially when its 
levels and patterns also differ between countries.

Consequences of missingness can to some extent be mitigated in the
analysis of data. The most general and powerful method for doing this is
\emph{multiple imputation}. It works by generating multiple copies of
the missing data, as random draws from an \emph{imputation model} for
the missing data points given the observed ones. Information from them
is then combined to obtain estimates and inference for whatever
\emph{analysis models} that are of interest. Since it was first
proposed by \cite{rubin1978multiple,Rubin_1987}, multiple imputation
has developed into a wide-ranging and much-used methodology with a large
literature (see e.g.\ book-length treatments of missing data in general
or multiple imputation in particular, in \citealt{schafer1997analysis},
\citealt{molenberghs+kenward07}, \citealt{van2018flexible},
\citealt{little2019statistical}, 
and
\citealt{carpenteretal23}) and has become particularly popular for analysing survey data \citep[see e.g.,][]{
he2010multiple,
kohler2017dealing,
grund2021treatment,axenfeld2024performance}. 



Any implementation of multiple imputation requires decisions
about two key elements: specification and use of the imputation
model, and a method for combining the imputations to obtain estimates
for analysis models. 
In this paper we propose a framework 
which combines a general latent variable model for the imputation with an efficient frequentist approach for the estimation \citep{wang1998large,robins2000inference}.

In general, the imputation model should be implicitly or explicitly
based on a specification of a joint distribution for (at least) the
variables in the analysis models (we refer to this joint distribution
also as the imputation model). Imputed values for missing observations
of any variable are then drawn from the conditional distribution of that
variable given others that is implied by the joint distribution. The
imputation model should be such that it does not imply constraints on
the parameters of the analysis model \citep{meng1994multiple}. When
many different analysis models may be fitted using the same imputations,
this implies that the imputation model should be as saturated as
practically feasible.
Many such distributions have been proposed, including
multivariate normal and other basic parametric distributions
\citep[]{schafer1997analysis}, copula models
\citep[]{zhao2020missing,christoffersen2021asymptotically,feldman2024nonparametric}, and latent variable models
\citep[]{song2002analysis,song2004imputation,lee2004maximum,song2007bayesian,cai2010bayesian,zhang2020imputed,bai2021matrix,pan2021joint,cahan2023factor}. Other approaches, such as ones
based on machine learning prediction models \citep{lall2022midas,zhang2025enhancing}, game-theory-based methods \citep{yoon2018gain,awan2021imputation} or the
widely used method of fully conditional specification
\citep{vanbuuren2007,van2018flexible} can be seen as pragmatic
approximations which specify the conditional distributions directly,
even when these do not necessarily imply a proper joint distribution for
all the variables.


We propose another general class of imputation models. It is based on
the idea of joint distributions for observed variables that are induced
by \emph{latent variables}. Elsewhere such models are most often used
when the observed variables are in some sense on an equal footing, such
as when they are measures of the same latent
constructs or when the latent variables are random effects which explain
dependence between repeated observations of the same variable (see e.g.\
many examples in \citealt{skrondal2004generalized} and
\citealt{bartholomew2011latent}). However, the observed variables do not
need to be inherently similar in this way. 
Instead, latent variables can
also be used purely as a device which can induce a joint distribution for any
set of variables. This is the idea that we will utilize here. It has the
advantage of flexibility and scalability. The model is specified by a
collection of univariate conditional distributions for each observed
variable individually, given the latent variables. These can be defined
for each variable in turn, and in a suitable form for each. The observed
variables can thus be of mixed types, including binary, ordinal and
continuous variables. The richness of the model is determined by the
number of latent variables, which can be selected empirically. 
As an imputation model, this is most closely related to the previously mentioned latent variable models for imputation, but our models are of a more general form, which makes them suitable for more general missing data settings.

A key element in models and methods for dealing with missing data is the
assumed \emph{missingness mechanism}.  A crucial distinction is whether
data are Missing at Random (MAR) or Missing Not at Random (MNAR), that
is whether the probabilities of missingness do not or do depend on data
which are themselves unobserved
\citep{rubin1978multiple,little2019statistical}. MAR missingness is
usually also \emph{ignorable} for likelihood-based estimation of
analysis models, meaning that the estimation can be done without an explicit model
for the probabilities of missingness, whereas MNAR missingness is
correspondingly \emph{non-ignorable}. Most statistical methods for
accounting for missing data, including most commonly used imputation
models, assume MAR missingness. MNAR is more challenging, because
allowing for it requires also a model for the indicator variables for
the missingness, the form of which is generally unverifiable.
Nevertheless, MNAR is likely to occur often in practice, so methods that
would allow for at least some forms of it are desirable.

Our latent-variable imputation model is defined in a way which also
accommodates a form of MNAR missingness. This is done by introducing in
the model two sets of latent variables, not only one for the observed
variables of interest but also another for the missingness indicators.
If these two are associated, the missingness mechanism is MAR
conditional on the latent variables but MNAR (and thus non-ignorable)
with respect to the observed variables. This situation is also known as
`latent ignorability'
\citep[]{frangakis1999addressing,harel2009partial}. Different instances
of it in other contexts are described in \cite{kuha2018latent} and the
previous literature cited therein. In the rest of this paper we use the
term \emph{non-ignorable} to refer to missingness which is only latently
ignorable in this sense. This form of MNAR missingness is still
identifiable from observable data, under the latent variable model. More
general forms, where missingness is driven directly by missing
variables, are not identifiable and would require untestable assumptions
about the missingness mechanism. They are not considered in this paper.

There are different ways of using multiple imputation to estimate
analysis models. The most common is the Bayesian framework
which was originally introduced by \cite{rubin1978multiple,Rubin_1987}.
In it, the imputations are drawn from the (exact or approximate) posterior distribution of
the missing data given the observed data,
integrated over the posterior distribution of the parameters of the
imputation model. Analysis models are then estimated separately for each imputed dataset, using standard
complete-data methods, and
the results re combined using simple formulas (``Rubin's rules'')
to obtain final estimates and inference.

In this paper we apply, instead, another general method of estimation
with multiple imputation. This is the frequentist approach proposed by
\cite{wang1998large} and \cite{robins2000inference}. It divides the
tasks of imputation and estimation between different steps in a slightly
different way (we will return to this comparison in more detail in
Section \ref{sec:analysis_infer}). First, multiple imputations are drawn
from the conditional distribution of missing data given observed data,
with the parameters of the imputation model fixed at their maximum
likelihood (ML) estimates. Second, point estimates of the parameters of
an analysis model are obtained as complete-data estimates for the pooled
dataset obtained by stacking all the imputed datasets together. Finally,
correct standard errors for these estimates are calculated in a way
which accounts for all the elements of uncertainty in this procedure.
\cite{wang1998large} and \cite{robins2000inference} show that these
estimates are generally more efficient than ones obtained using Rubin's
rules. There are also some computational differences. The imputation
step is simpler in the approach we use, in that it does not require
sampling of parameters from their posterior distribution, but the
formulas for variance estimation are correspondingly
more complex than Rubin's rules.


The generality and flexibility of the proposed approach come at the
price of computational demands. Estimation, imputation and inference
rely on a marginal likelihood, after integrating out the latent
variables. As our model can involve many latent variables, dealing
with the marginal likelihood becomes a computational challenge. To
tackle it, we propose an efficient stochastic approximation algorithm.
It employs a Robins-Monro update paired with MCMC sampling methods, to
iteratively update the latent factors and the missing data given the
observed data and current estimates of model parameters. The algorithm
is applied twice, first to obtain ML estimates of the parameters of the
imputation model, and then to generate the multiple imputations.
Estimates of analysis models are then obtained by applying the results
of \cite{robins2000inference}, here modified to allow also the inclusion
of (survey) weights in the estimation. Under certain mild conditions,
the procedure is guaranteed to converge to the true parameter values.

We analyze data from
 the European Social Survey, from its ``Justice  and Fairness'' module from 2018.
The goal of this module was to understand 
how people
perceive justice for themselves and others in various domains of life,
and how these perceptions vary between different European countries
\citep{ESS9justice2020}. 
We focus on views on distributions of wealth and income, examining individuals'
normative definitions of justice regarding them,  perceptions of how well justice
and fairness are actually achieved in their societies, and how these perceptions
correlate with their own financial status. There is a substantial amount of nonresponse in these data, to an extent which varies much between countries and variables, and clear evidence that the nonresponse is 
non-ignorable. Our proposed procedure is used to impute values for five outcome variables involved in this analysis, using also information from 25 other variables. The results show that accounting for 
the missingness in this way leads to non-trivially different conclusions in some cases than 
complete-case analysis, particularly for variables with higher rates of nonresponse.



The proposed imputation model is described in
Section~\ref{sec:imputation-model} and the framework for estimation and
inference  for analysis models in Section~\ref{sec:analysis_est_infer}. 
Section~\ref{sec:computation} gives details of the computational procedures. Section~\ref{sec:simulation} summarizes a
simulation study of the performance of the method. The survey
example is described in Section~\ref{sec:real_data}. The paper concludes
with a discussion in Section~\ref{sec:discussion}. Some additional
information is given in Supplementary Materials.

\section{Imputation model induced by latent variables}\label{sec:imputation-model}


Consider vectors $\bfy_i=(y_{i1},\ldots, y_{iJ})^\top$ and
$\mathbf{x}_i=(x_{i1},\ldots,x_{ip})^\top$ of data for units $i=1,\ldots,N$.
They can consist of mixed data types, including continuous, binary, and
ordinal variables. Here all $\mathbf{x}_{i}$ are taken to be fully
observed, but any of $\bfy_i$ may be missing. Let
$\bfz_i=(z_{i1},\ldots, z_{iJ})^\top$ denote the corresponding
response  indicators, where $z_{ij}=1$ if $y_{ij}$ is observed and
$z_{ij}=0$ if $y_{ij}$ is missing. We partition each $\bfy_i$ into observed
components $\bfy_{i,obs}=\{y_{ij}: z_{ij}=1;\, j=1,\ldots,J\}$ and missing
components $\bfy_{i,mis}=\{y_{ij}: z_{ij}=0;\, j=1,\ldots,J\}$. The
full but partially unobserved data thus consist of $(\bfy_{i},
\bfz_i, \bfx_i)$ and the observed data of $(\bfy_{i,obs}, \bfz_i,
\bfx_i)$, for $i=1,\ldots,N$. Between them, $\mathbf{y}_{i}$ and
$\mathbf{x}_{i}$ should include all the variables which will be involved
in any subsequent analysis models, but they may also include additional
variables which are used only for predicting (imputing) values of
$\mathbf{y}_{i,mis}$.

\begin{figure}
    \centering
    \includegraphics[width=0.5\textwidth]{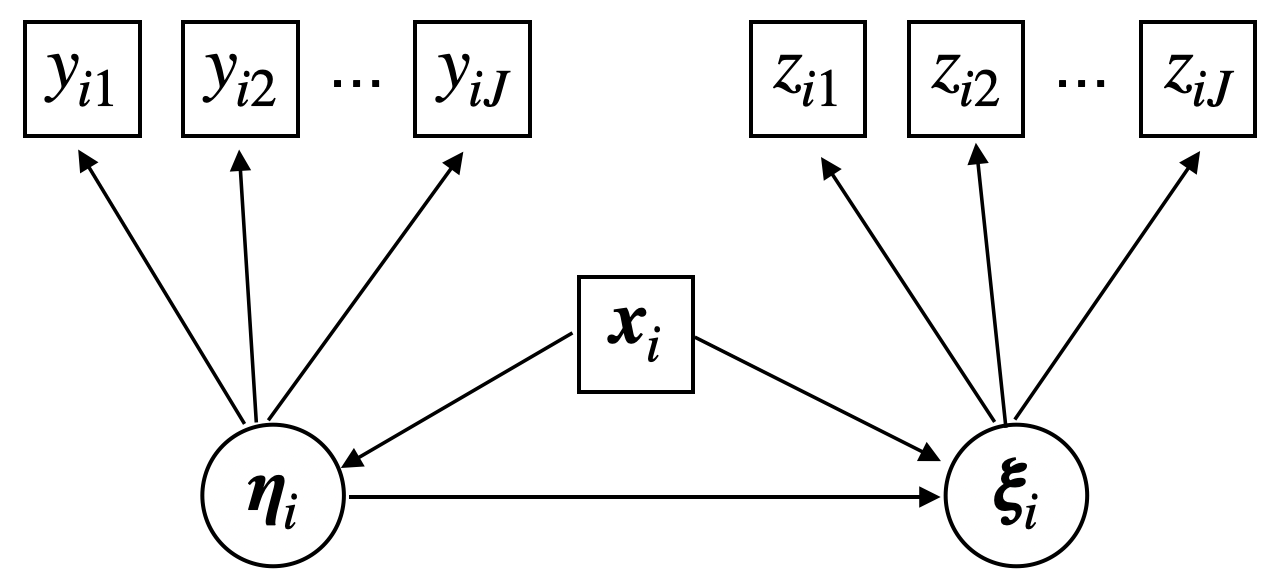}
    \caption{Path diagram of the imputation model. Here
    $\boldsymbol{\eta}_{i}$ and $\boldsymbol{\xi}_{i}$ are latent
    variables, and any of
    $\mathbf{x}_{i}$, $\boldsymbol{\eta}_{i}$ and $\boldsymbol{\xi}_{i}$
    can be vectors of variables.}
    \label{fig:path-diagram1}
\end{figure}

We specify the conditional joint distribution of $(\bfy_{i}, \bfz_i)$ given
$\bfx_i$ as being induced by
\begin{equation}\label{eq:joint_model}
    f(\bfy_i,\bfz_i,\bfeta_i,\bfxi_i\mid\bfx_i;\Psi) = \left[\prod_{j=1}^J g_j(y_{ij}\mid\bfeta_i;\Psi)\right]\left[\prod_{j=1}^J h_j(z_{ij}\mid\bfxi_i;\Psi)\right]\pi(\bfeta_i,\bfxi_i\mid\bfx_i;\Psi),
\end{equation}
where $f$, $g_j$, $h_j$ and $\pi$ are conditional density functions or
probability functions, $\Psi$ denotes all the free parameters, and
$\bfeta_i\in\mathbb{R}^{K_1}$ and $\bfxi_i\in\mathbb{R}^{K_2}$ are
latent variables that underlie the variables of interest and the
missingness indicators respectively. The model states that $\bfy_i$ is
conditionally independent of the other (latent and observed) variables
given $\bfeta_i$, and $\bfz_i$ is conditionally independent of the other
variables given $\bfxi_i$. It is also assumed that individual
variables $y_{ij}$ in $\bfy_i$ are independent of each other given
$\bfeta_i$, and different $z_{ij}$ in $\bfz_i$ are independent of
each other given $\bfxi_i$. Figure~\ref{fig:path-diagram1} shows the
path diagram of this model.


The joint distribution of the observable data
$(\bfy_{i}, \bfz_i)$ given $\bfx_i$ is obtained from
(\ref{eq:joint_model}) by integrating out the latent variables, as
\begin{equation}
    f(\bfy_{i},\bfz_i\mid\bfx_i;\Psi) =
    \int\!\!\int f(\bfy_i,\bfz_i,\bfeta_i,\bfxi_i\mid\bfx_i;\Psi)
   \; d\bfeta_i d\bfxi_i,
\label{eq:joint_allobs}
\end{equation}
and that of the actually observed data by also
integrating over the missing responses, as
\begin{equation}
    f(\bfy_{i,obs},\bfz_i\mid\bfx_i;\Psi) =
    \int f(\bfy_{i},\bfz_i\mid\bfx_i;\Psi)\,
   d\bfy_{i,mis}.
\label{eq:joint_obs}
\end{equation}
Even though (\ref{eq:joint_model}) involves various conditional
independencies given the latent variables, in (\ref{eq:joint_allobs})
all of $\bfy_i$ and $\bfz_i$ are associated given $\bfx_i$. In other
words, the dependence on the latent variables can induce associations
between all of the observable variables in their joint distribution
(\ref{eq:joint_allobs}). This also means that
conditional distributions of $\mathbf{y}_{i,mis}$, which will be used
for imputing values for them, will in general depend on all of
$(\mathbf{y}_{i,obs},\mathbf{z}_{i},\mathbf{x}_{i})$. The strengths and
patterns of these associations depend on the parameters of the models
and on the dimensions $K_{1}$ and $K_{2}$ of the latent variables
$\bfeta_i$ and $\bfxi_i$.

We take the latent variables to be jointly normally distributed given $\bfx_i$, so that $\pi(\bfeta_i,\bfxi_i\mid\bfx_i;\Psi)$ is defined by
\begin{equation}\label{eq:eta_xi}
   \bfeta_i\mid\bfx_i \sim \mathcal{N}\left(\bfbeta\bfx_i,\mathbf{I}\right)
\;\mbox{~and~} \;
 \bfxi_i\mid\bfeta_i,\bfx_i \sim \mathcal{N}\left(\bfzeta\bfx_i + \bfkappa\bfeta_i , \mathbf{I}\right),
\end{equation}
where $\bfbeta\in\mathbb{R}^{K_1\times p}$,
$\bfzeta\in\mathbb{R}^{K_2\times p}$ and
$\bfkappa\in\mathbb{R}^{K_2\times K_1}$ are coefficients.
For identifiability, covariance matrices of both 
distributions in \eqref{eq:eta_xi} are constrained to be identity
matrices.

The parameters $\bfkappa$ determine the associations between the two
sets of latent variables. If $\bfkappa\ne \mathbf{0}$, in
(\ref{eq:joint_allobs}) each $z_{ij}$ is associated (even given
$\bfx_i$) with all of $\bfy_i$, including the unobserved $\bfy_{i,mis}$. This is the `latent ignorability' MNAR
specification that we discussed in Section \ref{sec:intro}. The model
reduces to MAR when $\bfkappa=\mathbf{0}$, so both of these missingness
mechanisms are accommodated by (\ref{eq:joint_model}). If the analysis
is carried out under the MAR assumption, the imputation distributions
for $\mathbf{y}_{i,mis}$ will not depend on $\mathbf{z}_{i}$. All
elements of the models which involve $\boldsymbol{\xi}_{i}$ or
$\mathbf{z}_{i}$ could then be omitted from the specifications below and
from the model estimation (or even if they are included, they will play
no role in the imputation).


Different types of variables $y_{ij}$ can be accommodated by specifying
the models for them given $\boldsymbol{\eta}_{i}$ appropriately. We consider examples where $y_{ij}$ is continuous, binary or ordinal:



\begin{enumerate}
    \item \textbf{Normal linear model} for continuous response $y_{ij}\in\mathbb{R}$: 
    $g_j(y_{ij} \mid \bfeta_i; \Psi) = \sqrt{2\pi\sigma_j^2} \exp\left\{ - (y_{ij} - \alpha_{0j} - \bfalpha_j^\top \bfeta_i)^2 / (2\sigma_j^2) \right\}$.

    \item \textbf{Logistic model} for binary response $y_{ij}\in\{0,1\}$: 
    $g_j(y_{ij}=1| \bfeta_i; \Psi) = 
    \sigma(\alpha_{0j}+\bfalpha_j^\top\bfeta_i),$ where $\sigma(z) = 1 / (1 + \exp(-z)).$

    \item \textbf{Ordinal logistic model} for ordinal response $y_{ij} \in \{0, \ldots, M_j\}$: 
    $g_j(y_{ij} \mid \bfeta_i; \Psi) = \sigma(\bfalpha_j^\top \bfeta_i - \alpha_{0j,y_{ij}}) - \sigma(\bfalpha_j^\top \bfeta_i - \alpha_{0j,y_{ij}+1})$.
\end{enumerate}
Here $\boldsymbol{\alpha}_{j}=
(\alpha_{j1},\dots,\alpha_{jK_{1}})^{\top} $ are factor loadings,
$\alpha_{0j}$ and $\alpha_{0j,y}$ are intercept parameters,
and $\sigma_{j}$ in the normal linear model are conditional
variances. The model
can be defined in a comparable way also for other types of variables,
such as counts or nominal categorical variables. Accommodating them in the
imputation would require only corresponding modifications to the
estimation procedures that we describe in Section~\ref{sec:computation}.

As the model for each binary response indicator $z_{ij}\in\{0,1\}$, we assume
the logistic model $
    h_j(z_{ij}=1\mid\bfxi_i;\Psi) = \sigma(\gamma_{0j}+\bfgamma_j^\top\bfxi_i),$
where
$\boldsymbol{\gamma}_{j}=(\gamma_{j1},\dots,\gamma_{jK_{2}})^{\top} $
are factor loadings. These models are only identifiable up to
an orthogonal rotation of the loading vectors $\bfalpha_j$ and
$\bfgamma_j$, as usual for latent variable models \citep[see,
e.g.,][]{liu2023rotation}.
We fix this indeterminacy by imposing minimal zero constraints on the
loading parameters in the estimation, for example by
setting $\alpha_{jk}=0$ and
$\gamma_{jk}=0$ for all $k>j$.

We use maximum likelihood (ML)
estimation of $\mathbf{\Psi}$, assuming that  $(\bfy_{i}, \bfz_i)$ given $\bfx_i$ are independent across~$i$.
The ML estimate $\hat{\Psi}$
is obtained by maximizing the observed-data log-likelihood, as
\begin{equation}\label{eq:mml}
    \hat\Psi = \arg\max_\Psi \left\{
    \sum_{i=1}^N \log f(\bfy_{i,obs},\bfz_i\mid\bfx_i;\Psi)
    \right\},
\end{equation}
where $f(\bfy_{i,obs},\bfz_i\mid\bfx_i;\Psi)$ is
defined in \eqref{eq:joint_obs}. This is computationally nontrivial
because each term
in \eqref{eq:mml} involves a possibly high-dimensional integral. Based on this estimated imputation
model, we will want to draw multiple imputed values for
$\mathbf{y}_{i,mis}$ from the conditional distribution
$f(\mathbf{y}_{i,mis}|\mathbf{y}_{i,obs},\mathbf{z}_{i}, \mathbf{x}_{i};
\hat{\boldsymbol{\Psi}})$ implied by the joint model
\eqref{eq:joint_model}. How these computational steps can be carried out
is described in Section \ref{sec:computation}, after we have also
introduced the elements of estimation of analysis models in Section
\ref{sec:analysis_est_infer}. We note also that in maximizing
\eqref{eq:mml} we take the dimensions $K_1$ and $K_2$ of the latent variables
$\bfeta_i$ and $\bfxi_i$ to be known and fixed. In practice, however,
they are initially unknown and need to be chosen empirically. We
select them by first estimating $\Psi$ for different choices of $(K_1,K_2)$ and
choosing the solution which minimizes the Bayesian Information Criterion
\citep[BIC;][]{schwarz1978estimating}.

\section{Estimating analysis models using imputed data}\label{sec:analysis_est_infer}


\subsection{Analysis models}

Suppose that the aim of the analysis is the estimation of population parameters $\bftheta\in\mathbb{R}^q$ and that 
if there was no missing data, $\boldsymbol{\theta}$ would
be estimated by solving the estimating equations
\begin{equation}\label{eq:estimating_eq_nomissing}
    \sum_{i=1}^N w_i\, U(\bfy_i,\mathbf{x}_{i},\bftheta) = \boldsymbol{0},
\end{equation}
where  $U(\bfy_i,\mathbf{x}_{i},\bftheta)\in\mathbb{R}^q$ is an
estimating function and $w_i$ is a weight for unit $i$.
We assume that as the
sample size $N$ tends to infinity, the solution to
\eqref{eq:estimating_eq_nomissing} uniquely exists and converges to a
limit point $\bftheta_0$, which is taken to be the ``true value'' of the
parameters of interest.

The notation  indicates that $\bftheta$ are different from the
parameters $\boldsymbol{\Psi}$ of the imputation model defined in
Section \ref{sec:imputation-model}. For consistency we retain the notational
distinction between those variables which are subject to missingness in
observed data ($\mathbf{y}_{i}$) and those that are not
($\mathbf{x}_{i}$), but we note that this distinction is irrelevant for
analysis models where either kinds of variables may appear in any roles.
The weights $w_{i}$ are included in (\ref{eq:estimating_eq_nomissing}),
in particular, to allow for the possibility of using survey weights in
the analysis of survey data, as in our examples in Section
\ref{sec:real_data}. Unweighted estimators are obtained by setting all
$w_i = 1$ throughout.

To make this setting clearer, we give some examples of common analysis
models. When $\bftheta$ is the vector of means of functions $r(\bfy_i)$,
we have $U(\bfy_i,\mathbf{x}_{i},\bftheta) = r(\bfy_i) - \bftheta$ and the estimator
obtained from \eqref{eq:estimating_eq_nomissing} is the vector of
weighted means ${\sum_{i=1}^N w_i r(\bfy_i)}/{(\sum_{i=1}^N w_i)}$. 
When the parameter of interest is the correlation between two variables in
$\bfy_i$, its estimator is the standard
(weighted) sample correlation.
When the parameters are the
coefficients of the best linear predictor of one variable $y_{ij}$ given
the rest of $(\mathbf{y}_{i},\mathbf{x}_{i})$ (here denoted
$\mathbf{X}_{i}$ and taken to include a constant 1), least
squares estimates are obtained by setting
$U(\bfy_i,\mathbf{x}_{i},\bftheta)=(y_{ij}-\bftheta^{\top}\mathbf{X}_{i})\mathbf{X}_{i}$.
Still more generally, if the parameters of the analysis model are to be
estimated using maximum likelihood estimation, $U(\bfy_i,
\mathbf{x}_{i},\bftheta)$ is
the score function (first derivative of the log likelihood function) for
that model contributed by unit~$i$ (if survey weights $w_i$ are also included, this
gives pseudo-likelihood estimation in survey analysis).

\subsection{Multiple imputation  estimator}\label{sec:analysis_infer}

Suppose now that $M$ sets of imputed values have been generated for
the missing data $\bfy_{i,mis}$ for each unit $i$, drawn from the
conditional joint distributions
\begin{equation}\label{eq:implied_cond}
    \phi(\bfy_{i,mis},\bfeta_i,\bfxi_i\mid\bfy_{i,obs},\bfz_i,\bfx_i;\Psi) =
    \frac{f(\bfy_i,\bfz_i,\bfeta_i,\bfxi_i\mid\bfx_i;\Psi)}{ f(\bfy_{i,obs},\bfz_i\mid\bfx_i;\Psi)},
\end{equation}
implied by the latent variable model~\eqref{eq:joint_model}, and
with $\Psi$ set at their maximum likelihood estimates
$\hat{\Psi}$ from (\ref{eq:mml}) (the estimation and imputation will be described in Section \ref{sec:computation}).
The imputation procedure outputs
$(\mathbf{y}^{m},\mathbf{x},\boldsymbol{\eta}^{m},\boldsymbol{\xi}^{m})=
\{(\bfy_1^{m},\mathbf{x}_{1},\bfeta_1^{m},\bfxi_1^{m}),\ldots,
(\bfy_N^{m},\mathbf{x}_{N},\bfeta_N^{m},\bfxi_N^{m})\}$ as the imputed
dataset $m=1,\dots,M$, where $\bfy_i^m=(\bfy_{i,obs},\bfy_{i,mis}^m)$.
The $(\bfy^{m},\mathbf{x})$ are used for point estimation of the
parameters $\boldsymbol{\theta}$ of analysis models, and estimation of
their variances also involves the sampled latent variables
$(\bfeta^{m},\bfxi^{m})$.

Define $U_i^m(\hat\Psi,\bftheta) =
U(\bfy_i^m,\mathbf{x}_{i},\bftheta)$, where we include $\hat\Psi$ in the notation to
emphasize that $\bfy_i^m$ are drawn from the imputation model with
these parameter estimates.
The multiple imputation point estimate
$\hat\bftheta$ of $\bftheta$ is obtained by solving
\begin{equation}
\label{eq:analysis_estimating_equation}
    \sum_{i=1}^N \sum_{m=1}^M w_i\, U_i^m(\hat\Psi,\bftheta) =
    M\sum_{i=1}^N w_i\,\bar{U}_i(\hat\Psi,\bftheta) = \boldsymbol{0},
\end{equation}
where $\bar{U}_i(\hat\Psi,\bftheta) = M^{-1}\sum_{m=1}^M
U_i^m(\hat\Psi,\bftheta)$. The two expressions in
\eqref{eq:analysis_estimating_equation} show how this aggregates
information across the imputed datasets. The second expression shows
that the estimating function $\bar{U}_i(\cdot)$ is an average of the
$U_i^m(\cdot)$ across these datasets. The first expression has the form
of the original estimating function in~\eqref{eq:estimating_eq_nomissing}, but calculated for the dataset of
$NM$ observations that is obtained by pooling
$(\bfy^1,\mathbf{x}),\dots,(\bfy^M,\mathbf{x})$
together and treating them as if they were all independent observations.
This is convenient, because $\hat{\bftheta}$ can be
obtained by applying standard (complete-data) estimation to this
pooled dataset.

Here we pause to compare the formulas for this approach with the commonly used ``Rubin's rules'' for multiple-imputation
estimation. For the latter, point estimation is done by
first solving $\sum_{i} w_i\, U_i^m(\hat\Psi,\bftheta) = \mathbf{0}$ to
get $\hat{\boldsymbol{\theta}}^{m}$ for $m=1,\dots,M$, and the final
estimate is their mean $\hat{\boldsymbol{\theta}}=M^{-1}\sum_{m}
\hat{\boldsymbol{\theta}}^{m}$. Rubin's rules thus pool
estimates that have first been calculated separately for each imputed
dataset, whereas (\ref{eq:analysis_estimating_equation}) pools the
datasets first and carries out the estimation once for all of them
together. There is an analogous but somewhat larger difference in how
variance estimation should be carried out. With Rubin's rules this takes
the form
$\hat{\text{var}}(\hat{\boldsymbol{\theta}})=\mathbf{V}+\mathbf{B}$
where
$\mathbf{V}=M^{-1}\sum_{m}\hat{\text{var}}(\hat{\boldsymbol{\theta}}^{m})$
and $\mathbf{B}$ desribes the variability of
$\hat{\boldsymbol{\theta}}_{1},\dots,\hat{\boldsymbol{\theta}}_{M}$. The
term $\mathbf{B}$ captures the uncertainties arising both from the
imputations being random variables and from estimation of the parameters
$\Psi$ of the imputation model. It can do this because Rubin's rules are
designed to be used with imputations that are drawn from the posterior
distribution of the missing data, which thus reflects also uncertainty
about $\Psi$.
Our imputations, in contrast, are conditional on the point estimate
$\hat{\Psi}$, so the uncertainty in $\hat{\Psi}$ should be carried
forward in another way. How this is done is motivated by the asymptotic properties of our multiple-imputation estimator, as given by Theorem~\ref{thm:analysis_infer} below. 
Its assumptions and proof are given in Supplementary Section~A.1.
They are based on the theory in
\cite{robins2000inference}, with the minor extension to allow explicitly
also for weights in estimating equations of the analysis model.






\begin{theorem}\label{thm:analysis_infer}
    Under
    Assumptions~S1-S8,
    the estimator of the analysis model parameters $\hat\bftheta$ that
    solves \eqref{eq:analysis_estimating_equation} is asymptotically normally distributed, with
        $\sqrt{(\sum_{i=1}^N w_i)^2/(\sum_{i=1}^N w_i^2)}\;(\hat\bftheta - \bftheta^*)$ converging in distribution to $\mathcal{N}(0,\Sigma)$,
    where $\bftheta^*$ is the limiting point stated in Assumption~S3.
     The asymptotic variance-covariance matrix is given by $\Sigma = \tau^{-1}\Omega\tau^{-1},$ where
     \begin{footnotesize}
    \begin{gather*}
        \tau = -\mathbb{E}\left[\frac{\partial}{\partial\bftheta^\top}\bar{U}_i(\Psi^*,\bftheta)\right]\Bigg\vert_{\bftheta = \bftheta^*},\quad\kappa = \mathbb{E}\left[U_i^m(\Psi^*,\bftheta^*)S_{i,mis}^m(\Psi^*)^\top\right],\quad\Lambda(\Psi^*) = \mathbb{E}\left[D_i(\Psi^*)D_i(\Psi^*)^\top\right]\\
        \Omega = \mathbb{E}\left[\bar{U}_i(\Psi^*,\bftheta^*)\bar{U}_i(\Psi^*,\bftheta^*)^\top\right] + \kappa\Lambda(\Psi^*)\kappa^\top + \mathbb{E}\left[\kappa D_i(\Psi^*)\bar{U}_i(\Psi^*,\bftheta^*)^\top + \left(\kappa D_i(\Psi^*)\bar{U}_i(\Psi^*,\bftheta^*)^\top\right)^\top\right].
    \end{gather*}
    \end{footnotesize}The expectations above are taken over the joint density of observed and imputed variables, which is determined by the imputation model and the true data-generating process. This density, along with formal definitions of the missing information score $S_{i,mis}^m(\Psi^*)$ and the influence function $D_i(\Psi^*)$, is provided in Supplementary Section~A.1.
\end{theorem}

    The asymptotic variance matrix $\Sigma$ takes a sandwich form, where
the form of $\tau$ is standard for estimating equations, coming from the
Taylor expansion of the score equations. The inner term $\Omega$
computes the covariance matrix of $\bar{U}_i(\Psi^*,\bftheta^*) + \kappa
D_i(\Psi^*)$. Here, $\bar{U}_i(\Psi^*,\bftheta^*)$  captures both the
sampling variability and the uncertainty of multiple imputations.  In
addition,  the term $\kappa D_i(\Psi^*)$ quantifies the additional
variability brought by the estimation of the imputation model. As the
same data is used for estimating the imputation and analysis models,
$\bar{U}_i(\Psi^*,\bftheta^*)$  and $\kappa D_i(\Psi^*)$ are correlated,
resulting in the cross terms in the expression for $\Omega$. 
A consistent plug-in estimator for $\Sigma$ is given in Supplementary Section~A.2.

\section{Computation: Estimation and imputation}\label{sec:computation}

Computational implementation of the proposed method requires three
elements: calculating maximum likelihood estimates of the parameters of
the imputation model, generating multiple imputations  from the
estimated imputation model, and calculating point estimates and variance
estimates of the parameters of an analysis model. These steps are
discussed in Sections \ref{ss_comp_impmodel}--\ref{ss_comp_analysis}
respectively, with some details relegated to Supplementary
Material.

\subsection{Estimation of imputation model: Stochastic approximation}
\label{ss_comp_impmodel}


The maximum likelihood estimator for the imputation model is defined by
\eqref{eq:mml}. Direct evaluation of the marginal log-likelihood and its
gradient are computationally infeasible in general, due to the
intractable high-dimensional integrals.
To tackle this problem, we employ a stochastic approximation approach.
This is motivated by the fact that
\begin{equation*}
\label{eq:imputer_stochastic_gradient}
    \frac{\partial}{\partial\Psi} \left(\sum_{i=1}^N \log f(\bfy_{i,obs},\bfz_i\mid\bfx_i;\Psi)\right)  = \sum_{i=1}^N \mathbb{E}\left[ \frac{\partial}{\partial\Psi}\,
    \log f(\bfy_i,\mathbf{z}_{i},\bfeta_i,\bfxi_i\mid \bfx_i;\Psi)\mid
    \bfy_{i,obs},\bfz_i,\bfx_i;\Psi\right],
\end{equation*}
where the expectation is with respect to
$\mathbf{y}_{i,mis},\boldsymbol{\eta}_{i},\boldsymbol{\xi}_{i}$ under
the distribution defined in~\eqref{eq:implied_cond},
for $i=1,\dots, N$. By drawing samples
of $(\bfy_{i,mis},\bfeta_i,\bfxi_i)$ from this
conditional distribution, we can use $\sum_{i=1}^N \partial \log
f(\bfy_i,\mathbf{z}_{i},\bfeta_i,\bfxi_i\vert \bfx_i;\Psi)/\partial\Psi$ as
a stochastic gradient of the marginal likelihood in \eqref{eq:mml} and
solve the optimization problem by a stochastic-gradient-based  approach.

This procedure is described in Algorithm~\ref{alg:imputer_sa}. It
alternates between two steps: drawing samples of the missing
values and latent variables from the conditional distribution
{$\phi(\bfy_{i,mis},\bfeta_i,\bfxi_i|\bfy_{i,obs},\bfz_i,\bfx_i;\Psi)$}
evaluated under the current parameter value, and updating $\Psi$ based
on a stochastic gradient constructed using the sampled missing values
and latent variables. The final estimate of $\Psi$ is given by its
average over the iterations, following
\cite{ruppert1988efficient} and \cite{polyak1992acceleration}. The
sampling is done using a Gibbs sampler which generates values of $\bfeta_i$, $\bfxi_i$, and $\bfy_{i,mis}$ successively. This is straightforward for $\bfy_{i,mis}$ but less so for 
$\bfeta_i$ and $\bfxi_i$. To do it efficiently, we develop a Gibbs sampler by
extending the P{\'o}lya-Gamma augmentation scheme for binary response
data \citep{polson2013bayesian} to the current setting with mixed data
types. Its details are described in Supplementary Section~B.
In the stochastic gradient update, $\rho_t > 0$ is a
step size satisfying $\sum_{t}\rho_t =\infty$ and
$\sum_t\rho_t^2<\infty$, which holds if we let $\rho_t$ decay at the
rate $t^{-c}$ for some $c \in (0.5, 1]$. In the implementation, we set
$\rho_{t}=0.01\, t^{-0.51}$ where $c=0.51$ is chosen to ensure that the
step size does not decay too quickly. We
select the values of $T_0$ and $T$ by examining trace plots of the
latent variables and parameter values to ensure MCMC convergence and
mixing. In our simulation studies, setting $T_0=1000$ and $T=3000$
achieved stable convergence.

\begin{algorithm}[hbt!]
    \SetKwInOut{Output}{Output}
    \caption{Estimation of the parameters $\Psi$ of the imputation model
    using stochastic approximation}\label{alg:imputer_sa}
    \setstretch{1.2}
    \KwData{
    Observed data $(\mathbf{y}_{i,obs}, \mathbf{z}_{i}, \mathbf{x}_{i})$
    and initial values of the missing observations $\bfy_{i,mis}^{(0)}$
    and of latent variables $\bfeta_i^{(0)},\bfxi_i^{(0)}$ for
    $i=1,\ldots,N$; initial values of model parameters $\Psi^{(0)}$;
    number of MCMC iterations $T$ and of burn-in iterations $T_0$; step
    size $\rho_{t}=A\,t^{-c}$.}
    \For{each iteration $t = 1$ to $T$}{


    \For{each $i = 1$ to $N$}{
    Given $\bfy_{i,mis}^{(t-1)}, \bfy_{i,obs},\bfz_i, \bfxi_i^{(t-1)}, \Psi^{(t-1)},\bfx_i$, sample $\bfeta_i^{(t)}$ using an MCMC kernel with its invariance distribution being
    \begin{equation}\label{eq:posterior_eta}
        p(\bfeta_i\mid\bfy_{i,mis}^{(t-1)},\bfy_{i,obs},\bfz_i,\bfx_i,\bfxi_i^{(t-1)}; \Psi^{(t-1)})\propto \phi(\bfy_{i,mis}^{(t-1)},\bfeta_i,\bfxi_i^{(t-1)}\mid\bfy_{i,obs},\bfz_i,\bfx_i;\Psi^{(t-1)}).
    \end{equation}

    Given $\bfy_{i,mis}^{(t-1)},\bfy_{i,obs},\bfz_i,\bfeta_i^{(t)},\Psi^{(t-1)},\bfx_i$, sample $\bfxi_i^{(t)}$ using an MCMC kernel with its invariance distribution being
    \begin{equation}\label{eq:posterior_xi}
        p(\bfxi_i\mid\bfy_{i,mis}^{(t-1)},\bfy_{i,obs},\bfz_i,\bfx_i,\bfeta_i^{(t)};\Psi^{(t-1)})\propto \phi(\bfy_{i,mis}^{(t-1)},\bfeta_i^{(t)},\bfxi_i\mid\bfy_{i,obs},\bfz_i,\bfx_i;\Psi^{(t-1)}).
    \end{equation}

    Given $\bfeta_i^{(t)},\Psi^{(t-1)}$,
    sample $y_{ij}^{(t)}$ from $g_j(y_{ij}\mid\bfeta_i^{(t)};\Psi^{(t-1)})$,
    for missing observations, i.e.\ ones for which $z_{ij}=0$, $j=1,\ldots,J$.
    }

    Given
    $\{\bfy_{i}^{(t)},\bfz_i,\bfeta_i^{(t)},\bfxi_i^{(t)},\bfx_i:i=1,\ldots,N\}$,
where
    $\bfy_i^{(t)} = (\bfy_{i,obs},\bfy_{i,mis}^{(t)})$,
    update parameters $\Psi$ of the imputation model using stochastic approximation, i.e.,
    \begin{align}\label{eq:sgd}
            \Psi^{(t)} &= \Psi^{(t-1)} +
            \rho_t \sum_{i=1}^N \frac{\partial}{\partial\Psi}
            \log
            f(\bfy_i^{(t)},\bfz_i,\bfeta_i^{(t)},\bfxi_i^{(t)}\mid\bfx_i; \Psi)\bigg\vert_{\Psi=\Psi^{(t-1)}}.
    \end{align}
    }
    \Output{Trajectory average $\sum_{t=T_0+1}^T\Psi^{(t)}/(T-T_0)$.
    This is taken forward as the estimate $\hat{\Psi}$ of~$\Psi$.
    }
\end{algorithm}

\subsection{Multiple imputation {and additional computational details}}\label{sec:computation_auxiliary}


After obtaining $\hat \Psi$ using Algorithm~\ref{alg:imputer_sa}, we
generate multiple imputed datasets using the imputation model
with its parameters fixed at $\hat \Psi$. This is done by the same MCMC
algorithm as in Algorithm~\ref{alg:imputer_sa}, except that now $\Psi$
is fixed at $\hat{\Psi}$ in every iteration. For clarity, this is
described separately as Algorithm~\ref{alg:imputer_impute_aux}. Here the
burn-in and total iteration lengths $T_0$ and $T$ can differ from their
choices in Algorithm~\ref{alg:imputer_sa}. We determine appropriate
values by examining the trace plots of the Markov chain. Specifically,
the values of $T_0=1000$ and $T=3000$ worked well in our simulation
studies. To reduce the autocorrelation between the imputations,
it is also desirable to perform thinning by retaining only
every $k$th imputed dataset. The
number of imputed datasets from Algorithm~2 is then $M=(T-T_{0})/k$. In
our implementation, we set $k=100$ (and thus $M=20$), which achieved a
good balance between computational efficiency and reduction of the
autocorrelation. Further computational details for
Algorithm~\ref{alg:imputer_impute_aux} can be found in Supplementary Section~B.


\begin{algorithm}[hbt!]
    \SetKwInOut{Output}{Output}
    \caption{MCMC procedure for multiple imputation}\label{alg:imputer_impute_aux}
    \setstretch{1.2}
    \KwData{
    Observed data $(\mathbf{y}_{i,obs}, \mathbf{z}_{i}, \mathbf{x}_{i})$
    and initial values of the missing observations $\bfy_{i,mis}^{(0)}$
    and of latent variables $\bfeta_i^{(0)},\bfxi_i^{(0)}$ for
    $i=1,\ldots,N$;
estimate  $\hat{\Psi}$ of the parameters of the imputation model;
    number of MCMC iterations $T$ and of burn-in iterations $T_0$;
thinning interval $k$.}
    \For{each $i = 1$ to $N$}{

    {
    \For{each iteration $t = 1$ to $T$}{


    Given $\bfy_{i,mis}^{(t-1)}, \bfxi_i^{(t-1)}, \hat\Psi, \bfy_{i,obs},\bfx_i$, sample $\bfeta_i^{(t)}$ using an MCMC kernel with its invariance distribution being
    \begin{equation}\label{eq:posterior_eta1}
        p(\bfeta_i\mid\bfy_{i,mis}^{(t-1)},\bfy_{i,obs},\bfx_i,\bfxi_i^{(t-1)}; \hat\Psi)\propto \phi(\bfy_{i,mis}^{(t-1)},\bfeta_i,\bfxi_i^{(t-1)}\mid\bfy_{i,obs},\bfz_i,\bfx_i;\hat\Psi).
    \end{equation}

    Given $\bfz_i,\bfeta_i^{(t)},\hat\Psi,\bfx_i$, sample $\bfxi_i^{(t)}$ using an MCMC kernel with its invariance distribution being
    \begin{equation}\label{eq:posterior_xi1}
    p(\bfxi_i\mid\bfz_i,\bfeta_i^{(t)},\bfx_i;\hat\Psi)\propto \phi(\bfy_{i,mis}^{(t-1)},\bfeta_i^{(t)},\bfxi_i\mid\bfy_{i,obs},\bfz_i,\bfx_i;\hat\Psi).
    \end{equation}
    Given $\bfeta_i^{(t)},\hat\Psi$,
    sample $y_{ij}^{(t)}$ from $g_j(y_{ij}\mid\bfeta_i^{(t)};\hat{\Psi})$
    for missing observations, i.e.\ ones for which $z_{ij}=0$, $j=1,\ldots,J$.

    }
    Given
    $\bfy_{i,mis}^{(t)},\bfeta_i^{(t)},\bfz_i,\bfxi_i^{(t)},\bfy_{i,obs},\bfx_i,\hat\Psi,
    t=1,\ldots,T,$ compute the following quantities for $t > T_0$,
    $i=1,\dots,N$:
    \begin{align}
        S_i^{(t)} &= \frac{\partial}{\partial\Psi} \log f(\bfy_{i}^{(t)},\bfz_i,\bfeta_i^{(t)},\bfxi_i^{(t)}\mid\bfx_i;\Psi)\big\vert_{\Psi=\hat\Psi},\label{eq:s_it}\\
        H_i^{(t)} &= \frac{\partial^2}{\partial\Psi\partial\Psi^\top} \log f(\bfy_{i}^{(t)},\bfz_i,\bfeta_i^{(t)},\bfxi_i^{(t)}\mid\bfx_i;\Psi)\big\vert_{\Psi=\hat\Psi}.\label{eq:h_it}
    \end{align}
where
    $\bfy_i^{(t)} = (\bfy_{i,obs},\bfy_{i,mis}^{(t)})$.     Then calculate
    \begin{align}
        \bar{S}_{i,obs} &= \frac{1}{T-T_0}\sum_{t=T_0+1}^T S_{i}^{(t)},\label{eq:s_iobst}\\
        \bar{I}_{obs} &= \frac{1}{N}\sum_{i=1}^N \left\{ \bar{S}_{i,obs} \bar{S}_{i,obs}^\top
        - \frac{1}{T-T_0}\sum_{t=T_0+1}^T
        \left[H_i^{(t)}+S_i^{(t)}S_i^{(t)\top}\right]\right\}.
        \label{eq:l_it}
    \end{align}
     }
    } 
    \Output{$M=(T-T_{0})/k$ imputed datasets
    $\{\mathbf{y}_{i}^{m}: \, i=1,\dots,N\}$
    where $\bfy_i^m =
    \bfy_i^{(T_0 + mk)}$ for $m=1,\dots,M$;
    $\bar S_{i, obs}$ and $\bar I_{obs}$ that approximate
$S_{i,obs}(\hat \Psi)$ and $\hat I_{obs}$, respectively, for
$i=1,\dots,N$.}

\end{algorithm}

\subsection{Estimation of analysis model}
\label{ss_comp_analysis}

Apart from the imputed values $\mathbf{y}_{i,mis}^{m}$ for
$m=1,\dots,M$, Algorithm 2 produces also approximations $\bar
S_{i, obs}$ of $S_{i,obs}(\hat \Psi)$ ($i=1, ..., N$) and $\bar I_{obs}$
of $\hat I_{obs}$. As defined in Section
\ref{sec:analysis_infer}, these are quantities which are derived from
the imputation model but which are needed for calculation of estimated
variance matrices of the estimates of the parameters
$\boldsymbol{\theta}$ of (any) analysis models. Estimation of
$\boldsymbol{\theta}$ then proceeds as described in Section
\ref{sec:analysis_infer}, point estimation by solving
(\ref{eq:analysis_estimating_equation}) and variance estimation using the
quantities in Supplementary Section~A.2.


\section{Simulation studies}\label{sec:simulation}
We conducted simulation studies to evaluate the proposed method across various scenarios, which are summarized in Figure~\ref{fig:simu_settings} and detailed in Supplementary Section~C.1.
In each setting the data included a mixture of continuous and binary variables, and 
the estimands of interest were the marginal means of each variable (Supplementary Section~C.4 also reports results when the estimands were conditional means, with essentially similar conclusions). 
\begin{figure}[t!]
    \centering
    \includegraphics[width=.8\linewidth]{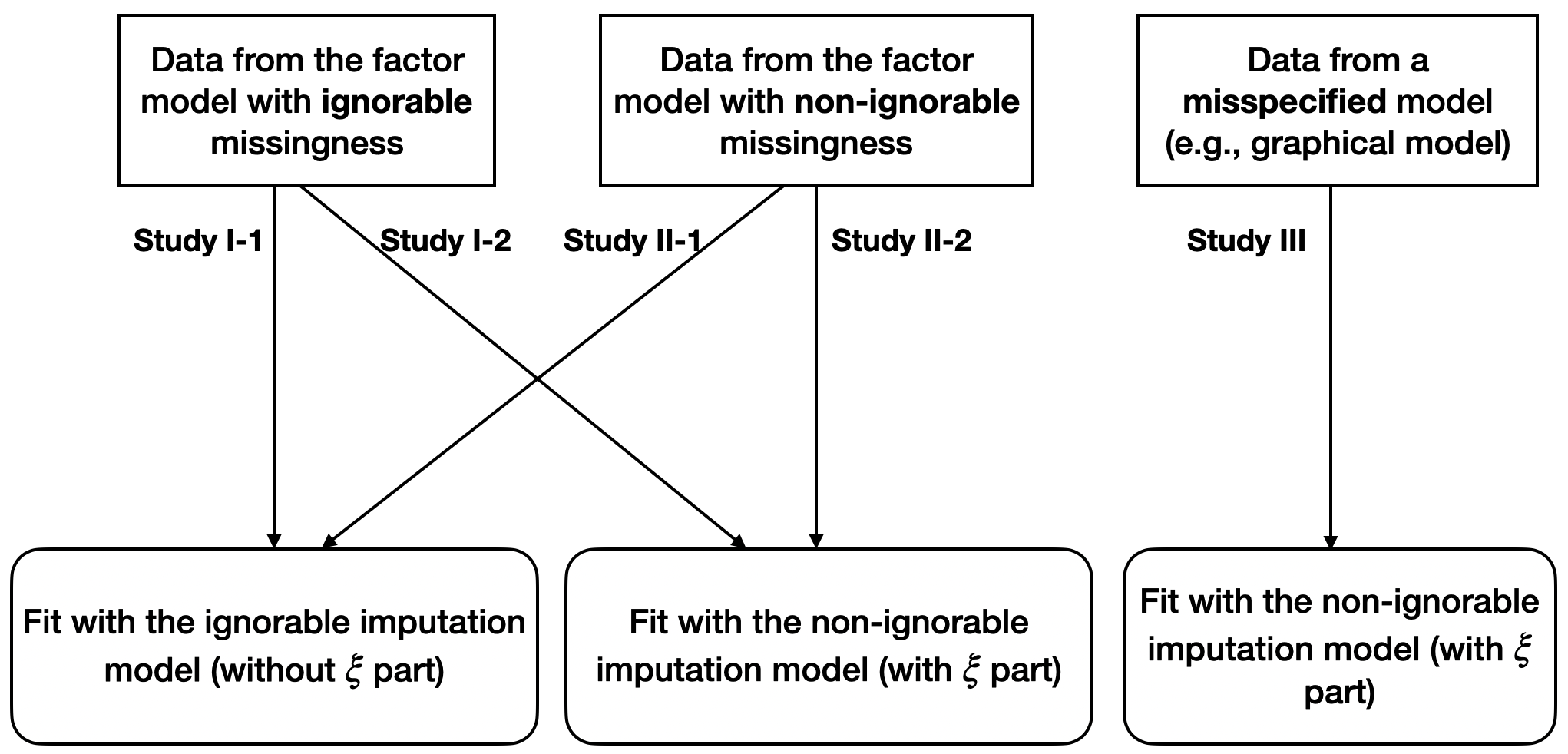}
    \caption{The overall simulation settings.}
    \label{fig:simu_settings}
\end{figure}

We first considered two settings where the true data-generating model is of the form of our latent variable model, and with ignorable missingness (i.e.\ the parameter $\boldsymbol{\kappa}$ is zero). The estimated imputation model is also of this form, in one case  assuming the missingness to ignorable (Study I-1) and in the other (Study I-2) allowing it to be non-ignorable. The proposed method performs very well, in that it gives unbiased point estimates, standard errors that accurately reflect sampling variability, and nominal coverage rates
(full results for these studies are given in Supplementary Section~C.2.
Crucially, these results are essentially the same in the two cases. This shows that applying a non-ignorable model to ignorable data (an over-specification) resulted in no discernible loss of efficiency or accuracy compared to a correctly specified ignorable model. This suggests that our method can be safely adopted as a default approach, even when the underlying missingness mechanism is unknown, without incurring a significant penalty if the data happen to be ignorable. 

\begin{figure}[p]\footnotesize
    \centering
    \begin{subfigure}{.4\textwidth}
        \centering
        \includegraphics[width=.8\linewidth]{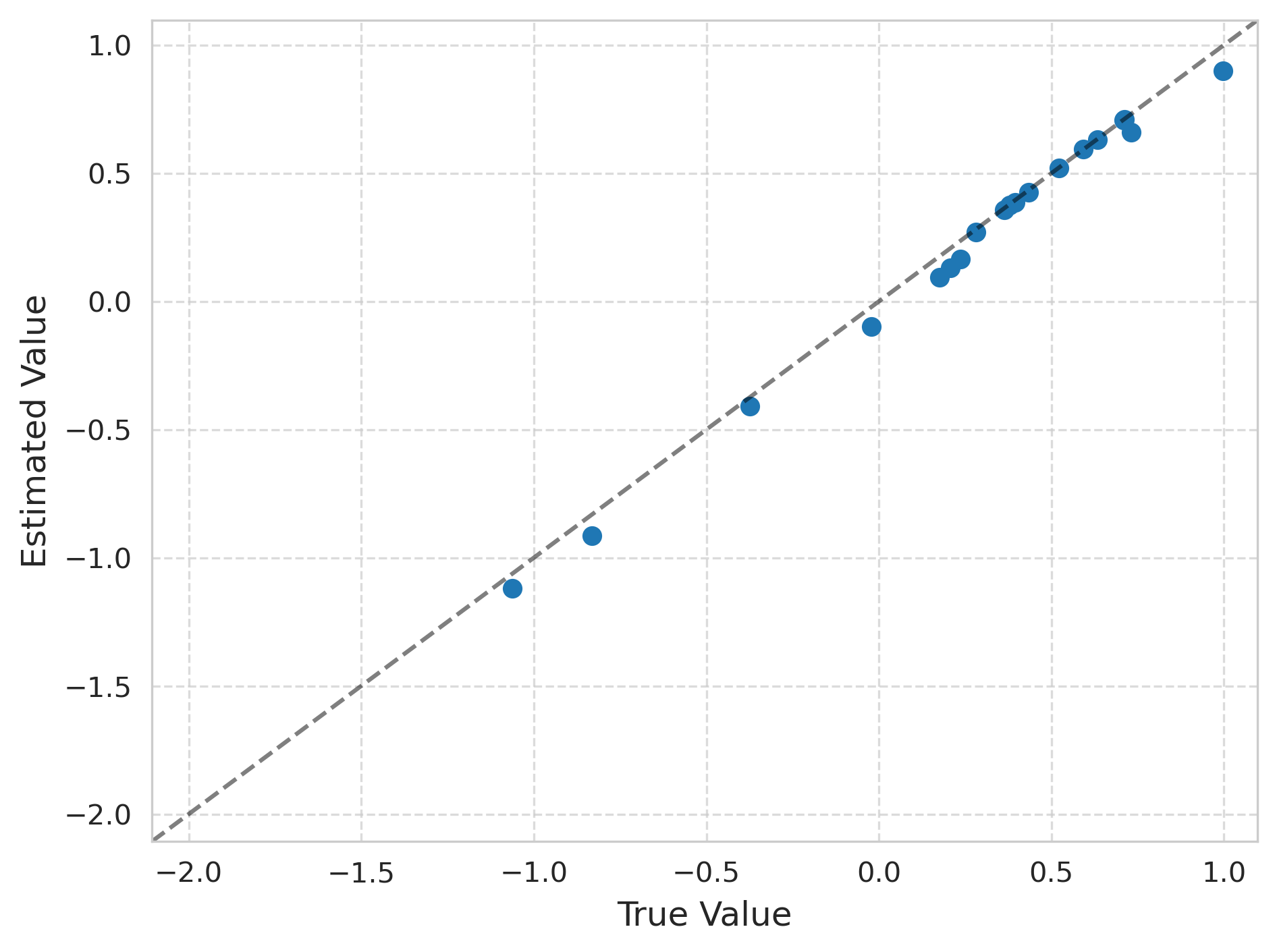}
        \caption{True values vs.\ mean estimates}
    \end{subfigure}%
    \begin{subfigure}{.55\textwidth}
        \centering
        \includegraphics[width=.9\linewidth]{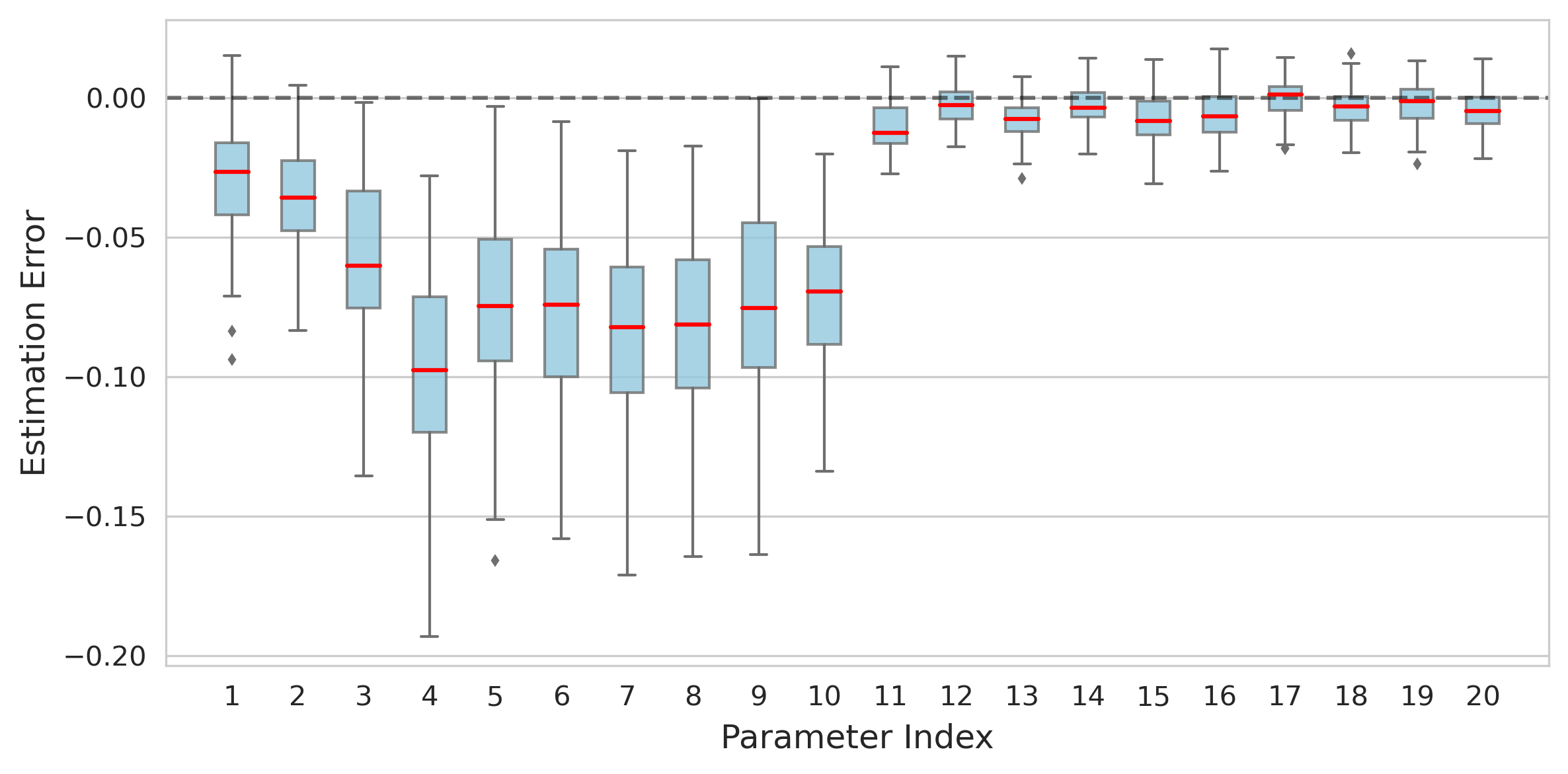}
        \caption{Distributions of point estimates$-$true values}
    \end{subfigure}
    
    \begin{subfigure}{.4\textwidth}
        \centering
        \includegraphics[width=.8\linewidth]{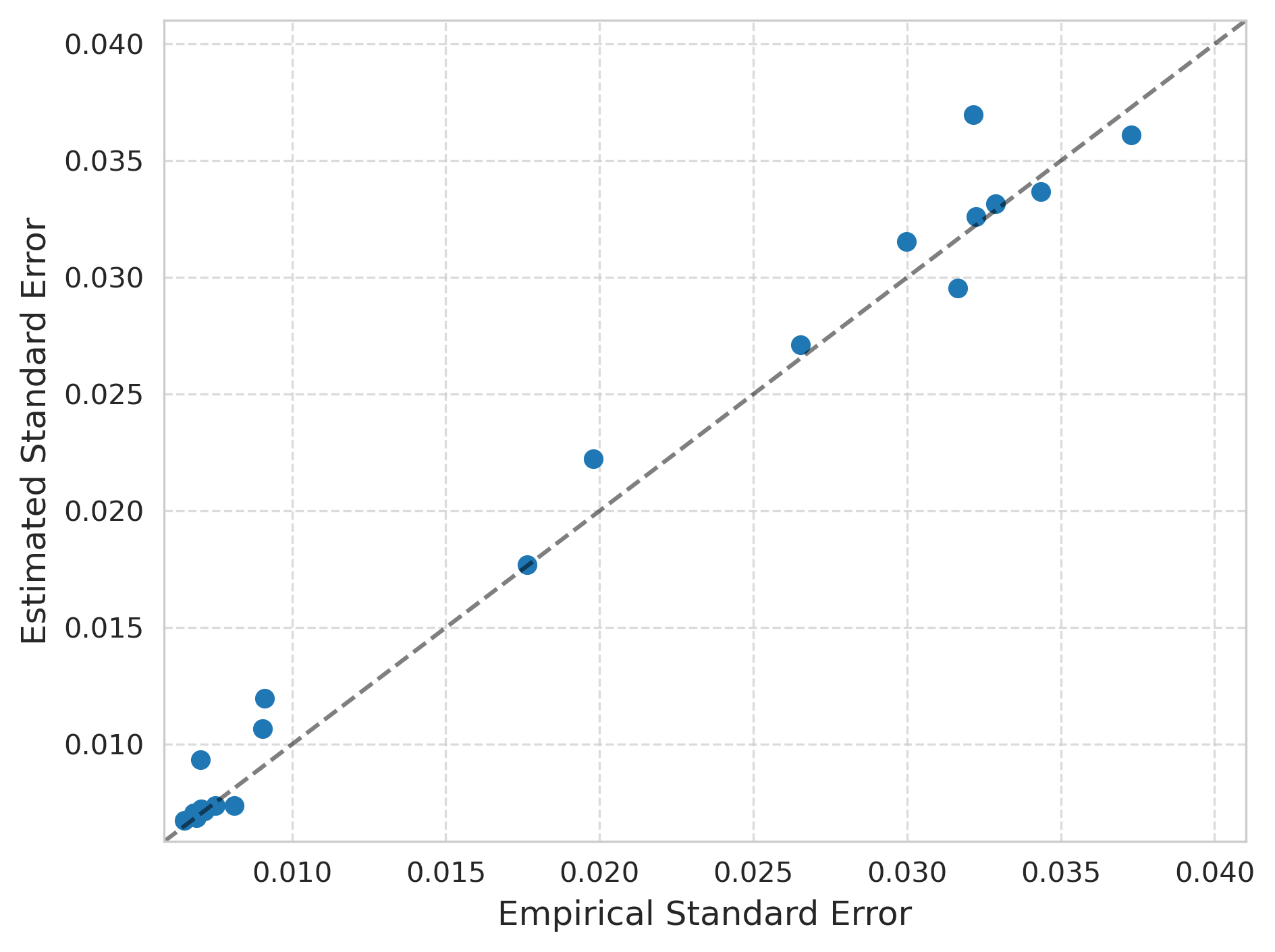}
                \caption{Simulation standard deviations vs.\ \\
        means of estimated standard errors}
    \end{subfigure}%
    \begin{subfigure}{.55\textwidth}
        \centering
        \includegraphics[width=.9\linewidth]{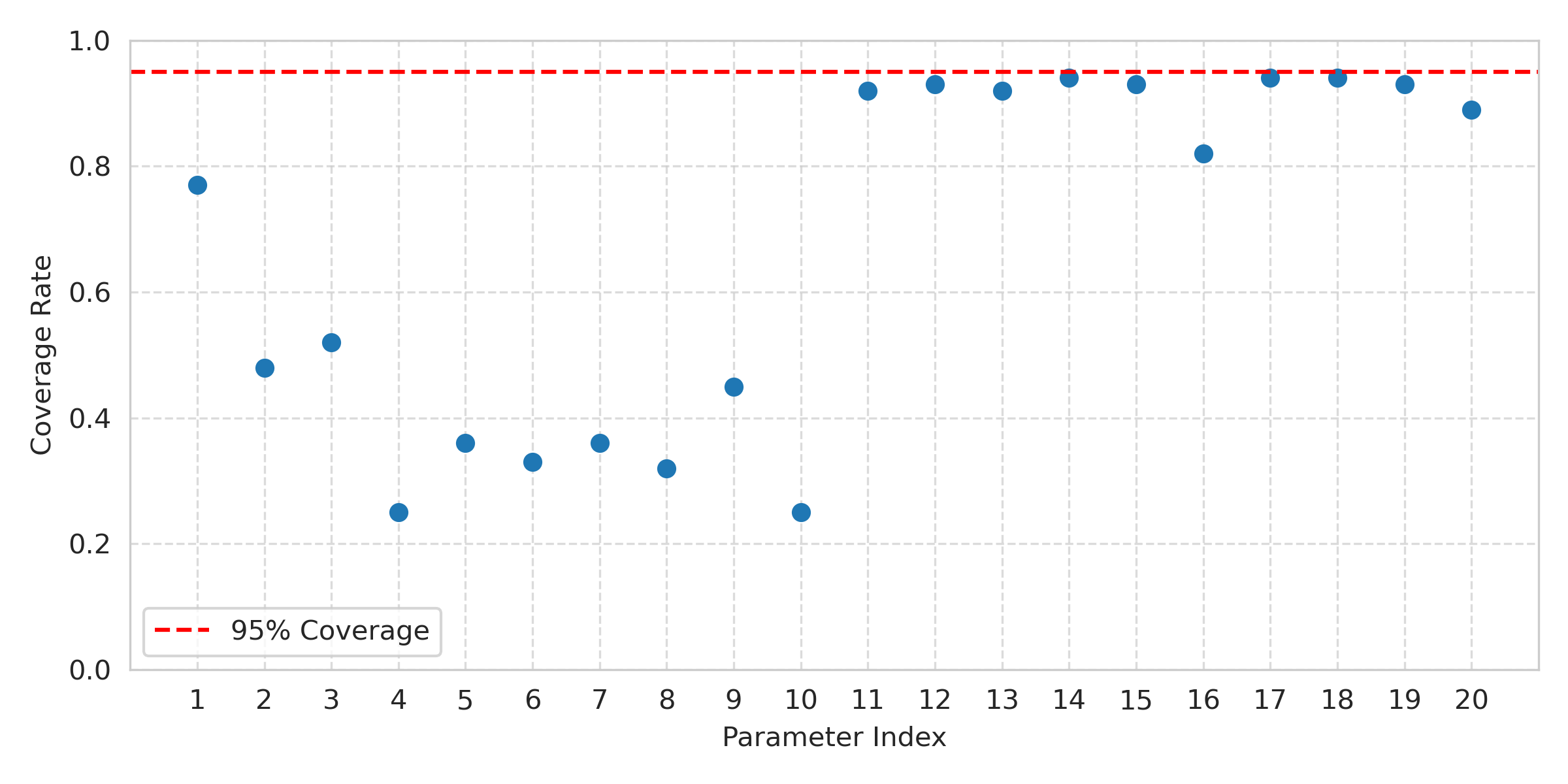}
        \caption{Coverage rates of 95\% confidence intervals}
    \end{subfigure}
    \caption{\small
    Simulation results when missingness is non-ignorable in the true model but ignorable in the estimated imputation model (Study II-1). Estimates for the means of 10 continuous (1--10) and 10 binary (11--20) variables show bias and poor coverage.
    }
    \label{f_sim_II_1}
\end{figure}

\begin{figure}[p]\footnotesize
    \centering
    \begin{subfigure}{.4\textwidth}
        \centering
        \includegraphics[width=.8\linewidth]{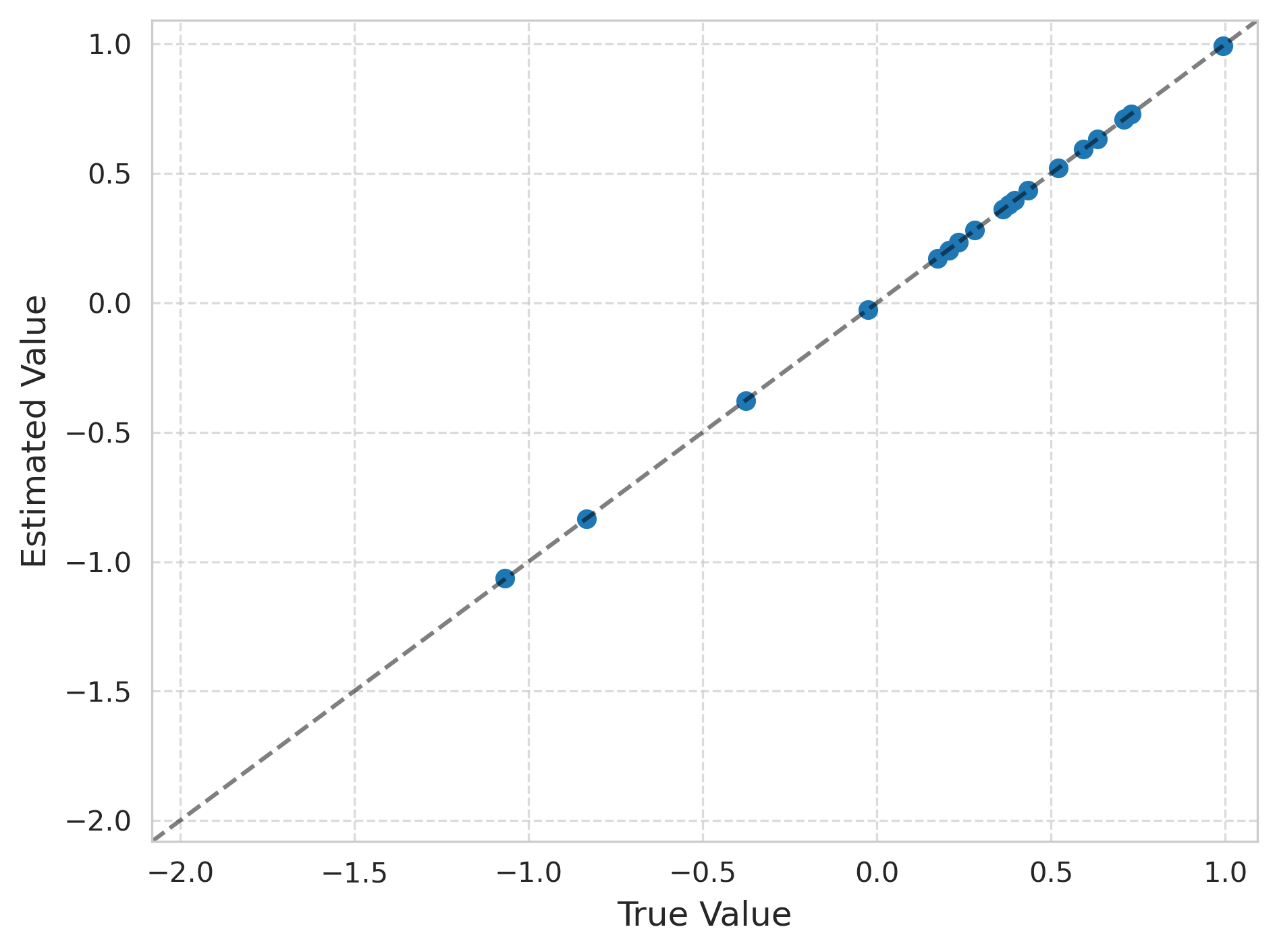}
        \caption{True values vs.\ mean estimates}
    \end{subfigure}%
    \begin{subfigure}{.55\textwidth}
        \centering
        \includegraphics[width=.9\linewidth]{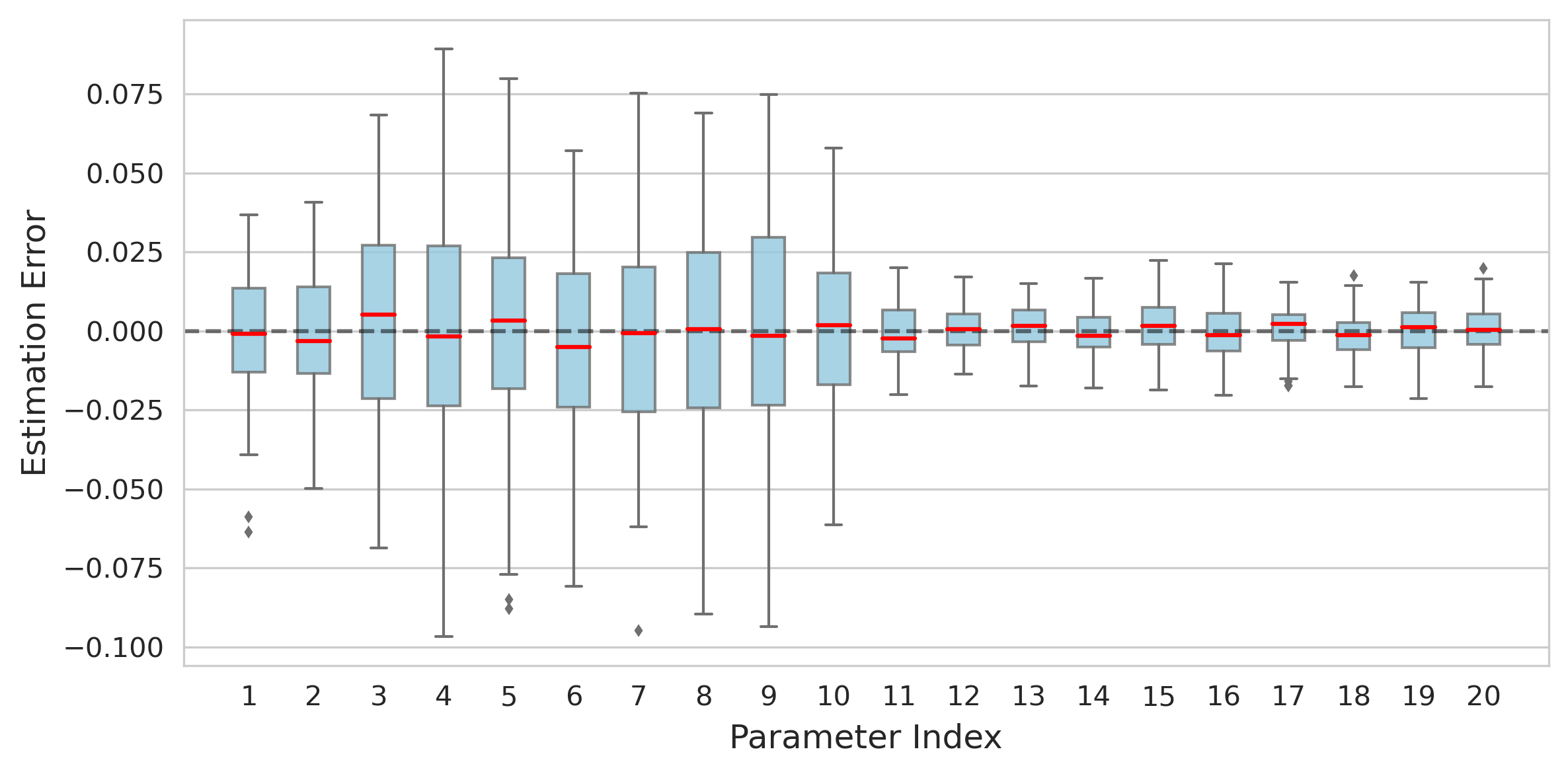}
        \caption{Distributions of point estimates$-$true values}
    \end{subfigure}
    
    \begin{subfigure}{.4\textwidth}
        \centering
        \includegraphics[width=.8\linewidth]{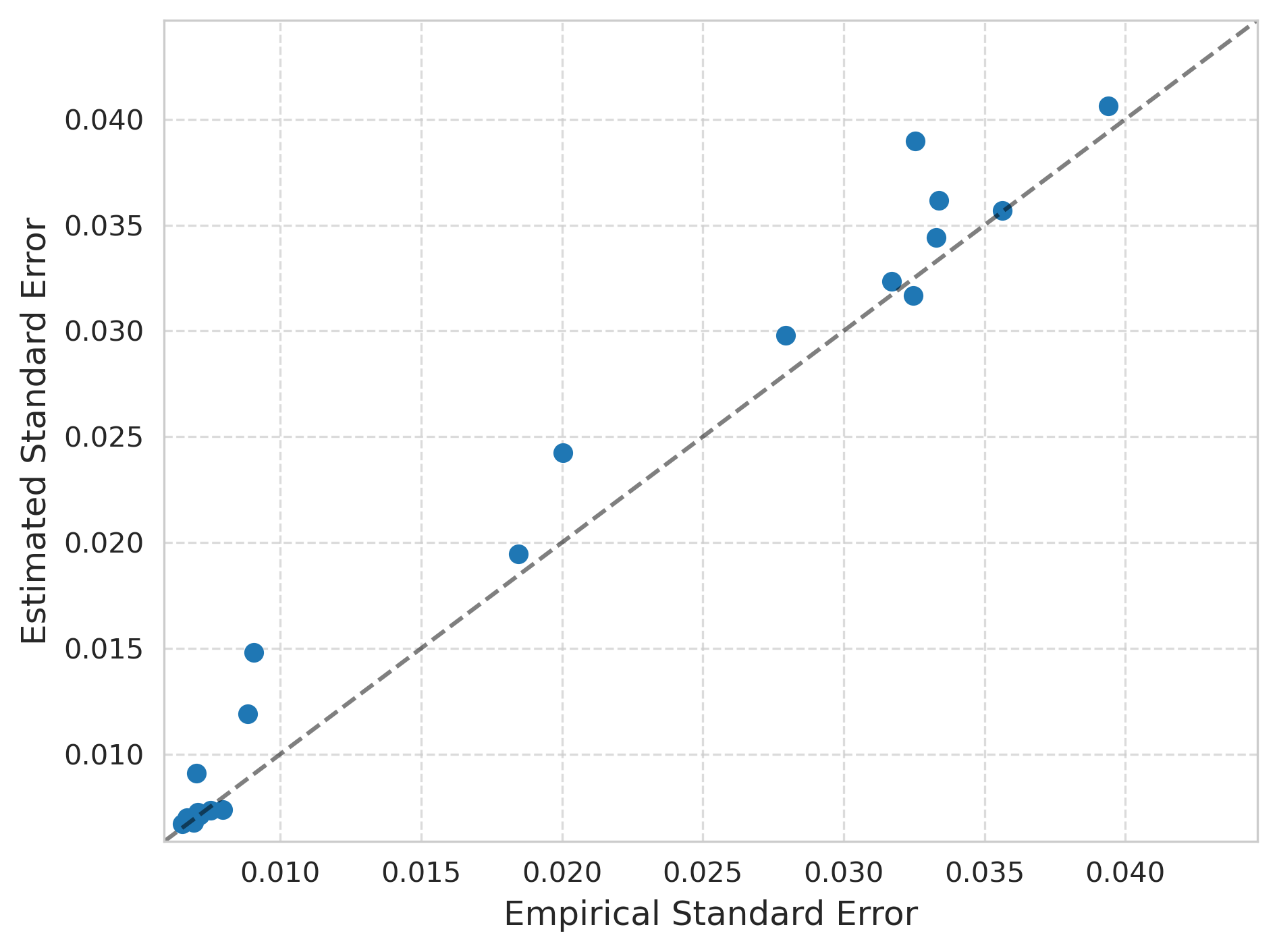}
                        \caption{Simulation standard deviations vs.\ \\
        means of estimated standard errors}
    \end{subfigure}%
    \begin{subfigure}{.55\textwidth}
        \centering
        \includegraphics[width=.9\linewidth]{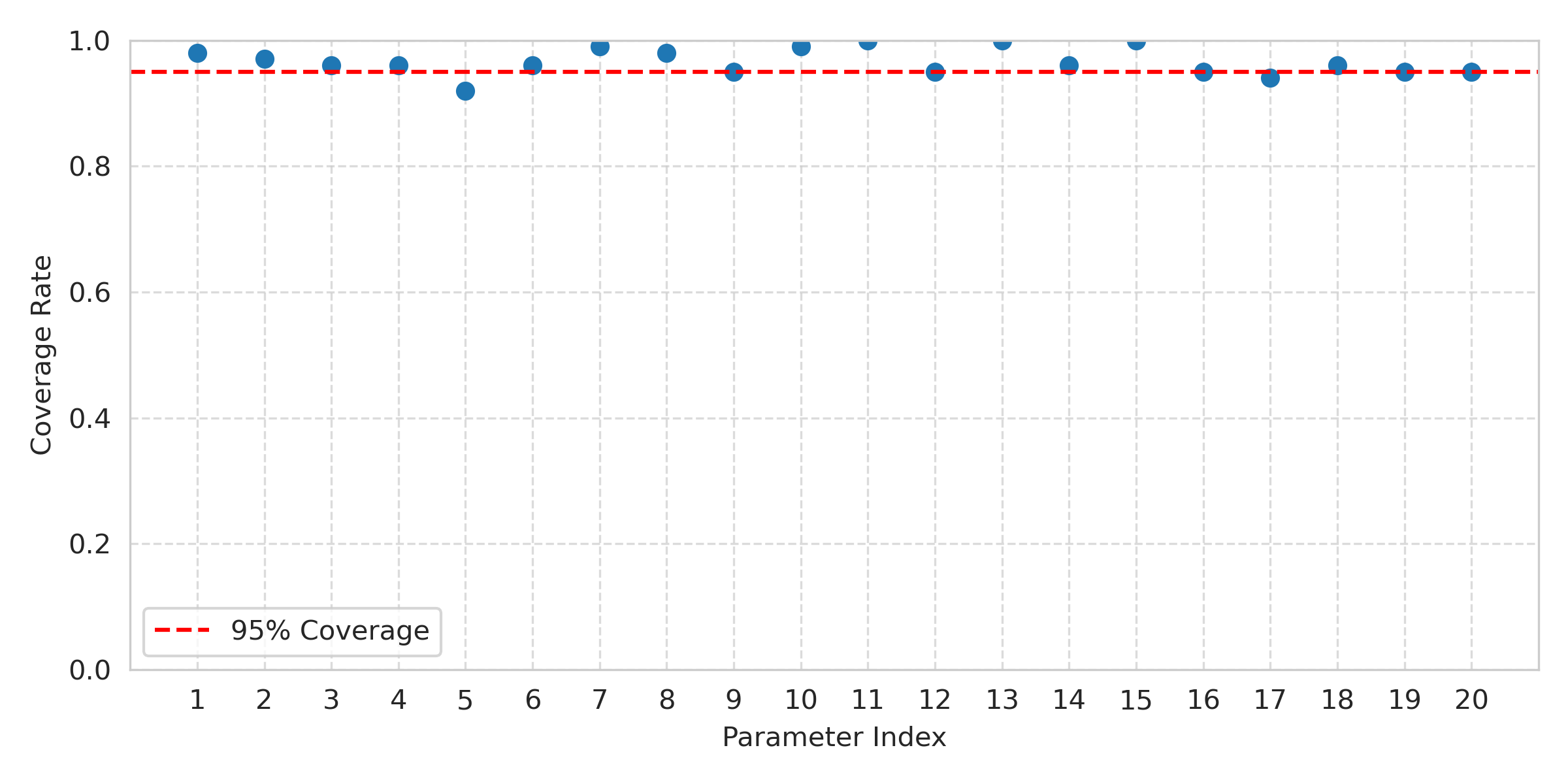}
        \caption{Coverage rates of 95\% confidence intervals}
    \end{subfigure}
    \caption{\small
    Simulation results when the missingness is non-ignorable in both the true model and the estimated imputation model (Study II-2). The proposed method yields accurate and well-calibrated estimates.
    }
    \label{f_sim_II_2}
\end{figure}

The next two settings are similar, except that the true missingness is non-ignorable. An imputation model that incorrectly assumes ignorability (Study II-1) now leads to systematically biased point estimates and severe under-coverage of confidence intervals, as shown in Figure~\ref{f_sim_II_1}. 
In contrast, allowing for the non-gnorability (Study II-2) again gives correct estimates (see Figure~\ref{f_sim_II_2}). This highlights the necessity of appropriately modeling non-ignorable missingness when it is present.

\begin{figure}[t]\footnotesize
    \begin{subfigure}{.5\textwidth}
    \centering
    \includegraphics[width=.9\linewidth]{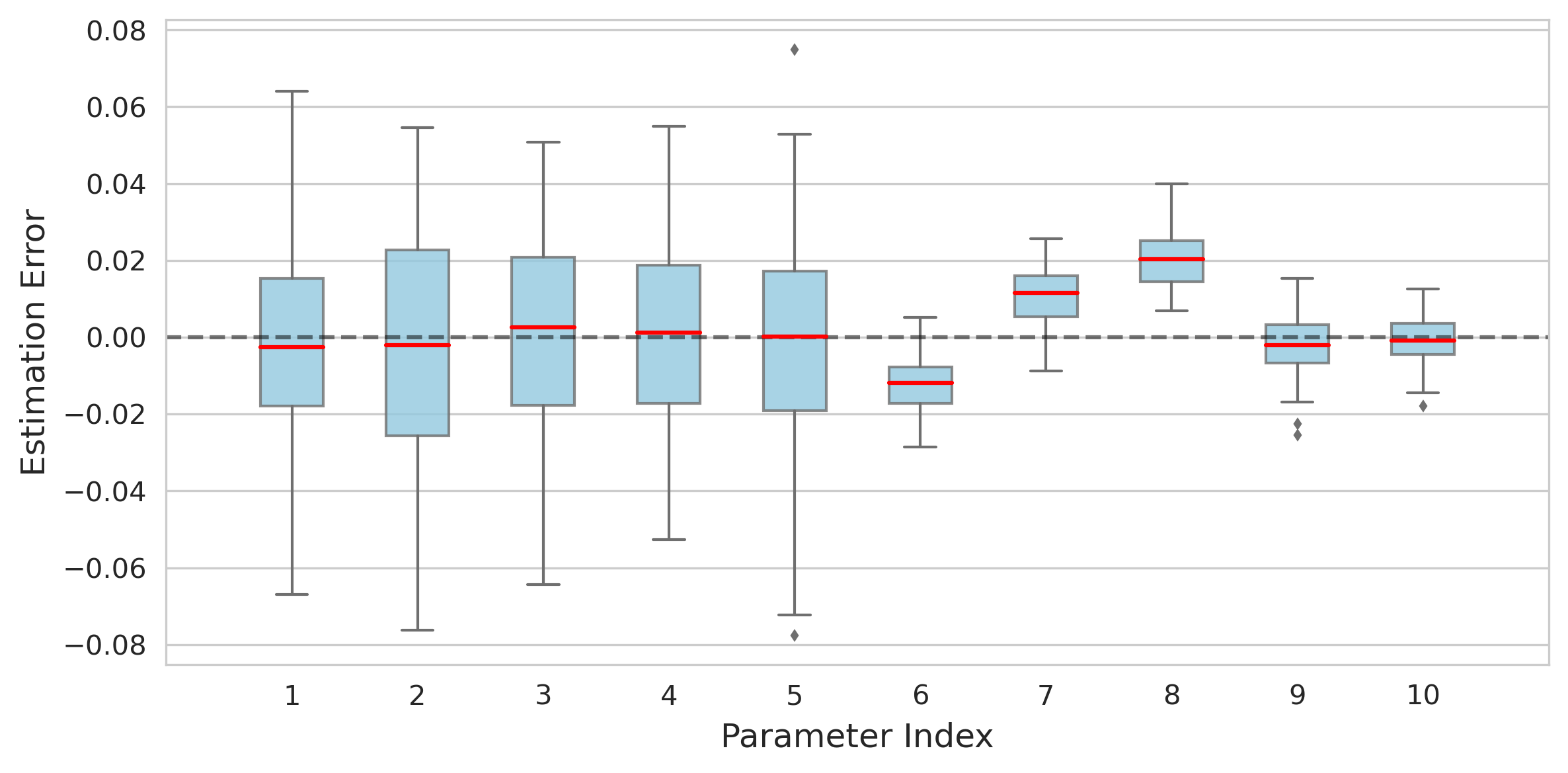}
          \caption{Distributions of point estimates$-$true values\\ \hspace*{2em}
          ($K_{1}=1$)}
  \end{subfigure}%
  \begin{subfigure}{.5\textwidth}
    \centering
    \includegraphics[width=.9\linewidth]{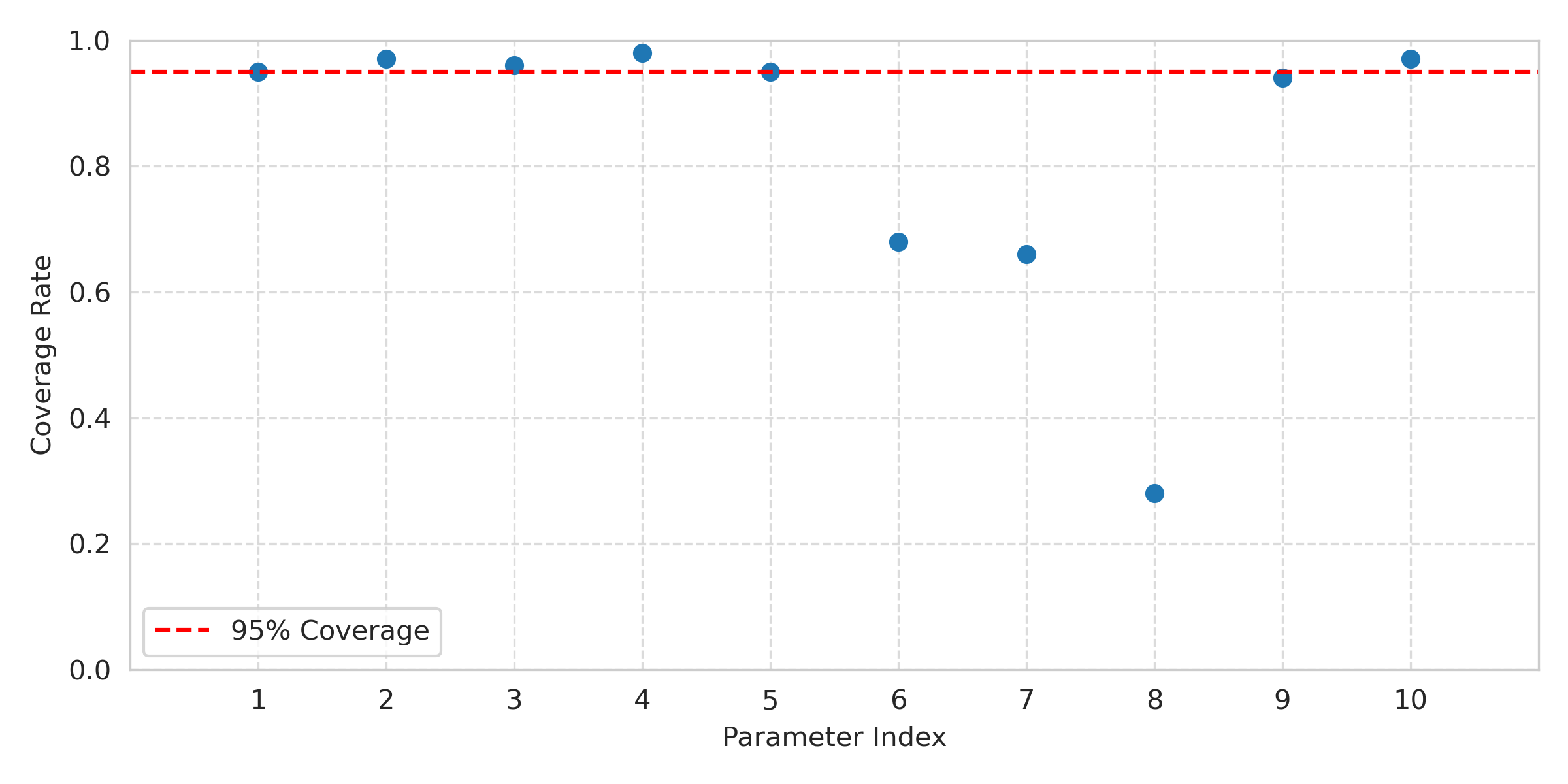}
    \caption{Coverage rates of 95\% confidence intervals\\ \hspace*{2em} ($K_1=1$).}
  \end{subfigure}
  \begin{subfigure}{.5\textwidth}
      \centering
      \includegraphics[width=.9\linewidth]{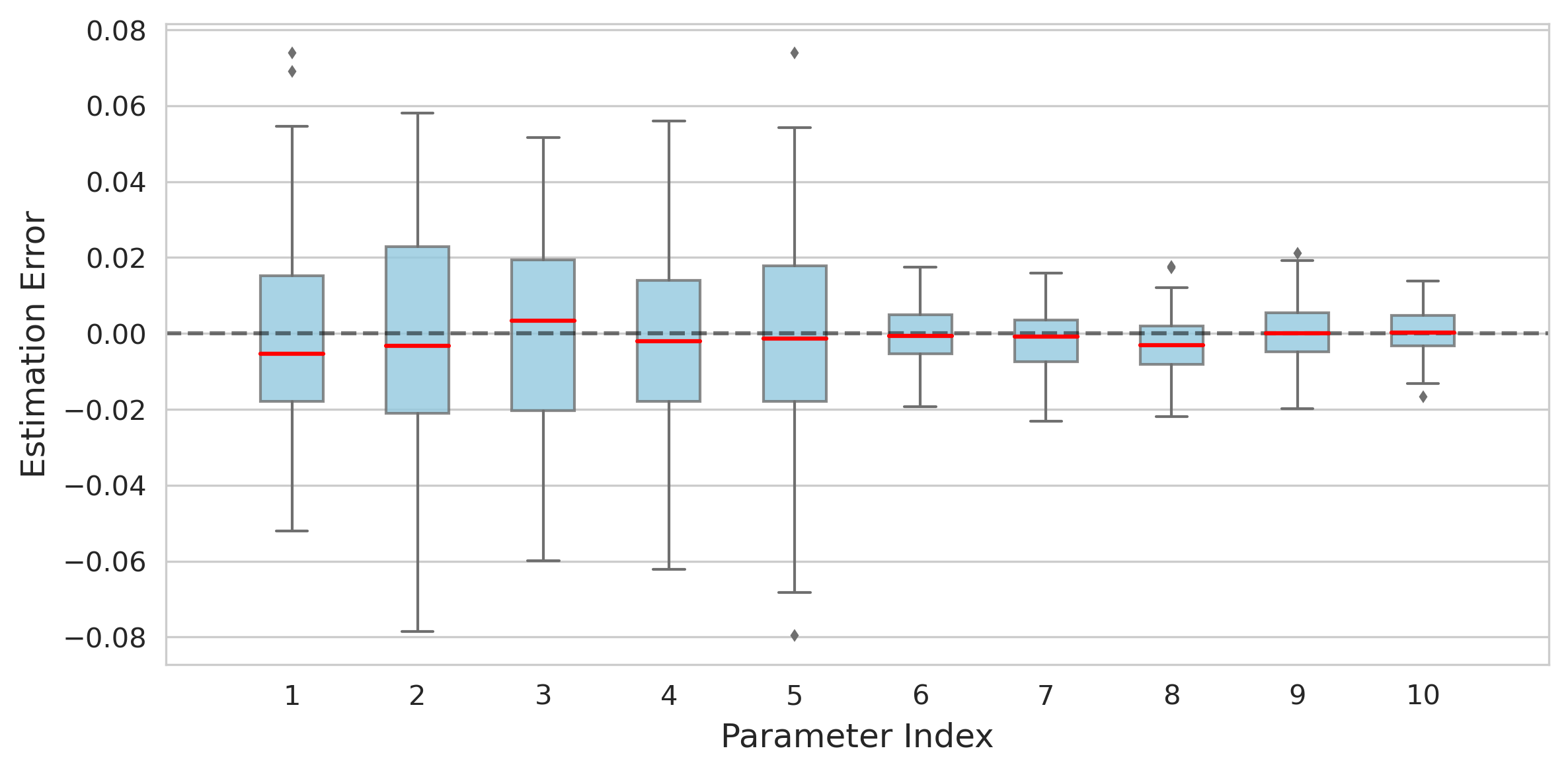}
          \caption{Distributions of point estimates$-$true values\\ \hspace*{2em}
          ($K_{1}=4$)}
    \end{subfigure}%
    \begin{subfigure}{.5\textwidth}
      \centering
      \includegraphics[width=.9\linewidth]{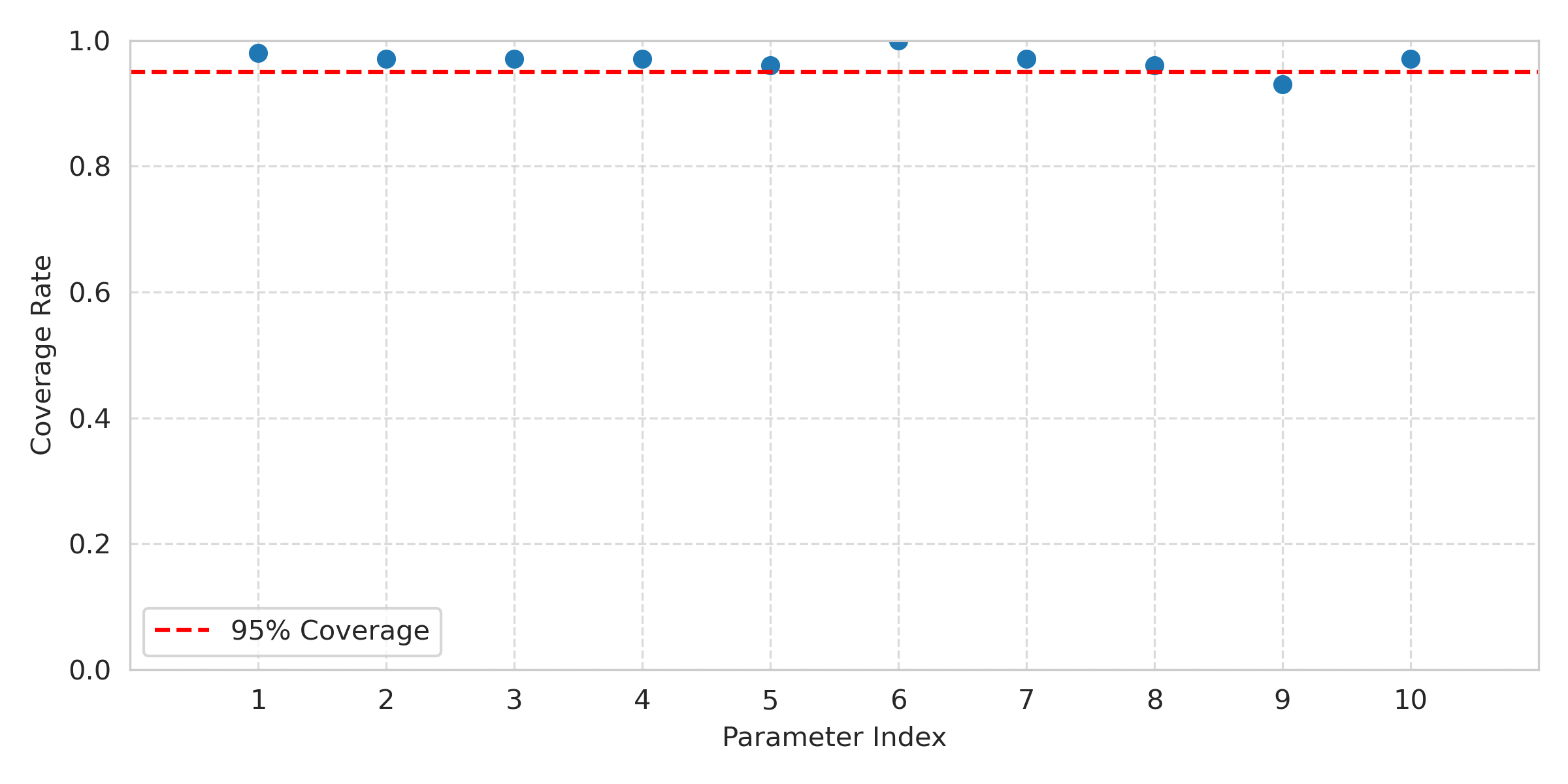}
    \caption{Coverage rates of 95\% confidence intervals\\ \hspace*{2em} ($K_1=4$).}
  \end{subfigure}
      \caption{\small Simulation results when the true data-genetating model is
      not of the assumed latent-variable form (Study III). The results are for means of 5 continuous variables (parameters 1--5) and 5 binary variables (6--10).
      Plots (a) and (b) are from simulations where the
      number of substantive factors $\boldsymbol{\eta}$ in the imputation
      model was $K_{1}=1$, and
      (c) and (d) from ones where it was $K_{1}=4$.
      }
  \label{f_sim_III}
\end{figure}

Finally, we examined robustness of the method
when the true joint distribution of the data were not of the form of the assumed latent variable model (Study III; the data-generating models are described in Supplementary Section~C.3). Here the crucial factor is that the assumed model can still
be a good approximation of the true distribution, as long as the dimensionality of the latent variables is large enough. 
The results presented in Figure~\ref{f_sim_III} show
that an imputation model that is too parsimonious 
(here with $K_1=1$ latent factor in $\boldsymbol{\eta}$)  yields biased estimates. 
However, increasing the model's flexibility with 
additional latent factors ($K_1=4$) effectively mitigates the bias and restores nominal coverage. This demonstrates that, provided the latent variable model is sufficiently flexible for underlying data structure, our method can serve as a robust imputation tool.


\section{An application to a cross-national survey}\label{sec:real_data}
\label{s_example}


\subsection{Data and research questions}
\label{ss_eample_data}

We apply the proposed method to data
from the European Social Survey (ESS). This is a 
repeated cross-sectional survey that is carried out every two years
since 2002 in around thirty European countries\footnote{See
\url{https://www.europeansocialsurvey.org/} for information about the
ESS. The data used here were obtained through
\url{https://www.europeansocialsurvey.org/data-portal};
\cite{ESS_Round_9_data}.}. It measures attitudes, beliefs and
behaviour among resident adult populations of the countries, in a standardized way
which facilitates cross-national comparisons. The ESS questionnaire
includes a set of core questions and two ``rotating modules''
which are selected anew for each round of the survey. We analyze data from round 9, conducted in 2018 in 29
countries, and focus on questions from its rotating module on
\emph{Justice and fairness in Europe}. The goal of this module was to measure
individuals' ``perceptions of justice for self and others regarding
different outcomes such as income, education and job chances''. The analyses that we
describe here are motivated by some of the questions considered also in
the topline findings report by \cite{ESS9justice2020}.

Here missing data arises because of survey nonresponse to individual
questions. Such nonresponse affects all surveys, and it may be
particularly problematic for cross-national surveys where nonresponse
rates can vary much between different countries. Designers of the ESS
have devoted extensive effort into understanding and reducing nonresponse in
the survey \citep{stoopetal2010}, but much of it inevitably remains. A
survey like the ESS thus presents a good case for the use of multiple imputation analysis.

Our analysis includes 30 variables, in two groups. First, we use
ten variables on background characteristics of the respondents such as
age and education. They are listed in Supplementary
Table~S4. These variables have some nonresponse in the
survey, and they could also be imputed. However, their rates of
nonresponse are much smaller than those of our main analysis variables.
We choose to treat the background variables in the role of fully
observed covariates $\mathbf{x}_{i}$, to provide an illustration of how
such variables can be included in the analysis. We accordingly omit
observations with nonresponse in them. This leaves a dataset of 48,577
respondents, which we use for all the analyses below.
It includes 767--2662 respondents from each of the 29
countries, as reported in Supplementary Table~S5.

Our focus is on twenty variables from the Justice and
Fairness module (household income is a core
question) which are considered in the role of analysis variables
$\mathbf{y}$. They are listed in Table~\ref{tab:responses}.
The rates of nonresponse overall (out of our base sample
of 48,577) are 18.9\% for household income and 1.7--8.9\%
for the other variables. There is a great deal of variation in
these rates between the countries, as shown in Supplementary Table S5.

The questions on the fairness of the respondent's own pay and that of
others in the same occupation (G14a and G17a in Table
\ref{tab:responses}) were not asked if the respondent  
reported having no income. We treat these absent values as inherently
undefined rather than missing, so their values are not imputed and such
respondents would be excluded from the population for any analysis model
for these variables. For imputation of other variables, a dummy variable
for not having income is also included as one of the observed covariates
$\mathbf{x}$.

\begin{table}[!htbp]
\footnotesize
\centering
\caption{
Description and summary statistics of the main analysis variables from
round 9 (2018) of the European Social Survey (ESS). The analysis also uses
ten background variables, which are listed in Table~S4 of the 
Supplementary Materials.
The variables highlighted in gray are used in the example analyses
reported in Section \ref{ss_example_analysis} or Supplementary Materials.
For these analyses, the role of the rest of these variables
(and the background variables) is to help impute missing values in
the analysis variables.}
\label{tab:responses}
\vspace*{1ex}
\begin{tabular}{llrrr}
\hline
 &  &   &  &   Missing \\
Variable$^{1}$ & Scale &  Mean & SD &   rate$^{7}$ \\
\hline
\textbf{Procedural justice of political system}       &&     &           &\\
~~fair chance for everyone to participate (G1)& Ordinal$^{2}$   & 1.72&1.05      & 5.3\% \\
~~government takes into account interests of all (G2) & Ordinal$^{2}$  & 1.51&0.91      & 3.4\% \\
~~decisions in politics are transparent (G3) & Ordinal$^{2}$ & 1.39&0.93      & 5.3\%\\[.5ex]
\textbf{Distributive justice: society is fair when...} &&     &           &\\
\rowcolor{lightgray}
~~income and wealth are equally distributed (G26) [equality] & Ordinal$^{3}$ &  1.71&1.15     & 2.1\%\\
\rowcolor{lightgray}
~~hard-working people earn more than others (G27) [equity] & Ordinal$^{3}$   &  1.03&0.84     & 1.7\%\\
~~it takes case of poor and needy unconditionally (G28) & Ordinal$^{3}$  &  1.14&0.88     & 2.0\%\\
~~people from high-status families have privileges (G29) & Ordinal$^{3}$ & 2.77&1.04 & 3.5\%\\[.5ex]
\textbf{Belief in a Just World}     &&          &           &\\
~~people get what they deserve (G30) & Ordinal$^{3}$ & 2.03&1.04 & 2.4\%\\
~~justice always prevails over injustice (G31) & Ordinal$^{3}$   & 2.08&1.08 & 2.4\%\\
~~people are eventually compensated for injustices (G32) & Ordinal$^{3}$  &
2.14&1.05     &  3.8\%\\[.5ex]
\textbf{Fairness of educational opportunities}       &&     &           &\\
~~for respondent  (G4)    & Cont.$^{4}$              &   8.09&2.83        &  3.9\%\\
~~for everyone in the country G6)     & Cont.$^{4}$   &   7.20&2.56        &  2.5\%\\[.5ex]
\textbf{Fairness of getting jobs}       &&     &           &\\
~~for respondent  (G5)         & Cont.$^{4}$           &   6.90&2.99   & 4.5\% \\
~~for everyone in the country (G7)     & Cont.$^{4}$    & 5.86&2.49
&  2.5\%\\[.5ex]
\textbf{Fairness of incomes}        &&  &           &\\
~~respondent's own (net) pay (G14a)         & Cont.$^{5}$                    & 3.38&1.65      & 5.2\% \\
~~pay for people in the same occupation as respondent (G17a)   & Cont.$^{5}$  & 3.38&1.66     & 8.9\%\\
\rowcolor{lightgray}
~~income of top 10\% earners among full-time employees (G18)  & Cont.$^{5}$  &  5.76&1.79  &   8.4\% \\
\rowcolor{lightgray}
~~income of bottom 10\% earners among full-time employees (G19)   & Cont.$^{5}$  &  2.63&1.71     &  5.2\% \\
~~wealth differences in the country (G20) & Cont.$^{5}$  &
6.05&2.39       &  6.6\%  \\[.5ex]
\rowcolor{lightgray}
\textbf{Respondent's household income from all sources (F41)}      &
Cont.$^{6}$               & 5.25&2.78 &  18.9\% \\
\hline
\end{tabular}
\begin{tabular}{llrrr}
\multicolumn{5}{l}{\footnotesize{\emph{Notes:}}}\\
\multicolumn{5}{l}{\footnotesize{\hspace*{0.4em}1. The code in parentheses is the number of the question in ESS questionnaire}}\\
\multicolumn{5}{l}{\scriptsize{\hspace*{0.2em}
(\texttt{https://stessrelpubprodwe.blob.core.windows.net/data/round9/fieldwork/source/ESS9\_source\_questionnaires.pdf})
}}\\
\multicolumn{5}{l}{\footnotesize{
\hspace*{0.2em}2. 5 response options coded from 0=``Not at all'' to 4=``A great
deal'', treated as ordinal.}}\\
\multicolumn{5}{l}{\footnotesize{
\hspace*{0.2em}3. 5 response options coded from 0=``Agree strongly'' to
4=``Disagree strongly'', treated as ordinal.}}\\
\multicolumn{5}{l}{\footnotesize{
\hspace*{0.2em}4. 11 response options (0=``Does not apply to at
all'' to
10=``Applies completely''), treated as continuous.}}\\
\multicolumn{5}{l}{\footnotesize{
\hspace*{0.2em}5. 9 response options (0=``Extremely unfairly low'' to
8=``Extremely unfairly high''), treated as continuous.}}\\
\multicolumn{5}{l}{\footnotesize{
\hspace*{0.2em}6. 10 income deciles in
respondent's country, from 0=Lowest to 9=Highest,
treated as continuous.}}\\
\multicolumn{5}{l}{\footnotesize{
\hspace*{0.2em}7. The base sample here is $N=48,577$ respondents for whom
all the background variables are observed. 
}}\\
\multicolumn{5}{l}{\footnotesize{
\hspace*{0.1em}
The  rates reported are the percentages of these respondents for whom each analysis
variable is missing.
}}\\ 
\end{tabular}
\end{table}

The ESS data are probability samples from the target populations in the
different countries. For correct population inference, the analysis
should include appropriate survey weights. We use the poststratification
weights that are included in the published ESS data. They combine
allowance for unequal sampling probabilities in the sampling design and final
calibration to known population distributions of some demographic
variables. The calibration step also serves as nonresponse weighting that
adjusts for imbalances arising from unit nonresponse, i.e.\
sampled respondents who did not reply to any of the survey. These weights
are used as the $w_{i}$ in the notation of Section
\ref{sec:analysis_est_infer} above. We note that these formulas
incorporate information only on the weights but not other relevant
features of the sampling design (specifically stratification and
multistage sampling).

\subsection{Multiple imputation analysis}
\label{ss_example_analysis}

We first select and estimate the imputation model for the
joint distribution of $\mathbf{y}_{i}$ given $\mathbf{x}_{i}$ for the
variables described above. This involves selecting the dimensions of the
substantive ($K_{1}$) and nonresponse ($K_{2}$) latent variables in the
model and examining
whether the association between these latent variables (the parameters
$\boldsymbol{\kappa}$) is non-zero, i.e.\ whether the nonresponse should
be treated as non-ignorable. 
This was done using data from Italy and Austria, the two countries with the largest sample sizes. Based on the BIC statistic, we adopted $K_1=3$ and $K_2=1$.
Furthermore, for all tested model configurations  likelihood ratio tests provided strong evidence ($p<0.001$) against the assumption of ignorable nonresponse. We therefore proceeded with the non-ignorable model for all subsequent analyses. Details of the model selection are provided in Supplementary Section~D.1.

The parameters of the imputation model were
estimated first using using Algorithm 1 with $T=3000$ iterations and
$T_0=1000$ burn-in steps.
Algorithm 2 was then used to
generate $M=20$ multiply imputed datasets, with $T=10,000$, $T_0=1000$
and a thinning interval of $k=450$. These steps were performed separately for each of the 29 countries.


Once the imputation has been done, we can carry out analyses of
substantive interest. We illustrate this with three examples, which
focus on different questions of interest for the Justice and Fairness
survey module. They also involve variables with different proportions of
missing data. The five variables that are involved in these
examples are highlighted in gray in Table \ref{tab:responses}, while for
the purposes of these three examples the rest of the variables (and the
background variables in Supplementary Table~S4) serve only to provide
information for imputing missing values in the analysis variables. In
each case estimated population parameters are compared between the 29
countries, which are identified in the plots by two-letter country codes
used by the ESS (see
Supplementary Table S5).

The first analysis concerns views on how resources should be
allocated between people in a society. This is a necessary starting
point, because people will judge the fairness of existing situation in
their society against the normative benchmarks that they hold. \cite{ESS9justice2020} examined (in their Figure 1) measures of two
distributive norms on income: \emph{equality}, measured by question G26
in the ESS (see our Table \ref{tab:responses}), and \emph{equity},
measured by question G27. The nonresponse rates for these variables are
fairly low, 2.1\% in our analysis sample overall (varying across the
countries from 0.2\% for Norway to 11.1\% for Bulgaria) for equality and
1.7\% (0.05\% for Estonia to 9.0\% for Bulgaria) for equity.

Results for the question on equality are shown in Figure
\ref{fig:equality} (and for equity in Supplementary Section~D.4.
What is shown here are the estimated percentages of adults who Agree
strongly or Agree with the equality principle in income distribution.
Substantively, we can see that there is much variation in this level of
agreement between countries, from over 60\% in some countries (most of
them in the Mediterranean and the Balkans) to less than 40\% in others,
mostly in Northern Europe. In contrast, support for the equity principle
is much higher and more uniform. Methodologically, the results suggest
that allowing for missing data makes little difference in this case.
Complete-case estimates (i.e.\ effectively assuming missingness
completely at random) are very similar to the multiple-imputation
estimates. For the standard errors, the effect of multiple imputation
can be in two directions: estimating the imputation model adds
uncertainty, but using information on the other variables for
respondents with missing data decreases it. Here the aggregate effect of
these is that the standard errors of the multiple-imputation estimates
are typically very slightly higher.



\begin{figure}[!htbp]
    \centering
    \includegraphics[width=0.9\linewidth]{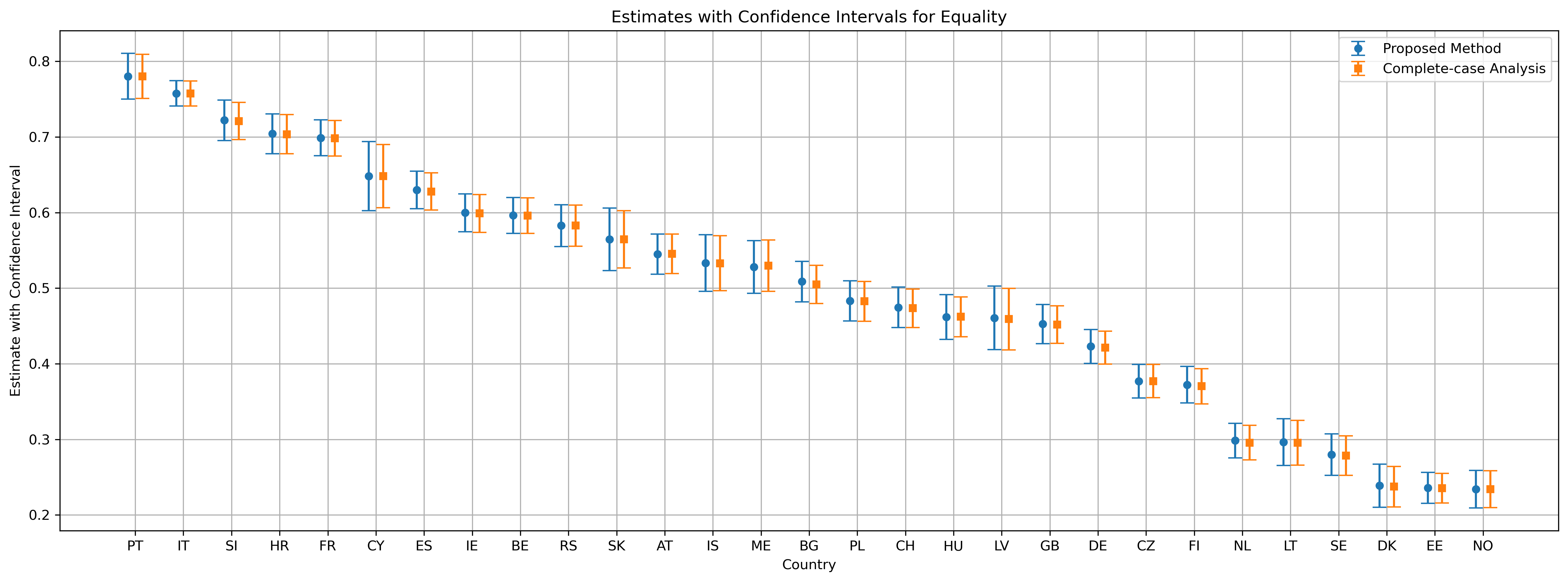}
    \caption{
    Estimates of normative perceptions about equality of income
    distributions. The figure shows proportions of people in adult
    populations 29 countries who would Agree Strongly or Agree with the
    statement that ``A society is fair when income and wealth are
    equally distributed among all people'', estimated from data from
    round 9 (2018) of the European Social Survey. Estimates are shown
    (with 95\% confidence intervals) based on only the observed
    responses to this question (in orange) and using the proposed multiple
    imputation procedure (in blue).
    }
    \label{fig:equality}
\end{figure}

\begin{figure}[!htbp]
    \centering
    \includegraphics[width=0.9\linewidth]{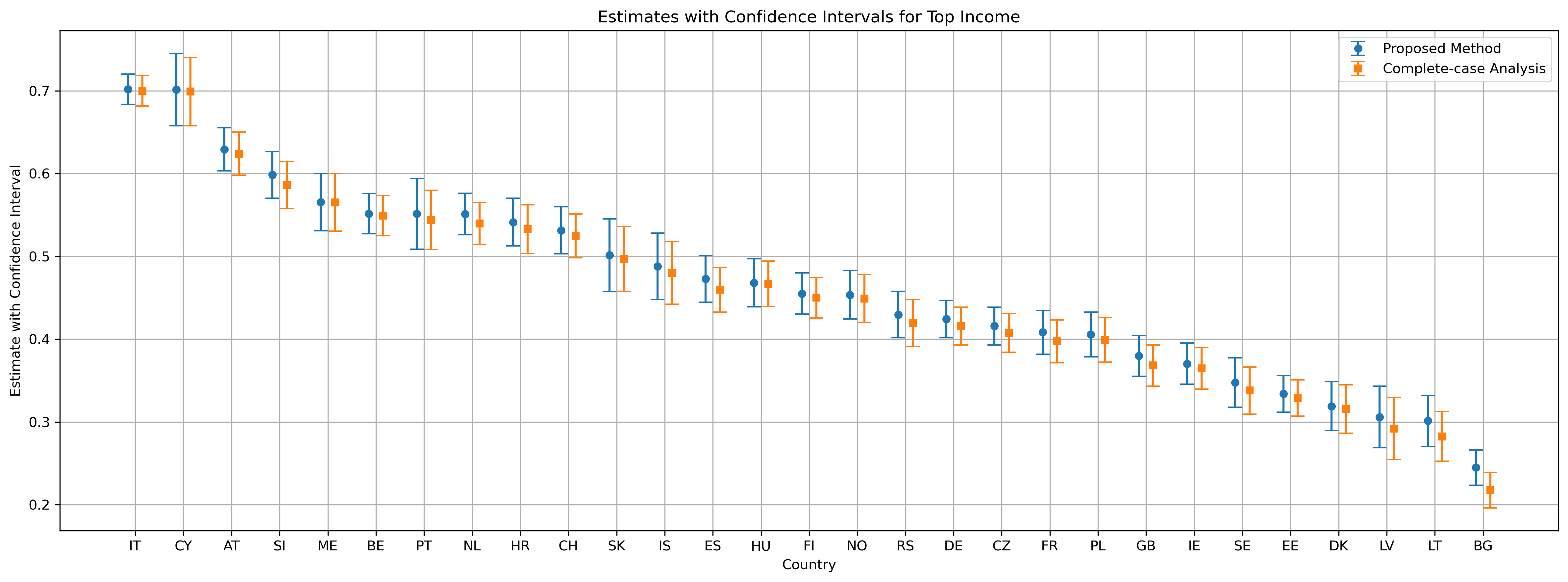}
    \caption{
    Estimates of perceived unfairness of top incomes. The figure shows
    proportions of people in adult populations of 29 countries who
    would give a response to
    the statement
``Please think about the top 10\% of employees working full-time in
[your country], earning more than [amount per month or per year]''
indicating that these incomes are unfairly high,
   estimated from data from
    round 9 (2018) of the European Social Survey. Estimates are shown
    (with 95\% confidence intervals) based on only the observed
    responses to this question (in orange) and using the proposed multiple
    imputation procedure (in blue).
    }
    \label{fig:income-fairness}
\end{figure}

The second analysis focuses on the respondents' assessment of the fairness
of actually existing income distributions. We again estimate average
responses to two related questions. They asked how people felt about the
incomes of the top 10\% of earners (Question G18) and the bottom
10\% of earners (G19) among full-time employees in their country. The
nonresponse rates are 8.4\% overall (1.3--17.1\% across the countries) for the top-10\% question and 5.2\% (1.3--11.8\%) for the bottom-10\% question. Figure \ref{fig:income-fairness}
shows estimated proportions of people who think that the incomes of the top 10\%
earners are unfairly high. The results again show much variation
between countries, with the proportions ranging from around 70\% in Italy to around
25\% in Bulgaria. In contrast, parallel results for the question on the
bottom 10\% of earners (see Supplementary Section~D.4) show that there is
much higher and much more uniform agreement that bottom incomes are too
low. Considering the treatment of missing data, multiple-imputation and
complete-case estimates are again fairly similar. However, the
differences are larger than in the first example, especially for
countries with  high rates of nonresponse for this question.
Furthermore, the differences between the estimates are consistently in
the same direction, with multiple-imputation estimates being
higher. In other words, the nonresponse seems to be systematically by
respondents who are somewhat more likely to think that top incomes are
unfairly high.


The third analysis explores associations between different constructs.
We estimate individual-level
correlations between respondents' own family
incomes and their perceptions of 
the fairness of top incomes.
The question of interest is thus whether individuals who are themselves
higher or lower in the income distribution tend to have different views on this aspect of collective fairness or unfairness. 
Here the survey nonresponse is potentially a major complication. 
The family income variable has a large amount of nonresponse on its own (18.9\% overall, 6.0--37.5\% across the countries), and the
bivariate nonresponse rate for this pair of variables is a correspondingly high 23.1\% (11.0-48.2\%). 
This level of missingness raises serious concerns about the validity of complete-case analysis, which would discard such large and potentially non-random portions of the sample.

\begin{figure}[!htbp]
    \centering
    \includegraphics[width=0.9\linewidth]{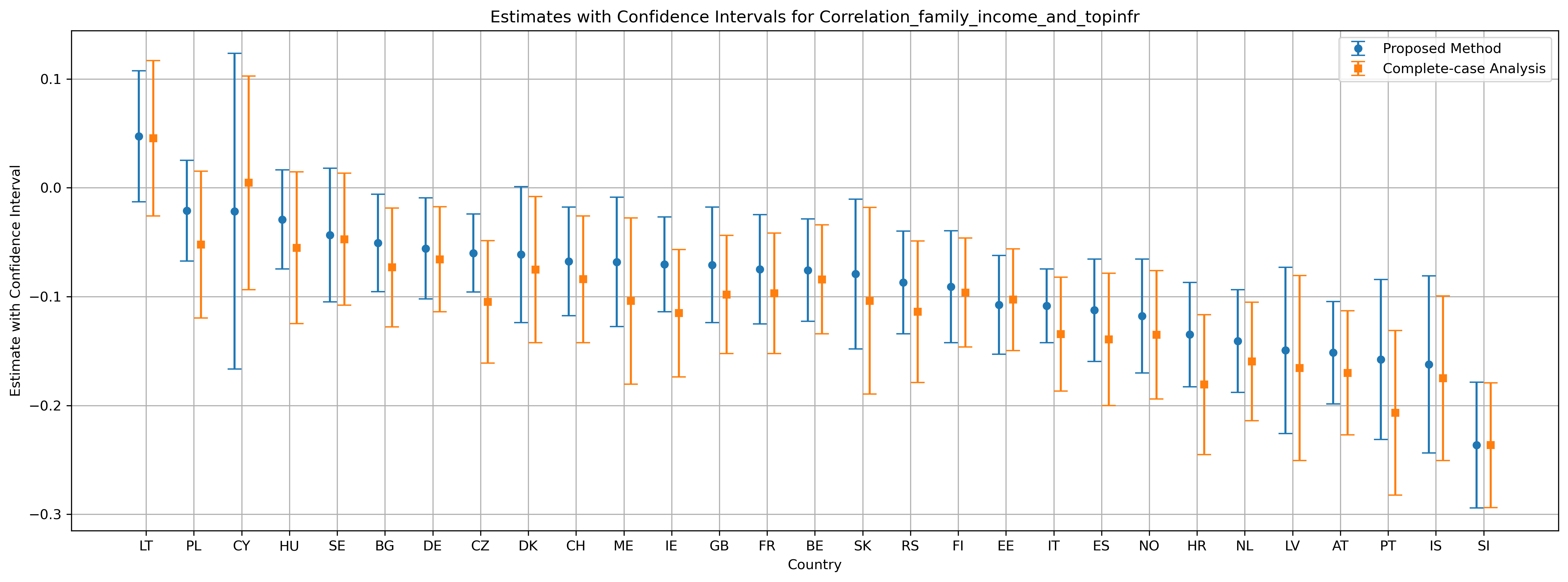}
    \caption{
    Correlations between individuals'
    own household income and 
their perceptions of the unfairness of top incomes in their countries
among adult populations of 29 countries,  
   estimated from data from
    round 9 (2018) of the European Social Survey. Estimates are shown
    (with 95\% confidence intervals) based on only the observed
    responses to this question (in orange) and using the proposed multiple
    imputation procedure (in blue).
    }
    \label{fig:corr}
\end{figure}

The estimated correlations, shown in Figure~\ref{fig:corr}, are mostly mildly but significantly negative. Individuals who have high incomes themselves tend to be somewhat less likely to view top incomes in their countries as unfair.
Complete-case and imputation-based estimates are non-trivially different. The sizes of their differences correlate with the nonresponse rates, but not very consistently: countries with the largest differences  are often not the ones with the most missing data. The differences are consistent in direction, in that for most countries the complete-case estimates are 
larger. This suggests that complete-case analysis exaggerates the strengths of these correlations, because of a selection bias that arises from the implausible assumption of missingness completely at random that it requires. The multiple imputation procedure offers 
a more flexible and realistic way of allowing for the missingness, while also using information from all of the available data.


\section{Discussion}\label{sec:discussion}

The most
challenging and consequential methodological task in setting up multiple imputation of missing data
is how to define and use 
the imputation model. The most theoretically coherent
approach is to define a joint distribution for all the analysis
variables and predictors of them, and then draw imputations from
conditional distributions implied by it. The question is how to
specify this distribution in a sufficiently comprehensive and
flexible way.

In this paper we proposed a general imputation model  based on
joint distributions that are induced by latent variable models for the
observable variables. This has the advantages of being transparent,
flexible and scalable. Individual observed variables, which can be of
any types, are added to the distribution one at a time, through
univariate models for them given the latent variables, and integration
over the latent variables then implies a joint distribution for the
observed variables. This gives a distribution where
observed variables are all associated with each other, but without need to specify these associations
directly for all pairs of the variables. The complexity of the
association structures can be adjusted simply by increasing or
decreasing the number of latent variables. With enough of them, the
latent variable model can flexibly approximate different kinds of true
joint distributions.

Our model allows missing-data mechanisms that are missing not at
random (MNAR) and non-ignorable. This is done by including additional
latent variables that determine probabilities of the missingness and
that are associated with the latent variables that underlie the
substantive variables. The missingness then has the proporty of `latent
ignorability' where it is missing at random given the latent
variables but MNAR with respect to the distribution of the
observed variables only. This form of MNAR is still identifiable from
observable data, under the latent variable model. Allowing for latent
ignorability is straightforward in our framework, 
because it means just adding more latent variables and more observed
variables (the binary missingness indicators) to the model. With it, imputation distributions for missing
observations will depend also on the pattern of observed and missing
values for other variables for the same unit. Our results suggest that
allowing for this possibility may often be important in practice. In our
applied example with the European Social Survey there was clear evidence
that nonresponse in the survey was non-ignorable, and the simulation
studies showed that not allowing for non-ignorability when it is present
can lead to non-trivial bias in subsequent analyses of interest.

For estimation and inference of analysis models, we proposed the approach developed by \cite{wang1998large} and
\cite{robins2000inference}. Procedurally, it differs from the more
commonly used Rubin's rules for multiple-imputation analysis  in
that the imputations are conditional on (maximum likelihood) estimates
of the parameters of the imputation model and that the uncertainty in
these estimates is accounted for in a correspondingly different way in
estimated analysis models. However, the key feature of multiple
imputation is unchanged, in that the imputation is done first and
separately, and what is carried forward from it 
can subsequently be used for estimation of any congenial analysis models. Our
simulations show that the procedure works well, as is also expected
on theoretical grounds.

Imputations from the latent variable model could also be
combined with Rubin's rules for final estimation. We could 
simply apply those rules with our imputations, although this would
result in some underestimation of variances of analysis parameters.
Alternatively, the imputation procedure could be modified to include
also posterior sampling of the parameters of the imputation model. This
extension remains to be developed.

We have proposed efficient computational procedures for estimating the
parameters of the model and for generating imputed
values from it. Further work on them will be useful. For
example, they can be extended without fundamentally new issues
to accommodate models for more types of variables  than the (continuous, binary and
ordinal) ones we considered explicitly. Limits of computational
feasibility also remain to be explored. This concerns both
dimensions of the latent variables, which determine the flexibility of
the distribution, and the numbers of observed variables. Here we used
up to five latent variables and twenty observed variables (plus ten
fully observed covariates) without difficulty, but still larger numbers
could be desirable in other contexts. From a theoretical point of view
there is no reason why the proposed method cannot be scaled up much further.

\section*{Supplementary Materials}
\paragraph{Supplementary material:} Document containing theoretical proofs, computational details, additional simulation results, and further analyses of the European Social Survey data, organized in four main sections A-D. (.pdf file)

\spacingset{1.5}
\bibliographystyle{apacite}
\bibliography{ref}

\end{document}


\maketitle

\setcounter{equation}{0}
\renewcommand{\theequation}{S\arabic{equation}}
\setcounter{remark}{0}
\renewcommand{\theremark}{S\arabic{remark}}
\setcounter{algocf}{0}
\renewcommand{\thealgocf}{S\arabic{algocf}}
\setcounter{figure}{0}
\renewcommand{\thefigure}{S\arabic{figure}}
\setcounter{table}{0}
\renewcommand{\thetable}{S\arabic{table}}
\setcounter{assumption}{0}
\renewcommand{\theassumption}{S\arabic{assumption}}

\newpage

\tableofcontents

\newpage

\appendix

\section{Supplement to Section 3: Further theoretical details}
\label{sec:theories}

\subsection{Assumptions and proof of Theorem~1}
\label{sec:proof}

We present the assumptions used to derive Theorem~1 in the main text.
\begin{assumption}\label{ass:iid_data}
  The observed data $(\bfy_{i,obs},\bfz_i,\bfx_i)$ are independent and
  identically distributed across units $i=1,\ldots,N$, with
  (unknown) density function $f_0(\bfy_{i,obs},\bfz_i,\bfx_i)$. We
  note that our imputation model may not be consistent with $f_0$, in
  the sense that there may not exist a parameter $\Psi$ such that
  $f(\bfy_{i,obs},\bfz_i\mid\bfx_i;\Psi)$ is consistent with the
  conditional distribution implied by
  $f_0(\bfy_{i,obs},\bfz_i,\bfx_i)$.
\end{assumption}

\begin{assumption}\label{ass:sampling_iid}
  Imputed values
  $(\bfy_{i,mis}^m,\bfxi_i^m,\bfeta_i^m)$ drawn from the conditional
  distribution
  $\phi(\bfy_{i,mis},\bfeta_i,\bfxi_i\mid\bfy_{i,obs},\bfz_i,\bfx_i;\hat\Psi)$
  are independent across $m=1,\ldots,M$ and $i=1,\dots,N$. Defining
  further
      \begin{equation}\label{eq:imputation_score_mis}
      S_{i,mis}^m(\Psi) = \frac{\partial}{\partial\Psi}\log
      \phi(\bfy_{i,mis}^m,\bfeta_i^m,\bfxi_i^m\mid\bfy_{i,obs},\bfz_i,\bfx_i;\Psi),
  \end{equation}
  the following holds:
  \begin{equation}
      S_{i,mis}^m(\Psi) = S_i^m(\Psi) - S_{i,obs}(\Psi),
  \end{equation}
  where $S_i^m(\Psi)=\partial \log
  f(\bfy_i^m,\bfz_i,\bfeta_i^m,\bfxi_i^m\mid\bfx_i;\Psi)/\partial
  \Psi$ and $S_{i,obs}(\Psi)=\partial \log
  f(\bfy_{i,obs},\bfz_i\mid\bfx_i;\Psi)/\partial \Psi$
  (see Supplementary Section \ref{ss_a_score_separation}).
\end{assumption}

\begin{assumption}\label{ass:imputation_influence}
As the sample size $N$ tends to infinity,
the probability that the optimization (5) and equation (8)
have unique solutions $\hat\Psi$ and $\hat\bftheta$
tends to 1, and $\hat\Psi$ and $\hat\bftheta$ converge to their limiting
points $\Psi^*$ and $\bftheta^*$.
Furthermore, $\hat\Psi$ is an
asymptotically linear estimator of $\Psi^*$ with influence function
$D_i(\Psi^*)=\left[I_{obs}(\Psi^*)\right]^{-1}S_{i,obs}(\Psi^*)$,
satisfying
  $$
  \sqrt{N} (\hat\Psi - \Psi^*) = \frac{1}{\sqrt{N}} \sum_{i=1}^{N} D_i(\Psi^*) + o_p(1),
  $$
  where
\begin{equation}
  I_{obs}(\Psi) =
  -\mathbb{E}\left[\frac{\partial}{\partial\Psi^\top}\,S_{i,obs}(\Psi)\right],
\end{equation}
in which the expectation is taken with respect to
$(\bfy_{i,obs},\bfz_i,\bfx_i)$ under the joint density $f_0$ in
Assumption~\ref{ass:iid_data}.
\end{assumption}

\begin{assumption}\label{ass:score_quant}
Define $\lambda(\Psi,\bftheta) = \mathbb{E}\left[U_i^m(\Psi,\bftheta)\right],$ where the expectation is taken with respect to the random variables $(\bfy_{i}^m, \bfz_i,\bfx_i)$ under the joint density
{
\begin{equation}\label{eq:combined_density}
  \prod_{m=1}^M \phi(\bfy_{i,mis}^m,\bfeta_i^m,\bfxi_i^m\mid\bfy_{i,obs},\bfz_i,\bfx_i;\Psi^*)f_0(\bfy_{i,obs},\bfz_i,\bfx_i).
\end{equation}
}
In a neighborhood of $(\Psi^*, \bftheta^*)$, we assume that both
$\partial \lambda(\Psi,\bftheta)/\partial \Psi^{\top}$ and $\partial
\lambda(\Psi,\bftheta)/\partial \bftheta^{\top}$ exist and are
continuous in $(\Psi,\bftheta)$, and the {matrix} $\partial
\lambda(\Psi,\bftheta)/\partial \bftheta^{\top}$ is invertible.

\end{assumption}

\begin{assumption}\label{ass:score_quant2}
  In a neighborhood of $(\Psi^*, \bftheta^*)$ and for all possible
  values of $\bfy_i^m$,  $\bfeta_i^m$, $\bfxi_i^m$, $\bfz_i$ and
  $\bfx_i$,  the partial derivatives $\partial \log
  \phi(\bfy_{i,mis}^m,\bfeta_i^m,\bfxi_i^m\mid\bfy_{i,obs},\bfz_i,\bfx_i;\Psi)/\partial
  \Psi$ and $\partial U_i^m(\Psi,\bftheta)/\partial
  \boldsymbol{\theta}^{\top}$ exist. In addition, these derivatives
  are bounded in $L^2$ with respect to the joint distribution
  \eqref{eq:combined_density}.
\end{assumption}

\begin{assumption}\label{ass:score_quant3}
  Let
  $$
  \mathcal{L}_{N,\theta}(\Psi_1,\Psi_2) = \frac{1}{\sqrt{\sum_{i=1}^N w_i^2}} \bigg\Vert \sum_{i=1}^N w_i \bar{U}_i(\Psi_1,\bftheta) - \sum_{i=1}^N w_i \bar{U}_i(\Psi_2,\bftheta) - \sum_{i=1}^N w_i\lambda(\Psi_1,\bftheta) + \sum_{i=1}^N w_i\lambda(\Psi_2,\bftheta)\bigg\Vert.
  $$
  There exists a positive constant $c$ such that, for any $\Psi_1,\Psi_2$ in a neighbourhood of $\Psi^*$,
  $$
  \sup_{\Vert\Psi_1-\Psi_2\Vert < c} \mathcal{L}_{N,\theta}(\Psi_1,\Psi_2) \rightarrow 0
  $$
  uniformly for all $\bftheta$ in a neighbourhood of $\bftheta^*$ as $N$ goes to infinity.
\end{assumption}

\begin{assumption}\label{ass:score_consist}
In a neighborhood of $(\Psi^*,\bftheta^*)$, there exists a positive constant $d$ such that $\mathbb{E}\left\{U_i^m(\Psi,\bftheta)^{2+d}\right\}$ is finite, where the expectation is with respect to
$(\bfy_{i}^m, \bfz_i,\bfx_i)$ under the distribution implied by the joint density
\eqref{eq:combined_density}.
\end{assumption}

\begin{assumption}\label{ass:score_quant4}
  There exists a constant $C >0 $ that is independent of sample size
  $N$, such that the weights $w_i$ satisfy
  $C^{-1} \leq w_i \leq C$
  for all $i=1,\ldots,N$.
\end{assumption}

Assumptions S1-S7 are standard regularity conditions that are
also made in \cite{robins2000inference}, while Assumption S8
is added for completeness, to require boundedness of the weights associated
with the units in the
analysis model (in a survey context it will be satisfied automatically, as
survey weights are in practice always finite by construction).
Assumption
\ref{ass:iid_data} allows the imputation model to be misspecified, in
the sense that there may not be a value of $\Psi$ such that the
conditional density function $f(\bfy_{i,obs},\bfz_i\mid\bfx_i; \Psi)$
coincides with the true conditional density
$f_0(\mathbf{y}_{i,obs},\mathbf{z}_{i} | \bfx_i)$, for all possible
values of $\mathbf{y}_{i,obs}$, $\mathbf{z}_{i}$ and $\bfx_i$. Under
standard assumptions for the asymptotic theory for maximum likelihood
estimation, the limiting point $\Psi^*$
for $\Psi$  in Assumption S3
minimizes the Kullback-Leibler divergence between the true and specified
distributions for the observed data \citep{white1982maximum}, i.e.,
    \begin{equation}
        \Psi^* = \arg\min_{\Psi}
        \mathbb{E}\left(\log \left[\frac{f_0(\bfy_{i,obs},\bfz_i|\bfx_i)}{f(\bfy_{i,obs},\bfz_i\mid\bfx_i; \Psi)}\right]\right),
    \end{equation}
    where the expectation is with respect to the true density $f_0(\bfy_{i,obs},\bfz_i,\bfx_i)$,
    and $\hat\Psi$ is asymptotically normal with
    $\sqrt{N}(\hat\Psi - \Psi^*) \rightarrow
    \mathcal{N}(0,\Lambda(\Psi^*))$, where
    $\Lambda(\Psi^*) = \mathbb{E}\left[D_i(\Psi^*)D_i(\Psi^*)^\top\right]$.
{When the imputation model is correctly specified,
$f_0(\bfy_{i,obs},\bfz_i|\bfx_i) = f(\bfy_{i,obs},\bfz_i\mid\bfx_i;
\Psi^*)$ for all possible values of $\mathbf{y}_{i,obs}$,
$\mathbf{z}_{i}$ and $\bfx_i$. In that case, with additional regularity
conditions,  the limiting point $\bftheta^*$ will coincide with the
``true value'' $\bftheta_0$ defined based on complete data.} If
the imputation model is misspecified,  $\bftheta^*$ may differ from
$\bftheta_0$.

\begin{proof}
  The proof establishes the asymptotic distribution of the two-step
  M-estimator $\hat{\bftheta}$, defined as the solution to
  $\sum_{i=1}^N w_i \bar{U}_i(\hat{\bfPsi}, \hat{\bftheta}) =
  \mathbf{0}$. We derive an asymptotic representation for
  $\sqrt{W_N}(\hat{\bftheta}-\bftheta^*)$, where $W_N = (\sum_i
  w_i)^2 / (\sum_i w_i^2)$ is the effective sample size.

  We first perform a Taylor expansion of the estimating function
  around $\bftheta^*$, which gives
  \[
    \mathbf{0} = \sum_{i=1}^N w_i \bar{U}_i(\hat{\bfPsi}, \bftheta^*)
    + \left(\sum_{i=1}^N w_i \frac{\partial \bar{U}_i(\hat{\bfPsi},
    \tilde{\bftheta})}{\partial\bftheta^\top}\right)(\hat{\bftheta} -
    \bftheta^*),
  \]
  where $\tilde{\bftheta}$ lies on the line segment between
  $\hat{\bftheta}$ and $\bftheta^*$. Rearranging for $(\hat{\bftheta}
  - \bftheta^*)$ and multiplying by $\sqrt{W_N}$ yields
  \[
    \sqrt{W_N}(\hat{\bftheta} - \bftheta^*) = -\left(\frac{1}{\sum_i
      w_i}\sum_{j=1}^N w_j \frac{\partial \bar{U}_j(\hat{\bfPsi},
    \tilde{\bftheta})}{\partial\bftheta^\top}\right)^{-1}
    \left(\frac{\sqrt{W_N}}{\sum_i w_i}\sum_{k=1}^N w_k
    \bar{U}_k(\hat{\bfPsi}, \bftheta^*)\right).
  \]
  By the law of the large numbers for weighted averages (justified by
  Assumptions~3, 5, and 8), the first inverse matrix term converges
  in probability to $(-\boldsymbol{\tau})^{-1}$, where
  $\boldsymbol{\tau}$ is defined in Theorem 1 in the main
  text. The scaling factor in the second term simplifies as
  $\sqrt{W_N} / (\sum_i w_i) = 1 / \sqrt{\sum_i w_i^2}$. Thus, we
  obtain the key representation:
  \begin{equation}\label{eq:proof_rep1}
    \sqrt{W_N}(\hat{\bftheta} - \bftheta^*) = \boldsymbol{\tau}^{-1}
    \frac{1}{\sqrt{\sum_{k=1}^N w_k^2}}\sum_{k=1}^N w_k
    \bar{U}_k(\hat{\bfPsi}, \bftheta^*) + o_p(1).
  \end{equation}

  Next, we analyze the sum on the right-hand side of
  \eqref{eq:proof_rep1}. Under Assumption~6 of stochastic
  equicontinuity, we can expand this sum around $\bfPsi^*$:
  \begin{equation}\label{eq:proof_rep2}
    \sum_{k=1}^N w_k \bar{U}_k(\hat{\bfPsi}, \bftheta^*) \approx
    \sum_{k=1}^N w_k \bar{U}_k(\bfPsi^*, \bftheta^*) +
    \left(\sum_{k=1}^N w_k\right) \frac{\partial
    \boldsymbol{\lambda}(\bfPsi^*, \bftheta^*)}{\partial \bfPsi^\top}
    (\hat{\bfPsi} - \bfPsi^*).
  \end{equation}
  The derivative matrix is shown to be $\boldsymbol{\kappa}$ from
  Theorem 1 in the main text by applying the Leibniz rule
  to $\boldsymbol{\lambda}(\bfPsi, \bftheta^*) =
  \mathbb{E}[\bar{U}_i(\bfPsi, \bftheta^*)]$ under Assumption~5. The
  conditional independence of imputations (Assumption~2) ensures that
  cross-product terms between different imputations vanish, yielding
  $\mathbb{E}[(\frac{1}{M}\sum_m U_i^m) (\sum_j
  \mathbf{S}_{i,mis}^j)^\top] = \mathbb{E}[U_i^m
  (\mathbf{S}_{i,mis}^m)^\top] = \boldsymbol{\kappa}$.

  Let $\boldsymbol{\psi}_i = \bar{U}_i(\bfPsi^*, \bftheta^*) +
  \boldsymbol{\kappa}\mathbf{D}_i(\bfPsi^*)$. We substitute the
  influence function representation for $\hat{\bfPsi}$ from
  Assumption~3, $(\hat{\bfPsi} - \bfPsi^*) = \frac{1}{N}\sum_j
  \mathbf{D}_j(\bfPsi^*) + o_p(N^{-1/2})$, into
  \eqref{eq:proof_rep2}. Standard arguments for two-step M-estimators
  show that we can approximate the correction term as a sum over
  individual influence components:
  \[
    \left(\sum_k w_k\right) \boldsymbol{\kappa} (\hat{\bfPsi} -
    \bfPsi^*) \approx \sum_{k=1}^N w_k \boldsymbol{\kappa}
    \mathbf{D}_k(\bfPsi^*).
  \]
  Therefore, the sum in \eqref{eq:proof_rep2} has the asymptotic representation:
  \[
    \sum_{k=1}^N w_k \bar{U}_k(\hat{\bfPsi}, \bftheta^*) =
    \sum_{k=1}^N w_k \left( \bar{U}_k(\bfPsi^*, \bftheta^*) +
    \boldsymbol{\kappa} \mathbf{D}_k(\bfPsi^*) \right) +
    o_p(\sqrt{N}) = \sum_{k=1}^N w_k \boldsymbol{\psi}_k + o_p(\sqrt{N}).
  \]
  The terms $\boldsymbol{\psi}_k$ are i.i.d. by Assumption~1 and have
  mean zero, since $\mathbb{E}[\bar{U}_k]=\mathbf{0}$ by definition
  of $\bftheta^*$ and $\mathbb{E}[\mathbf{D}_k]=\mathbf{0}$ as it is
  an influence function. Substituting this back into
  \eqref{eq:proof_rep1}, we arrive at the final influence function
  representation for $\hat{\bftheta}$:
  \begin{equation}\label{eq:proof_final_rep}
    \sqrt{W_N}(\hat{\bftheta} - \bftheta^*) = \boldsymbol{\tau}^{-1}
    \frac{1}{\sqrt{\sum_{k=1}^N w_k^2}}\sum_{k=1}^N w_k
    \boldsymbol{\psi}_k + o_p(1).
  \end{equation}

  The term on the right-hand side of \eqref{eq:proof_final_rep} is a
  scaled sum of i.i.d. random variables with mean zero. By the
  Lindeberg-Feller Central Limit Theorem for weighted sums, which
  holds under the moment condition of Assumption~7 and bounded
  weights of Assumption~8,
  \[
    \frac{1}{\sqrt{\sum_{k=1}^N w_k^2}} \sum_{k=1}^N w_k
    \boldsymbol{\psi}_k \xrightarrow{d} \calN(\mathbf{0},
    \mathrm{Var}(\boldsymbol{\psi}_k)).
  \]
  The variance of the influence contribution is $\boldsymbol{\Omega}
  = \mathrm{Var}(\boldsymbol{\psi}_k) =
  \mathbb{E}[\boldsymbol{\psi}_k\boldsymbol{\psi}_k^\top]$. Expanding
  this variance gives
  \begin{align*}
    \boldsymbol{\Omega} & =
    \mathrm{Var}\left(\bar{U}_i(\bfPsi^*,\bftheta^*) +
    \boldsymbol{\kappa} \mathbf{D}_i(\bfPsi^*)\right)             \\
    & = \mathbb{E}[\bar{U}_i\bar{U}_i^\top] +
    \boldsymbol{\kappa}\mathbb{E}[\mathbf{D}_i\mathbf{D}_i^\top]\boldsymbol{\kappa}^\top
    + \mathbb{E}[\boldsymbol{\kappa}\mathbf{D}_i\bar{U}_i^\top +
    \bar{U}_i\mathbf{D}_i^\top\boldsymbol{\kappa}^\top],
  \end{align*}
  where $\bar{U}_i = \bar{U}_i(\bfPsi^*,\bftheta^*)$ and
  $\mathbf{D}_i = \mathbf{D}_i(\bfPsi^*)$. Using the definition
  $\boldsymbol{\Lambda}(\bfPsi^*) =
  \mathbb{E}[\mathbf{D}_i\mathbf{D}_i^\top]$ as in
  Theorem 1, we recover the expression for
  $\boldsymbol{\Omega}$. From
  \eqref{eq:proof_final_rep}, the asymptotic distribution is therefore
  \[
    \sqrt{W_N}(\hat{\bftheta} - \bftheta^*) \xrightarrow{d}
    \calN(\mathbf{0},
    \boldsymbol{\tau}^{-1}\boldsymbol{\Omega}(\boldsymbol{\tau}^{-1})^\top),
  \]
  which completes the proof.
\end{proof}

\subsection{A consistent estimator of the asymptotic variance matrix $\Sigma$}
\label{ss_a_consistent_variance}
A consistent estimate of
    $\Sigma$ is then $\hat\Sigma = (\hat\tau)^{-1}\hat\Omega(\hat\tau)^{-1}$, where
\begin{equation}\label{eq:analysis_infer_consis}
    \begin{aligned}
    \hat\tau &= -\frac{1}{N}\sum_{i=1}^N \frac{\partial}{\partial\bftheta^\top}\bar{U}_i(\hat\Psi,\bftheta)\Big\vert_{\bftheta=\hat\bftheta}\\
    \hat\Omega &= \hat\Omega_c + \hat\kappa\hat\Lambda\hat\kappa^\top + \frac{1}{N}\sum_{i=1}^N\left[\hat\kappa \hat{D}_i\bar{U}_i(\hat\Psi,\hat\bftheta)^\top + (\hat\kappa \hat{D}_i\bar{U}_i(\hat\Psi,\hat\bftheta)^\top)^\top\right],\\
    \hat\Omega_c &= \frac{1}{N}\sum_{i=1}^N \bar{U}_i(\hat\Psi,\hat\bftheta)\bar{U}_i(\hat\Psi,\hat\bftheta)^\top,\quad \hat\kappa = \frac{1}{MN}\sum_{i=1}^N\sum_{m=1}^M \left[U_i^m(\hat\Psi,\hat\bftheta)(S_i^m(\hat\Psi) - S_{i,obs}(\hat\Psi))^\top\right],\\
    \hat\Lambda &= \frac{1}{N}\sum_{i=1}^N \hat D_i \hat D_i^\top, \mbox{~and~~}\hat D_i = \left[\hat I_{obs}\right]^{-1}S_{i,obs}(\hat\Psi).\\
    \end{aligned}
\end{equation}
In \eqref{eq:analysis_infer_consis}, $\hat I_{obs}$ is a consistent
estimate of $I_{obs} (\Psi^*)$, derived based on results in
\cite{louis1982finding}. That is, define
\begin{eqnarray*}
S_i(\Psi) &=& \frac{\partial}{\partial\Psi} \log f(\bfy_{i,obs},\bfy_{i,mis},\bfz_i,\bfeta_i,\bfxi_i\mid\bfx_i;\Psi)
\hspace*{2em}\text{and} \\
H_i(\Psi) &=& \frac{\partial^2}{\partial\Psi\partial\Psi^\top} \log f(\bfy_{i,obs},\bfy_{i,mis},\bfz_i,\bfeta_i,\bfxi_i\mid\bfx_i;\Psi),
\end{eqnarray*}
the first and second derivatives of the
complete-data log-likelihood contribution for observation unit $i$ under
the imputation model (1).
Then, $\hat I_{obs}$ is computed as
$$\hat I_{obs} = \frac{1}{N} \sum_{i=1}^N \left\{S_{i,obs}(\hat\Psi)
S_{i,obs}(\hat\Psi)^\top -\mathbb{E}\left[H_i(\hat\Psi) + S_i(\hat
\Psi)S_i(\hat \Psi)^\top\mid\bfy_{i,obs},\bfz_i,\bfx_i;\hat
\Psi\right]\right\},$$
and $S_{i,obs}(\hat\Psi)$ can be expressed
(see Supplementary Section \ref{ss_a_observed_score})
as
$$S_{i,obs}(\hat\Psi) = \mathbb{E}\left[S_i(\hat
\Psi)\mid\bfy_{i,obs},\bfz_i,\bfx_i;\hat \Psi\right],$$
where the expectations in both are with respect to the
conditional density $\phi(\bfy_{i,mis},\bfeta_i,\bfxi_i|\bfy_{i,obs},\bfz_i,\bfx_i;\hat\Psi)$.
Because of these conditional expectations, $S_{i,obs}(\hat\Psi)$ and $\hat
I_{obs}$ cannot be immediately computed analytically. However,
they can be obtained as a byproduct of an algorithm for multiple
imputation via Monte Carlo integration. This is described in
Section~4.2. The quantities in
\eqref{eq:analysis_infer_consis} can then be computed, giving the
estimated variance matrix $\hat \Sigma$ of the parameter estimates
$\hat{\boldsymbol{\theta}}$ of any analysis model
estimated using estimating equations of the form (8).

\subsection{Derivation of the observed score function}
\label{ss_a_observed_score}
According to Equation~(7), we have
\begin{equation}
  \begin{aligned}
    \mathbb{E} & (S_i(\Psi)\mid\bfy_{i,obs},\bfz_i,\bfx_i;\Psi)
    \\
    & = \int
    S_i(\Psi)\phi(\bfy_{i,mis},\bfeta_i,\bfxi_i\mid\bfy_{i,obs},\bfz_i,\bfx_i;\Psi)d\bfy_{i,mis}d\bfeta_i
    d\bfxi_i
    \\
    & = \int \frac{\partial}{\partial\Psi}\log
    f(\bfy_{i},\bfz_i,\bfeta_i,\bfxi_i\mid\bfx_i;\Psi)\phi(\bfy_{i,mis},\bfeta_i,\bfxi_i\mid\bfy_{i,obs},\bfz_i,\bfx_i;\Psi)d\bfy_{i,mis}d\bfeta_i
    d\bfxi_i
    \\
    & = \int \frac{\frac{\partial}{\partial\Psi}
    f(\bfy_i,\bfz_i,\bfeta_i,\bfxi_i\mid\bfx_i;\Psi)}{f(\bfy_i,\bfz_i,\bfeta_i,\bfxi_i\mid\bfx_i;\Psi)}\phi(\bfy_{i,mis},\bfeta_i,\bfxi_i\mid\bfy_{i,obs},\bfz_i,\bfx_i;\Psi)
    d\bfy_{i,mis}d\bfeta_i d\bfxi_i
    \\
    & = \int \frac{\frac{\partial}{\partial\Psi}
    f(\bfy_{i,mis},\bfeta_i,\bfxi_i\mid\bfy_{i,obs},\bfz_i,\bfx_i;\Psi)f(\bfy_{i,obs},\bfz_i\mid\bfx_i;\Psi)}{f(\bfy_{i,obs},\bfz_i\mid\bfx_i;\Psi)}d\bfy_{i,mis}d\bfeta_i
    d\bfxi_i
    \\
    & \quad + \int
    \frac{f(\bfy_{i,mis},\bfeta_i,\bfxi_i\mid\bfy_{i,obs},\bfz_i,\bfx_i;\Psi)\frac{\partial}{\partial\Psi}f(\bfy_{i,obs},\bfz_i\mid\bfx_i;\Psi)}{f(\bfy_{i,obs},\bfz_i\mid\bfx_i;\Psi)}d\bfy_{i,mis}d\bfeta_i
    d\bfxi_i
    \\
    & = 0 +
    \frac{\frac{\partial}{\partial\Psi}f(\bfy_{i,obs},\bfz_i\mid\bfx_i;\Psi)}{f(\bfy_{i,obs},\bfz_i\mid\bfx_i;\Psi)}
    \\
    & = \frac{\partial}{\partial\Psi}\log
    f(\bfy_{i,obs},\bfz_i\mid\bfx_i;\Psi) = S_{i,obs}(\Psi).
  \end{aligned}
\end{equation}

\subsection{Derivation of score function separation}
\label{ss_a_score_separation}

\begin{equation}
  \begin{aligned}
    S_i^m(\Psi) & = \frac{\partial}{\partial\Psi}\log
    f(\bfy_{i,obs},\bfy_{i,mis}^m,\bfz_i,\bfeta_i^m,\bfxi_i^m\mid
    \bfx_i;\Psi)
    \\
    & =
    \frac{\frac{\partial}{\partial\Psi}f(\bfy_{i,mis}^m,\bfeta_i^m,\bfxi_i^m\mid\bfy_{i,obs},\bfz_i,\bfx_i;\Psi)f(\bfy_{i,obs},\bfz_i\mid\bfx_i;\Psi)}{f(\bfy_{i,obs},\bfy_{i,mis}^m,\bfz_i,\bfeta_i^m,\bfxi_i^m\mid
    \bfx_i;\Psi)}
    \\
    & \quad +
    \frac{f(\bfy_{i,mis}^m,\bfeta_i^m,\bfxi_i^m\mid\bfy_{i,obs},\bfz_i,\bfx_i;\Psi)\frac{\partial}{\partial\Psi}f(\bfy_{i,obs},\bfz_i\mid\bfx_i;\Psi)}{f(\bfy_{i,obs},\bfy_{i,mis}^m,\bfz_i,\bfeta_i^m,\bfxi_i^m\mid
    \bfx_i;\Psi)}
    \\
    & =
    \frac{\frac{\partial}{\partial\Psi}f(\bfy_{i,mis}^m,\bfeta_i^m,\bfxi_i^m\mid\bfy_{i,obs},\bfz_i,\bfx_i;\Psi)}{f(\bfy_{i,mis}^m,\bfeta_i^m,\bfxi_i^m\mid\bfy_{i,obs},\bfz_i,\bfx_i;\Psi)}
    +
    \frac{\frac{\partial}{\partial\Psi}f(\bfy_{i,obs},\bfz_i\mid\bfx_i;\Psi)}{f(\bfy_{i,obs},\bfz_i\mid\bfx_i;\Psi)}
    \\
    & = S_{i,mis}^m(\Psi) + S_{i,obs}(\Psi).
  \end{aligned}
\end{equation}

\section{Supplement to Section 4: Further computational details}
\label{s_a_computations}

\subsection{{P{\'o}lya}-Gamma augmentation for mixed
responses}\label{sec:polya_gamma}

The proposed procedures outlined in Algorithms~1 and
2 requires drawing the latent variables
$\bfeta_i$ and $\bfxi_i$ from their full conditional posterior distributions. These posteriors are complex and lack standard forms due to the non-Gaussian likelihoods arising from binary and ordinal responses.
To circumvent the need for inefficient samplers like random-walk Metropolis-Hastings, we employ a data augmentation strategy based on the {P{\'o}lya}-Gamma augmentation \citep{polson2013bayesian}. We detail the procedure for  $\bfeta_i$ here; the corresponding derivations for $\bfxi_i$ are similar and areprovided in Section~\ref{sec:imputer-conditionals}.

Let the response vector be partitioned according to its data type, with $\mathcal{C}, \mathcal{B}, \mathcal{O} \subset
\{1,\ldots,J\}$ being disjoint index sets for continuous, binary, and ordinal variables, respectively, such that $\mathcal{C}\cup\mathcal{B}\cup\mathcal{O}=\{1,\ldots,J\}$. The posterior distribution of $\bfeta_i$ is given by
\begin{equation}\label{eq:mixed_posterior_eta}
  \begin{aligned}
    f(\bfeta_i\mid\bfy_{i},\bfxi_i,\bfx_i;\Psi) & \propto
    \prod_{j\in\mathcal{C}} g_c(y_{ij}\mid\bfeta_i;\Psi)\times
    \prod_{j\in\mathcal{B}}
    g_b(y_{ij}\mid\bfeta_i;\Psi)\times\prod_{j\in\mathcal{O}}g_o(y_{ij}\mid\bfeta_i;\Psi)\times
    \pi(\bfxi_i,\bfeta_i\mid\bfx_i;\Psi),
  \end{aligned}
\end{equation} 
where $g_c, g_b, g_o$ are the densities of continuous, binary, and
ordinal responses given in Section~2, and $\pi(\cdot)$ is the prior. 
The core of our approach is the {P{\'o}lya}-Gamma data augmentation scheme, which simplifies logistic-type likelihoods. A random variable $\omega$ follows the
{P{\'o}lya}-Gamma distribution $PG(b,c)$ where $b>0$ and
$c\in\mathbb{R}$, which can be described as a weighted sum of
independent Gamma random variables:
\begin{equation}
  \omega = \frac{1}{2\pi^2}\sum_{k=1}^\infty \frac{g_k}{(k-1/2)^2+c^2/(4\pi^2)},
\end{equation}
where $g_k\sim\Gamma(b,1)$, the Gamma distribution with shape and
rate parameters as $b$ and $1$, respectively.
The augmentation technique relies on the following key identity:
\begin{equation}\label{eq:polya_gamma1}
  \frac{\exp(\varphi)^{a}}{(1 + \exp(\varphi))^b}
  =
  2^{-b}\exp(\kappa\varphi)\mathbb{E}_{\omega}\left[\exp(-\omega\varphi^2/2)\right],
\end{equation}
where $\kappa = a - b/2$, $\omega\sim PG(b,0)$. This identity allows us to express the logistic function as a scale mixture of normals. A crucial property for Gibbs sampling is that the full conditional distribution of the auxiliary variable $\omega$ is also a {P{\'o}lya}-Gamma distribution: $\omega\mid
\varphi\sim PG(b, \varphi)$.

\subsection{Augmentation by Response Type}\label{sec:polya_gamma_augmentation}
We now apply this framework to each non-Gaussian response type.

\paragraph{Binary Responses.}
For binary responses $y_{ij}\in\{0,1\}$, 
the logistic likelihood is $p(y_{ij} \mid \boldsymbol{\eta}_i; \Psi) = \exp(\psi_{ij})^{y_{ij}}/(1+\exp(\psi_{ij}))$, where $\psi_{ij} = \alpha_{0j}+\boldsymbol{\alpha}_j^\top\boldsymbol{\eta}_i$. Applying \eqref{eq:polya_gamma1} with $a=y_{ij}$ and $b=1$, we introduce an auxiliary variable $\omega_{ij}^b \sim \text{PG}(1, 0)$. Conditional on $\omega_{ij}^b$, the likelihood becomes Gaussian in $\psi_{ij}$. The joint contribution of all binary responses, conditional on $\boldsymbol{\omega}_i^b = \{\omega_{ij}^b\}_{j \in \mathcal{B}}$, is proportional to:
\begin{align}
\prod_{j\in\mathcal{B}} p(y_{ij}, \omega_{ij}^b \mid \boldsymbol{\eta}_i; \Psi) 
&\propto \prod_{j\in\mathcal{B}} \exp\left(-\frac{\omega_{ij}^b}{2}(\alpha_{0j}+\boldsymbol{\alpha}_j^\top\boldsymbol{\eta}_i)^2 + \kappa_{ij}(\alpha_{0j}+\boldsymbol{\alpha}_j^\top\boldsymbol{\eta}_i)\right) \\
&\propto \exp\left[-\frac{1}{2}\left(\boldsymbol{\eta}_i^\top\mathbf{A}_b^\top \mathbf{D}_{\omega_i^b}\mathbf{A}_b\boldsymbol{\eta}_i - 2(\boldsymbol{\kappa}_{ib}-\mathbf{D}_{\omega_i^b}\boldsymbol{\alpha}_{0b})^\top\mathbf{A}_b\boldsymbol{\eta}_i\right)\right],
\end{align}
where $\kappa_{ij} = y_{ij}-1/2$, $\mathbf{A}_b$ is the matrix with rows $\boldsymbol{\alpha}_j^\top$ for $j \in \mathcal{B}$, $\boldsymbol{\alpha}_{0b}$ is the vector of intercepts, $\boldsymbol{\kappa}_{ib}$ is the vector of $\kappa_{ij}$ terms, and $\mathbf{D}_{\omega_i^b} = \text{diag}(\boldsymbol{\omega}_i^b)$.

\paragraph{Ordinal Responses.}
For an ordinal response $y_{ij} \in \{0, \dots, M_j\}$, the Graded Response Model (GRM) defines the probability of observing category $k$ as a difference of cumulative logistic probabilities:
\begin{equation}
    P(y_{ij}=k \mid \boldsymbol{\eta}_i; \Psi) = P(y_{ij} \ge k \mid \boldsymbol{\eta}_i) - P(y_{ij} \ge k+1 \mid \boldsymbol{\eta}_i),
\end{equation}
where $P(y_{ij} \ge k \mid \boldsymbol{\eta}_i) = \text{logit}^{-1}(\boldsymbol{\alpha}_j^\top\boldsymbol{\eta}_i - \alpha_{0j,k})$ and the thresholds satisfy $-\infty = \alpha_{0j,0} < \alpha_{0j,1} < \dots < \alpha_{0j,M_j+1} = \infty$.
This model is equivalent to assuming a continuous latent variable $W_{ij} \sim \text{Logistic}(\boldsymbol{\alpha}_j^\top\boldsymbol{\eta}_i, 1)$, where $y_{ij}=k$ if and only if $\alpha_{0j,k} < W_{ij} \le \alpha_{0j,k+1}$. The density of $W_{ij}$ is
\begin{equation}
p(W_{ij} \mid \boldsymbol{\eta}_i) = \frac{\exp(-(W_{ij} - \boldsymbol{\alpha}_j^\top\boldsymbol{\eta}_i))}{(1+\exp(-(W_{ij} - \boldsymbol{\alpha}_j^\top\boldsymbol{\eta}_i)))^2}.
\end{equation}
Letting $\varphi_{ij} = W_{ij} - \boldsymbol{\alpha}_j^\top\boldsymbol{\eta}_i$, this density is $e^{-\varphi_{ij}} / (1+e^{-\varphi_{ij}})^2 = e^{\varphi_{ij}} / (1+e^{\varphi_{ij}})^2$. This matches the form of \eqref{eq:polya_gamma1} with $a=1$ and $b=2$, yielding $\kappa = a-b/2=0$. We can therefore augment this density with a P{\'o}lya-Gamma variable $\omega_{ij}^o \sim \text{PG}(2,0)$, leading to a two-stage augmentation. The complete-data likelihood for a single ordinal response is:
\begin{align}
p(y_{ij}, W_{ij}, \omega_{ij}^o \mid \boldsymbol{\eta}_i; \Psi) &= p(y_{ij} \mid W_{ij}) p(W_{ij} \mid \boldsymbol{\eta}_i, \omega_{ij}^o) p(\omega_{ij}^o) \\
&\propto I(\alpha_{0j, y_{ij}} < W_{ij} \le \alpha_{0j, y_{ij}+1}) \times \exp\left[-\frac{\omega_{ij}^o}{2}(W_{ij} - \boldsymbol{\alpha}_j^\top\boldsymbol{\eta}_i)^2\right].
\end{align}
The joint likelihood contribution from all ordinal responses, conditional on the auxiliary variables $\mathbf{W}_i=\{W_{ij}\}_{j\in\mathcal{O}}$ and $\boldsymbol{\omega}_i^o=\{\omega_{ij}^o\}_{j\in\mathcal{O}}$, is thus proportional to:
\begin{equation}
\prod_{j\in\mathcal{O}} p(y_{ij}, W_{ij}, \omega_{ij}^o \mid \boldsymbol{\eta}_i; \Psi) \propto \exp\left[-\frac{1}{2}\left(\boldsymbol{\eta}_i^\top\mathbf{A}_o^\top \mathbf{D}_{\omega_i^o}\mathbf{A}_o\boldsymbol{\eta}_i - 2\mathbf{W}_i^\top \mathbf{D}_{\omega_i^o}\mathbf{A}_o\boldsymbol{\eta}_i\right)\right],
\end{equation}
where $\mathbf{A}_o$ is the $|\mathcal{O}| \times d_\eta$ matrix with rows $\boldsymbol{\alpha}_j^\top$ for $j \in \mathcal{O}$, and $\mathbf{D}_{\omega_i^o} = \text{diag}(\boldsymbol{\omega}_i^o)$.

\paragraph{Continuous Responses.}
For continuous responses, the standard linear model $y_{ij} \sim \mathcal{N}(\alpha_{0j} + \boldsymbol{\alpha}_j^\top\boldsymbol{\eta}_i, \sigma_j^2)$ is assumed. The likelihood contribution is already Gaussian in $\boldsymbol{\eta}_i$:
\begin{equation}
\prod_{j\in\mathcal{C}} p(y_{ij} \mid \boldsymbol{\eta}_i; \Psi) \propto \exp\left[-\frac{1}{2}\left(\boldsymbol{\eta}_i^\top\mathbf{A}_c^\top\boldsymbol{\Sigma}_c^{-1}\mathbf{A}_c\boldsymbol{\eta}_i - 2(\mathbf{y}_{ic}-\boldsymbol{\alpha}_{0c})^\top\boldsymbol{\Sigma}_c^{-1}\mathbf{A}_c\boldsymbol{\eta}_i\right)\right],
\end{equation}
where $\mathbf{A}_c$ is the $|\mathcal{C}| \times d_\eta$ matrix with rows $\boldsymbol{\alpha}_j^\top$ for $j \in \mathcal{C}$, $\boldsymbol{\alpha}_{0c}$ is the vector of intercepts, $\mathbf{y}_{ic}$ is the vector of observed continuous values, and $\boldsymbol{\Sigma}_c = \text{diag}(\{\sigma_j^2\}_{j \in \mathcal{C}})$.

\paragraph{Gibbs Sampling of Augmented Variables.}
The augmentation scheme defines a Gibbs sampler that iterates between drawing the auxiliary and primary latent variables. For each individual $i$, the sampling steps for the augmented variables are:
\begin{enumerate}
    \item \textbf{Sample P{\'o}lya-Gamma variables for binary data:} For each $j \in \mathcal{B}$, draw
    \begin{equation}
        \omega_{ij}^b \mid \boldsymbol{\eta}_i, \Psi \sim \PG\left(1, |\alpha_{0j}+\boldsymbol{\alpha}_j^\top\boldsymbol{\eta}_i|\right).
    \end{equation}
    \item \textbf{Sample latent variables for ordinal data:} For each $j \in \mathcal{O}$:
    \begin{enumerate}
        \item To draw the latent response $W_{ij}$, we examine its full conditional posterior. From the complete-data likelihood, the terms involving $W_{ij}$ are $I(\alpha_{0j, y_{ij}} < W_{ij} \le \alpha_{0j, y_{ij}+1})$ and $\exp[-\frac{\omega_{ij}^o}{2}(W_{ij} - \boldsymbol{\alpha}_j^\top\boldsymbol{\eta}_i)^2]$. Combining these reveals the posterior:
        \begin{equation}
            W_{ij} \mid \boldsymbol{\eta}_i, y_{ij}, \omega_{ij}^o, \Psi \sim \TN\left(\boldsymbol{\alpha}_j^\top\boldsymbol{\eta}_i, (\omega_{ij}^o)^{-1}; (\alpha_{0j, y_{ij}}, \alpha_{0j, y_{ij}+1}]\right).
        \end{equation}
        This draw is performed using inverse transform sampling. Let $U_{ij} \sim U(0,1)$, $\mu_{W} = \boldsymbol{\alpha}_j^\top\boldsymbol{\eta}_i$, $\sigma_{W}^2 = 1/\omega_{ij}^o$, and let $q(x) = \Phi((x - \mu_W)/\sigma_W)$ be the standard normal CDF evaluated at the scaled argument. Then $W_{ij}$ is sampled as:
        \begin{equation}
        W_{ij} = \mu_W + \sigma_W \Phi^{-1}\left(U_{ij}(q(\alpha_{0j,y_{ij}+1}) - q(\alpha_{0j,y_{ij}})) + q(\alpha_{0j,y_{ij}})\right).
        \end{equation}
        \item Draw the P{\'o}lya-Gamma variable $\omega_{ij}^o$ conditional on the newly sampled $W_{ij}$:
        \begin{equation}
            \omega_{ij}^o \mid \boldsymbol{\eta}_i, W_{ij}, \Psi \sim \PG\left(2, |\boldsymbol{\alpha}_j^\top\boldsymbol{\eta}_i - W_{ij}|\right).
        \end{equation}
    \end{enumerate}
\end{enumerate}

\paragraph{Full Conditional Posterior of $\boldsymbol{\eta}_i$.}
Augmenting the likelihood with the full set of auxiliary variables $\boldsymbol{\Omega}_i = \{\boldsymbol{\omega}_i^b, \mathbf{W}_i, \boldsymbol{\omega}_i^o\}$ and combining with the priors $\boldsymbol{\eta}_i \mid \mathbf{x}_i \sim \mathcal{N}(\boldsymbol{\beta}\mathbf{x}_i, \mathbf{I})$ and $\boldsymbol{\xi}_i \mid \boldsymbol{\eta}_i, \mathbf{x}_i \sim \mathcal{N}(\boldsymbol{\zeta}\mathbf{x}_i+\boldsymbol{\kappa}\boldsymbol{\eta}_i, \mathbf{I})$, the full conditional for $\boldsymbol{\eta}_i$ is a multivariate normal distribution:
\begin{equation}
\boldsymbol{\eta}_i \mid \mathbf{y}_i, \boldsymbol{\xi}_i, \mathbf{x}_i, \boldsymbol{\Omega}_i, \Psi \sim \mathcal{N}(\boldsymbol{\mu}_{\eta_i}, \boldsymbol{\Sigma}_{\eta_i}),
\end{equation}
with posterior covariance and mean given by:
\begin{align}
\boldsymbol{\Sigma}_{\eta_i} &= (\mathbf{A}^\top \mathbf{D}_{\Omega_i} \mathbf{A} + \boldsymbol{\kappa}^\top\boldsymbol{\kappa} + \mathbf{I})^{-1}, \\
\boldsymbol{\mu}_{\eta_i} &= \boldsymbol{\Sigma}_{\eta_i}\left(\mathbf{A}^\top\boldsymbol{\mu}_{\text{data}} + \boldsymbol{\kappa}^\top(\boldsymbol{\xi}_i-\boldsymbol{\zeta}\mathbf{x}_i) + \boldsymbol{\beta}\mathbf{x}_i\right).
\end{align}
The block matrices are defined as:
\begin{gather}
\mathbf{A} = 
\begin{bmatrix} 
\mathbf{A}_c \\ \mathbf{A}_b \\ \mathbf{A}_o 
\end{bmatrix}, \quad
\mathbf{D}_{\Omega_i} = 
\begin{bmatrix} 
\boldsymbol{\Sigma}_c^{-1} & \mathbf{0} & \mathbf{0} \\ 
\mathbf{0} & \mathbf{D}_{\omega_i^b} & \mathbf{0} \\ 
\mathbf{0} & \mathbf{0} & \mathbf{D}_{\omega_i^o} 
\end{bmatrix}, \quad
\boldsymbol{\mu}_{\text{data}} = 
\begin{bmatrix} 
\boldsymbol{\Sigma}_c^{-1}(\mathbf{y}_{ic}-\boldsymbol{\alpha}_{0c}) \\ 
\boldsymbol{\kappa}_{ib}-\mathbf{D}_{\omega_i^b}\boldsymbol{\alpha}_{0b} \\ 
\mathbf{D}_{\omega_i^o}\mathbf{W}_i 
\end{bmatrix}.
\end{gather}
Thus, conditional on the augmented data $\boldsymbol{\Omega}_i$, we can draw $\boldsymbol{\eta}_i$ directly from this multivariate normal distribution, ensuring an efficient Gibbs sampling step.

\subsection{P{\'o}lya-Gamma Augmentation for the Missing Data Mechanism}\label{sec:imputer-conditionals}

Complementary to the sampling of the response model's latent variables $\boldsymbol{\eta}_i$, we now detail the procedure for updating the latent variables $\boldsymbol{\xi}_i$, which govern the missing data mechanism. The sampling of $\boldsymbol{\xi}_i$ follows a similar P{\'o}lya-Gamma augmentation strategy, as the missingness model is also based on a logistic link.

The full conditional posterior distribution for $\boldsymbol{\xi}_i$ depends on the likelihood of the missingness indicators, $\mathbf{z}_i = (z_{i1}, \dots, z_{iJ})^\top$, and its own structural prior. The model is specified as:
\begin{align}
    \text{Likelihood: } & p(\mathbf{z}_i \mid \boldsymbol{\xi}_i; \Psi) = \prod_{j=1}^J \frac{\exp((\gamma_{0j}+\boldsymbol{\gamma}_j^\top\boldsymbol{\xi}_i)z_{ij})}{1+\exp(\gamma_{0j}+\boldsymbol{\gamma}_j^\top\boldsymbol{\xi}_i)}, \label{eq:missing_likelihood} \\
    \text{Prior: } & p(\boldsymbol{\xi}_i \mid \boldsymbol{\eta}_i, \mathbf{x}_i; \Psi) \sim \mathcal{N}(\boldsymbol{\zeta}\mathbf{x}_i + \boldsymbol{\kappa}\boldsymbol{\eta}_i, \mathbf{I}). \label{eq:xi_prior}
\end{align}
Here, $\gamma_{0j}$ is the intercept and $\boldsymbol{\gamma}_j$ is the vector of loadings for the $j$-th missingness indicator.

The logistic form of the likelihood in \eqref{eq:missing_likelihood} makes direct sampling of $\boldsymbol{\xi}_i$ intractable. We apply the P{\'o}lya-Gamma identity from \eqref{eq:polya_gamma1} by setting the linear predictor $\phi_{ij} = \gamma_{0j}+\boldsymbol{\gamma}_j^\top\boldsymbol{\xi}_i$, with $a=z_{ij}$ and $b=1$. This introduces a set of auxiliary variables $\boldsymbol{\omega}_i^z = \{\omega_{ij}^z\}_{j=1}^J$, where each $\omega_{ij}^z \sim \PG(1,0)$.

\paragraph{Gibbs Sampling of Augmented Variables.}
Within each main Gibbs iteration, the first step is to sample the auxiliary variables for the missingness model. For each individual $i$ and for each variable $j \in \{1, \dots, J\}$, we draw:
\begin{equation}
    \omega_{ij}^z \mid \boldsymbol{\xi}_i, \Psi \sim \PG\left(1, |\gamma_{0j}+\boldsymbol{\gamma}_j^\top\boldsymbol{\xi}_i|\right).
\end{equation}

\paragraph{Full Conditional Posterior of $\boldsymbol{\xi}_i$.}
Conditional on the auxiliary variables $\boldsymbol{\omega}_i^z$, the posterior for $\boldsymbol{\xi}_i$ becomes Gaussian. The augmented posterior is proportional to the product of the augmented likelihood and the prior:
\begin{align}
p(\boldsymbol{\xi}_i \mid \mathbf{z}_i, \boldsymbol{\eta}_i, \mathbf{x}_i, \boldsymbol{\omega}_i^z, \Psi) &\propto \left( \prod_{j=1}^J \exp\left[-\frac{\omega_{ij}^z}{2}\phi_{ij}^2 + \tilde{z}_{ij}\phi_{ij}\right] \right) \times \exp\left[-\frac{1}{2}(\boldsymbol{\xi}_i - (\boldsymbol{\zeta}\mathbf{x}_i + \boldsymbol{\kappa}\boldsymbol{\eta}_i))^\top(\boldsymbol{\xi}_i - (\boldsymbol{\zeta}\mathbf{x}_i + \boldsymbol{\kappa}\boldsymbol{\eta}_i))\right],
\end{align}
where $\tilde{z}_{ij} = z_{ij} - 1/2$, and $\phi_{ij} = \gamma_{0j}+\boldsymbol{\gamma}_j^\top\boldsymbol{\xi}_i$. By collecting all terms involving $\boldsymbol{\xi}_i$, we can derive the parameters of its multivariate normal posterior $\boldsymbol{\xi}_i \mid \cdot \sim \mathcal{N}(\boldsymbol{\mu}_{\xi_i}, \boldsymbol{\Sigma}_{\xi_i})$, with posterior covariance and mean given by:
\begin{align}
\boldsymbol{\Sigma}_{\xi_i} &= (\boldsymbol{\Gamma}^\top \mathbf{D}_{\omega_i^z} \boldsymbol{\Gamma} + \mathbf{I})^{-1}, \\
\boldsymbol{\mu}_{\xi_i} &= \boldsymbol{\Sigma}_{\xi_i}\left( \boldsymbol{\Gamma}^\top(\boldsymbol{\tilde{z}}_i - \mathbf{D}_{\omega_i^z}\boldsymbol{\gamma}_0) + \boldsymbol{\zeta}\mathbf{x}_i + \boldsymbol{\kappa}\boldsymbol{\eta}_i \right),
\end{align}
where $\boldsymbol{\Gamma}$ is the $J \times d_\xi$ matrix with rows $\boldsymbol{\gamma}_j^\top$, $\boldsymbol{\gamma}_0$ is the $J \times 1$ vector of intercepts $\{\gamma_{0j}\}$, $\mathbf{D}_{\omega_i^z} = \text{diag}(\boldsymbol{\omega}_i^z)$, and $\boldsymbol{\tilde{z}}_i$ is the vector of $\{\tilde{z}_{ij}\}$ terms. Given the auxiliary variables, $\boldsymbol{\xi}_i$ can be sampled directly from this multivariate normal distribution, ensuring an efficient update.

\newpage

\section{Supplement to Section 5: More on the simulation studies}
\label{s_a_simulations}

\subsection{Simulation Settings}\label{ss:simu_settings_supp}

\begin{figure}[htbp]
    \centering
    \includegraphics[width=.8\linewidth]{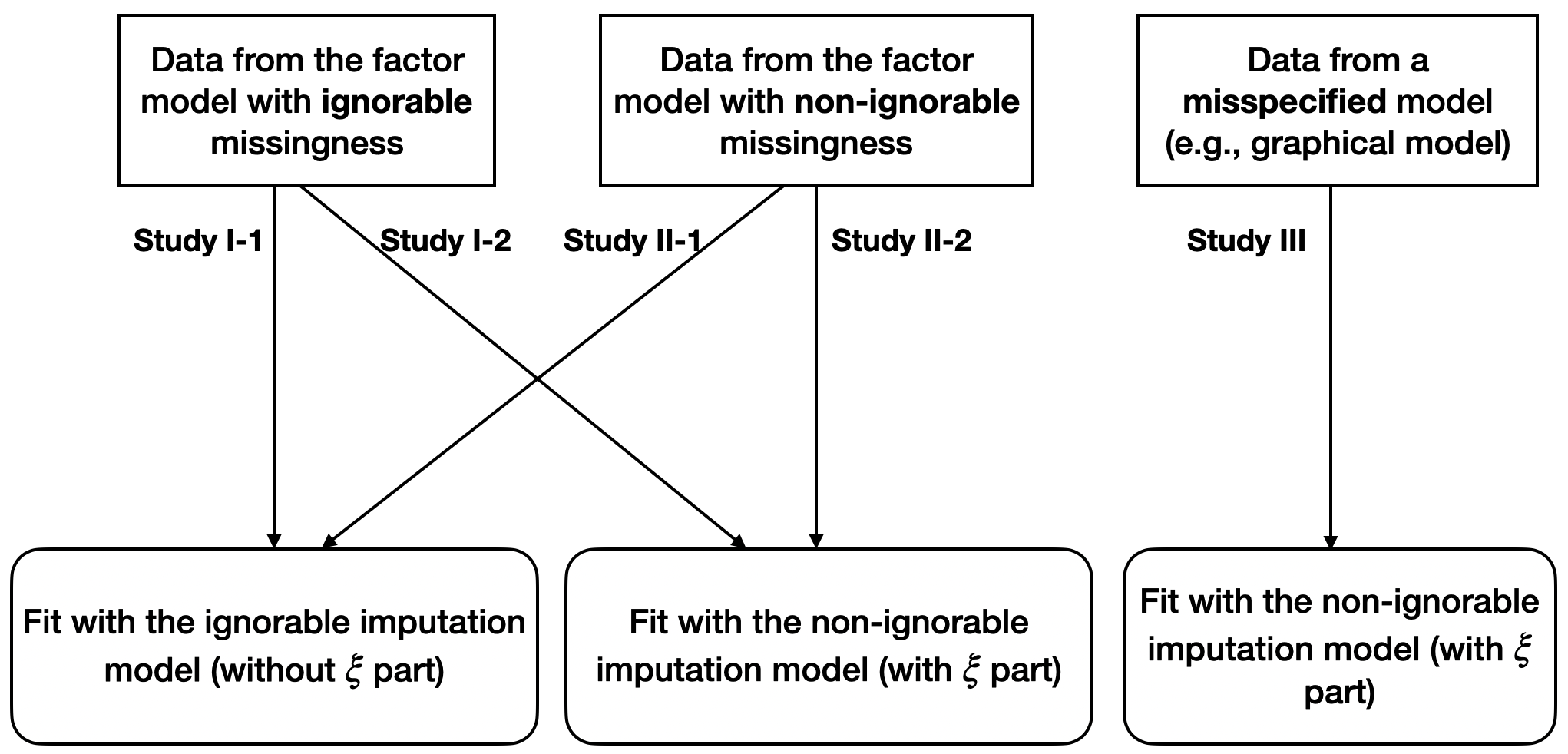}
    \caption{The overall simulation settings.}
    \label{fig:simu_settings_supp}
\end{figure}

To evaluate the performance and robustness of the proposed method, we conduct simulation studies. We consider five broad settings which differ in the nature of true and assumed joint distributions of the data, as summarised in Figure \ref{fig:simu_settings_supp}. In each study we generate 100 replicate datasets with $N=5000$ observations. Each observation comprises a data vector $\mathbf{y}_i$ and a vector $\mathbf{z}_i$ of response indicators, where $y_{ij}$ is treated as missing when $z_{ij}=0$. Here $\mathbf{y}_i$ includes $J_C$ continuous and $J_B$ binary variables. Parameters of the imputation model for each dataset are estimated using Algorithm~1, with $T=3000$ stochastic optimization iterations of which the first $T_{0}=1000$ are discarded as burn-in. The imputation is then carried out using Algorithm~2 with $T=3000$, $T_{0}=1000$ and $k=100$, to generate $M=20$ multiply imputed datasets for estimation.

In the first four settings (Studies I and II) the true data-generating process is of the form of the latent variable model (1) in the main text. The model has $K_{1}=4$ substantive factors in $\boldsymbol{\eta}_{i}$ and $K_{2}=1$ missingness factor in $\boldsymbol{\xi}_{i}$, generated from (2) with no $\mathbf{x}_{i}$. For the parameters $\boldsymbol{\kappa}$ we consider two settings: one where the true missingness is ignorable, so that we set $\boldsymbol{\kappa}=\mathbf{0}$, and one where it is non-ignorable, with the elements of $\boldsymbol{\kappa}$ drawn independently from $U(1.0, 2.0)$. We then generate $J_{C}=10$ continuous variables from (3) and $J_{B}=10$ binary variables from (4). For model parameters, we draw intercepts $\alpha_{0j}$ independently from standard normal distributions and factor loading parameters $\alpha_{jk}$ from uniform distributions $U(0.5,1.5)$, for $j=1,\ldots,J=J_{C}+J_{B}$ and $k=1,\ldots,K_1=4$. The residual standard deviations for the continuous variables are all set to $\sigma_j=0.5$. Missingness indicators $z_{ij}$ are then generated from model (5), with the intercept parameters $\gamma_{0j}$ drawn independently from $N(-3, 0.5^2)$ and the loading parameters $\gamma_{jk}$ independently from $U(0.2, 1.2)$, for $j=1,\ldots,J$ and $k=1,\ldots,K_2=1$. Variables $y_{ij}$ are then coded as missing when $z_{ij}=0$. This yields an average missingness rate of 7\%.

The estimated imputation model is correctly specified for these first four settings, except that in two cases it does not match the data-generating model in whether it takes the missingness to be ignorable ($\boldsymbol{\kappa}$ fixed at $\mathbf{0}$) or non-ignorable ($\boldsymbol{\kappa}$ is estimated). These combinations of settings in these studies are as follows:
\begin{itemize}
\item Study I-1: Both true model and estimated imputation model are ignorable.
\item Study I-2: True model is ignorable, estimated imputation model is non-ignorable.
\item Study II-1: True model is non-ignorable, estimated imputation model is ignorable.
\item Study II-2: Both true model and estimated imputation model are non-ignorable.
\end{itemize}

The fifth simulation (Study III) examines the robustness of the approach in a situation where the true joint distribution of the data does not match the latent variable form assumed by our imputation model. Specifically, we generate $J_{C}=5$ continuous variables from a sparse Gaussian graphical model and $J_{B}=5$ binary variables from a sparse Ising model. The specific values of the precision matrices are given in Supplementary Section \ref{ss_a_simIII}. The continuous and binary components are taken to be independent of each other. Missing data are introduced through an ignorable mechanism where the missingness probability for the first nine variables depends on the tenth (binary) variable which is fully observed, through $P(z_{ij}=0\mid y_{i,10}=1)=0.1$ and $P(z_{ij}=0\mid y_{i,10}=0)=0.4$ for $j=1,\ldots,9$. This results in an overall missing rate of approximately 17\%. In this study the estimated imputation model is still of the latent variable form, allowing for non-ignorability ($\boldsymbol{\kappa}\ne \mathbf{0}$) for additional flexibility (results for estimation under $\boldsymbol{\kappa}=\mathbf{0}$ are shown for comparison in Supplementary Section \ref{ss_a_IIIignorable}). We consider two settings for it, a parsimonious model with $K_{1}=1$ and a more flexible one with $K_{1}=4$ substantive latent factors, with $K_{2}=1$ nonresponse factor in both cases.

As the analysis model, in all the studies we focus on estimating the marginal means of each of the $J_{B}+J_{C}$ variables in $\mathbf{y}$. In addition, in Supplementary Section \ref{ss_a_sim_condmean} we report results for another analysis model, the conditional means of the rest of the variables given one of the binary variables. Conclusions for them are essentially the same as those for the marginal means discussed in the main text.

\subsection{Additional Results for Studies I-1 and I-2}\label{ss:add_sim_results}

This section provides the results for Studies I-1 and I-2, where the true data-generating mechanism has ignorable missingness.

\begin{figure}[p]\footnotesize
    \centering
    \begin{subfigure}{.4\textwidth}
        \centering
        \includegraphics[width=.78\linewidth]{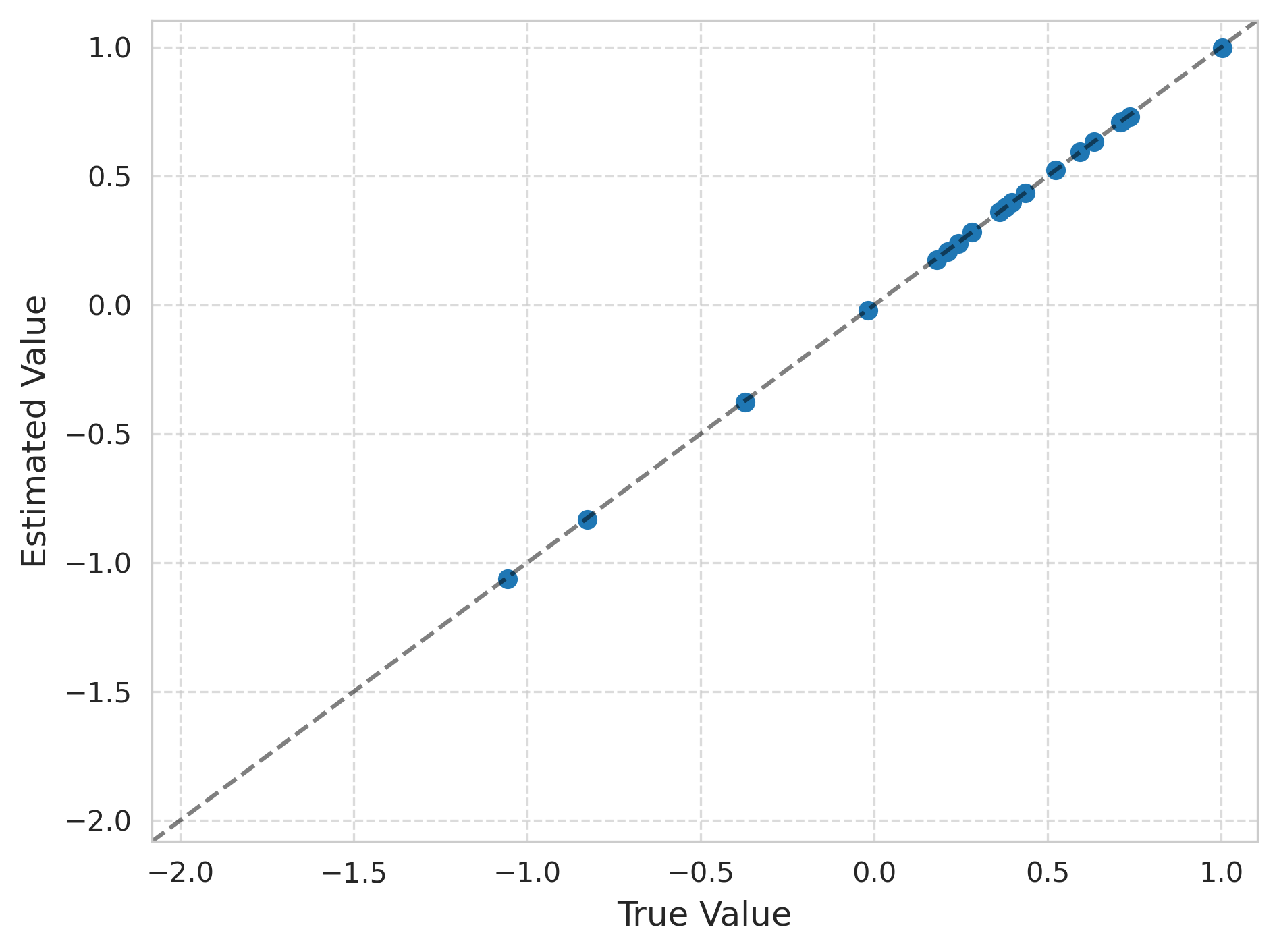}
        \caption{True vs. means of point estimates}
    \end{subfigure}%
    \begin{subfigure}{.55\textwidth}
        \centering
        \includegraphics[width=.9\linewidth]{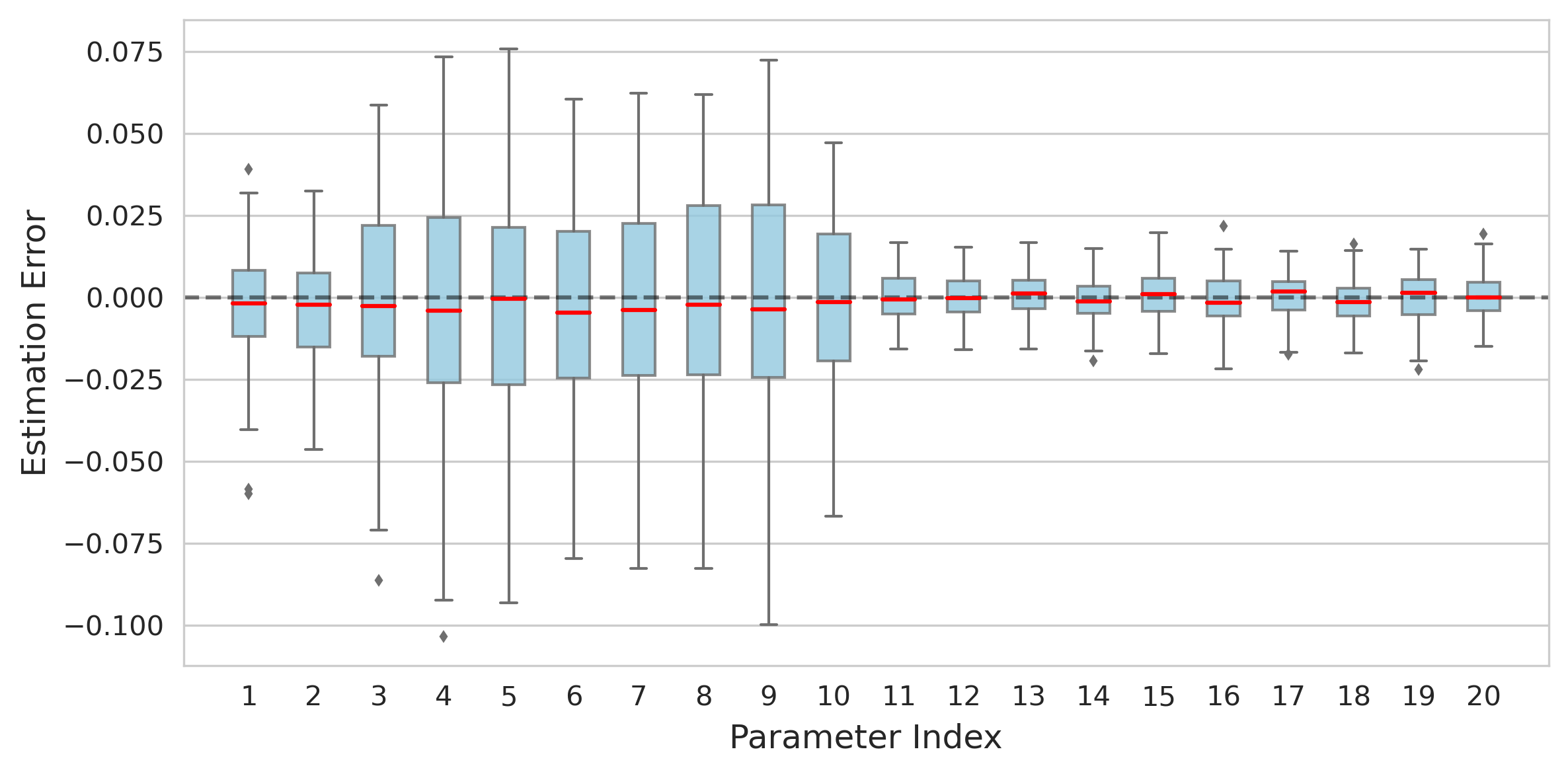}
        \caption{Distributions of point estimates $-$ true values}
    \end{subfigure}
    \begin{subfigure}{.4\textwidth}
        \centering
        \includegraphics[width=.78\linewidth]{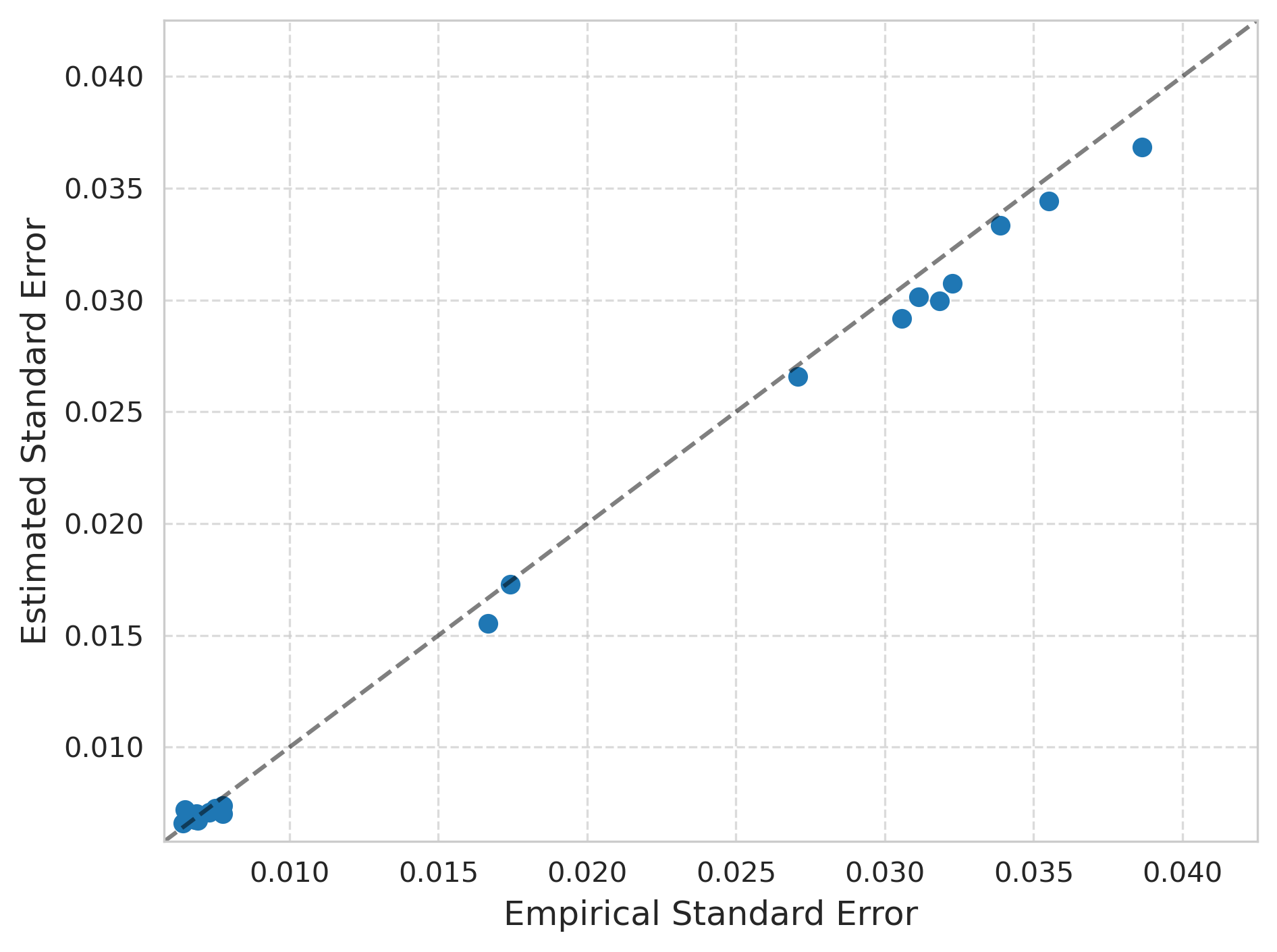}
        \caption{Empirical vs. means of estimated standard errors}
    \end{subfigure}%
    \begin{subfigure}{.55\textwidth}
        \centering
        \includegraphics[width=.9\linewidth]{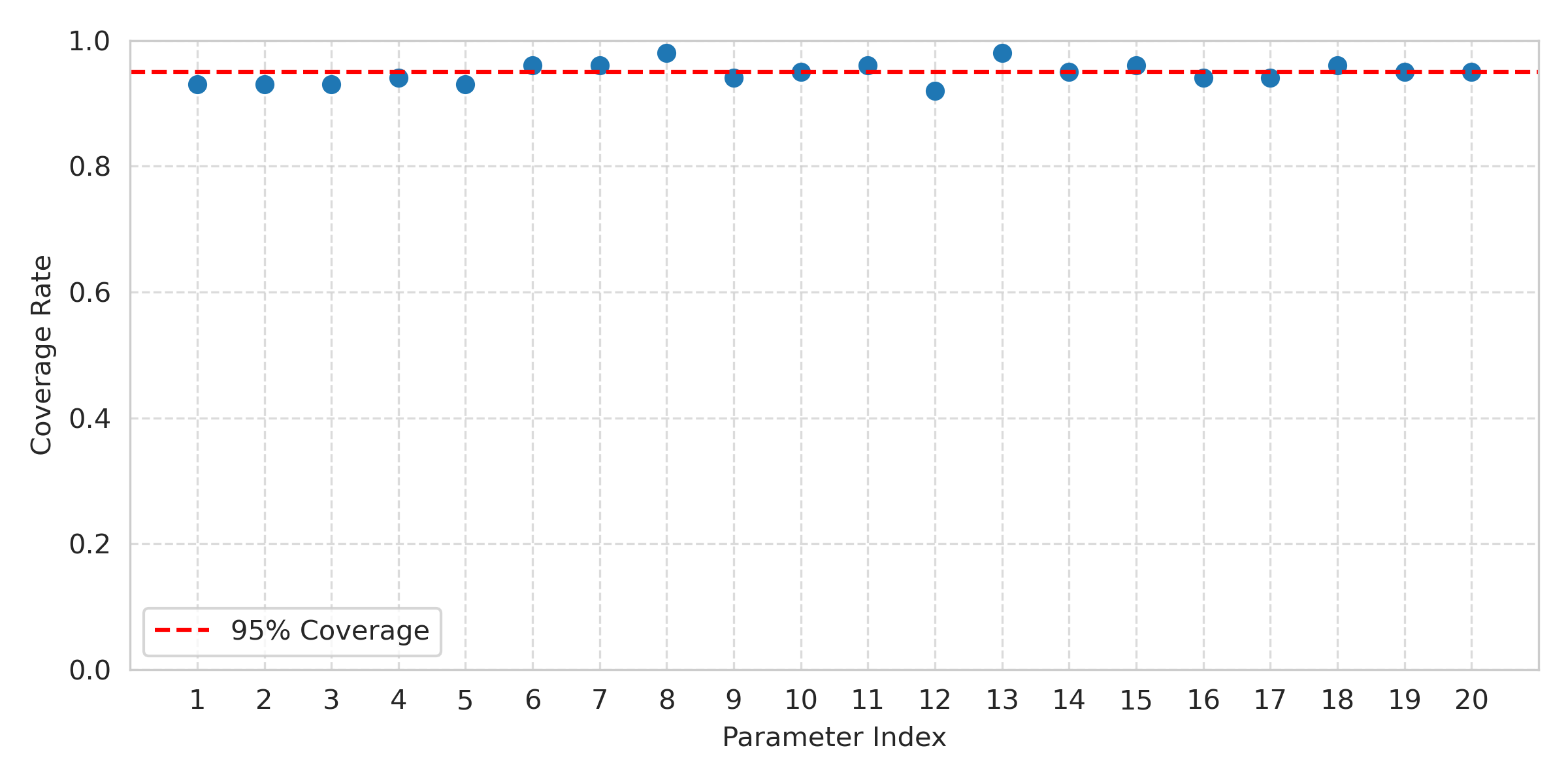}
        \caption{Coverage rates of 95\% confidence intervals}
    \end{subfigure}
    \caption{\small
    Estimation results for marginal means in Study I-1, where both the true and assumed imputation model have ignorable missingness.
    }
    \label{f_sim_I_1}
\end{figure}

In Study I-1 (Figure \ref{f_sim_I_1}) the estimated imputation model correctly assumes ignorability. The results show that the method works very well, providing accurate point estimates, correct estimates of their standard errors, and achieving coverage rates for confidence intervals that align closely with the nominal 95\% level.

\begin{figure}[p]\footnotesize
    \centering
    \begin{subfigure}{.4\textwidth}
        \centering
        \includegraphics[width=.8\linewidth]{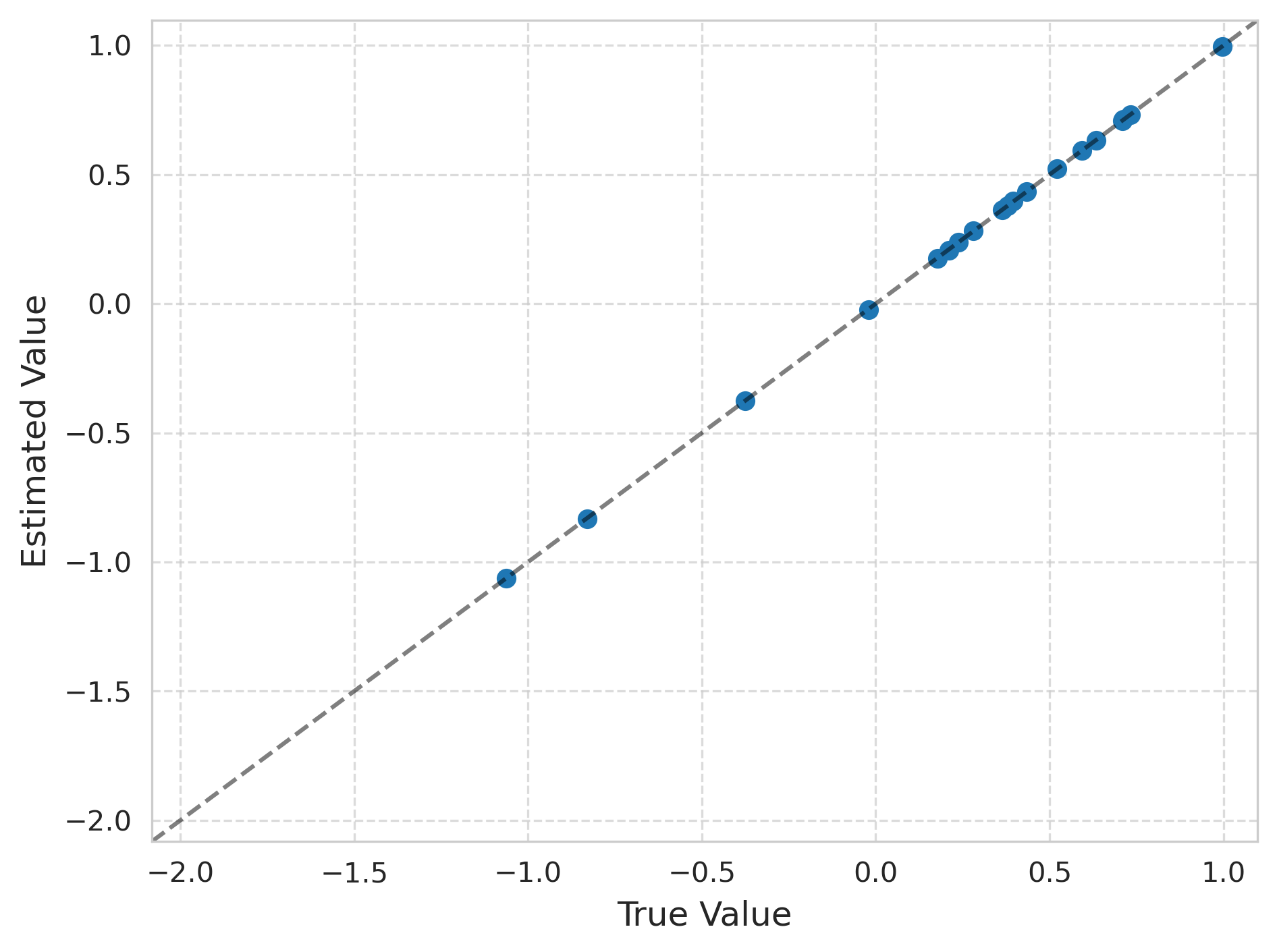}
        \caption{True vs. means of point estimates}
    \end{subfigure}%
    \begin{subfigure}{.55\textwidth}
        \centering
        \includegraphics[width=.9\linewidth]{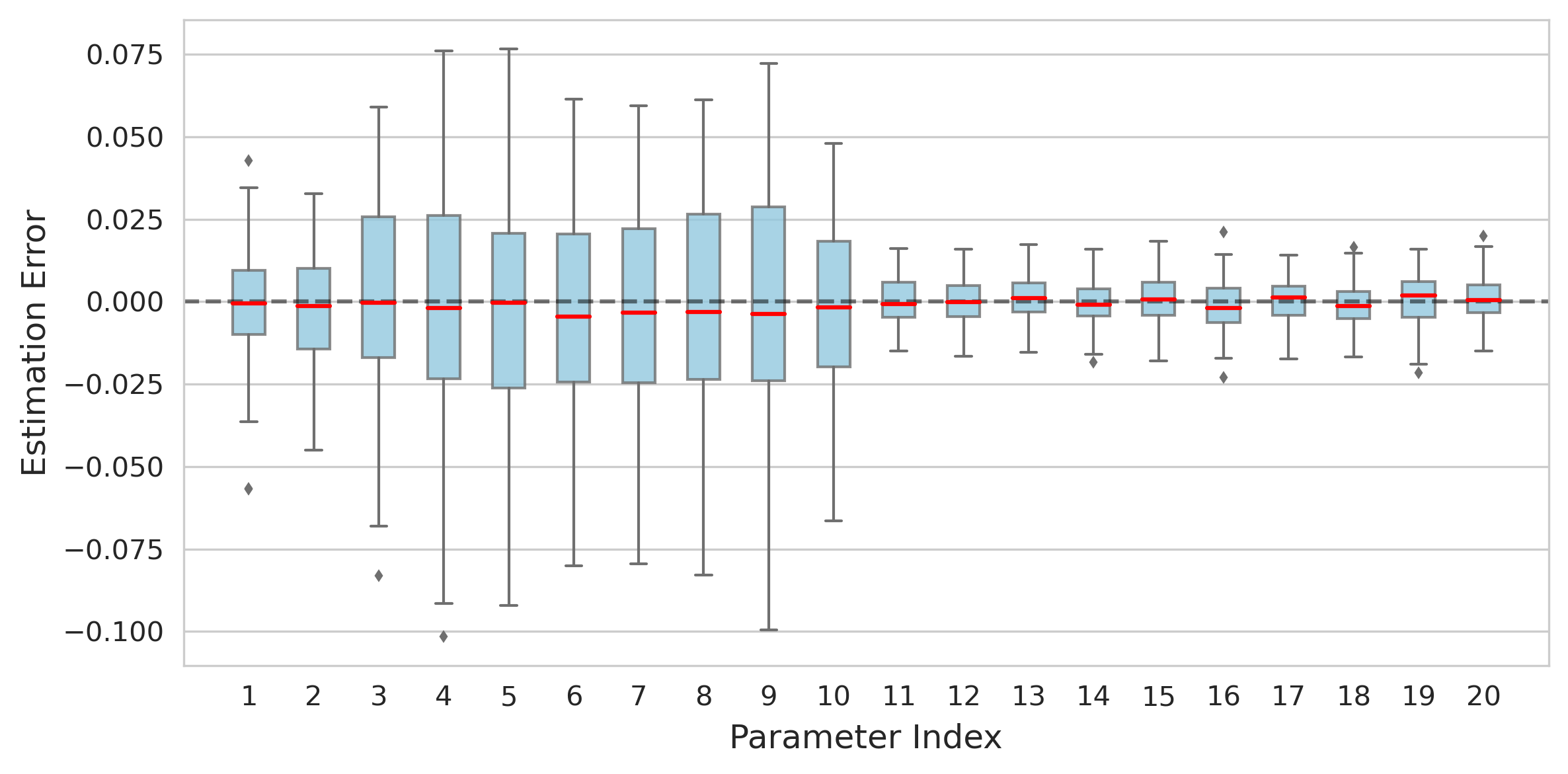}
        \caption{Distributions of point estimates $-$ true values}
    \end{subfigure}
    \begin{subfigure}{.4\textwidth}
        \centering
        \includegraphics[width=.8\linewidth]{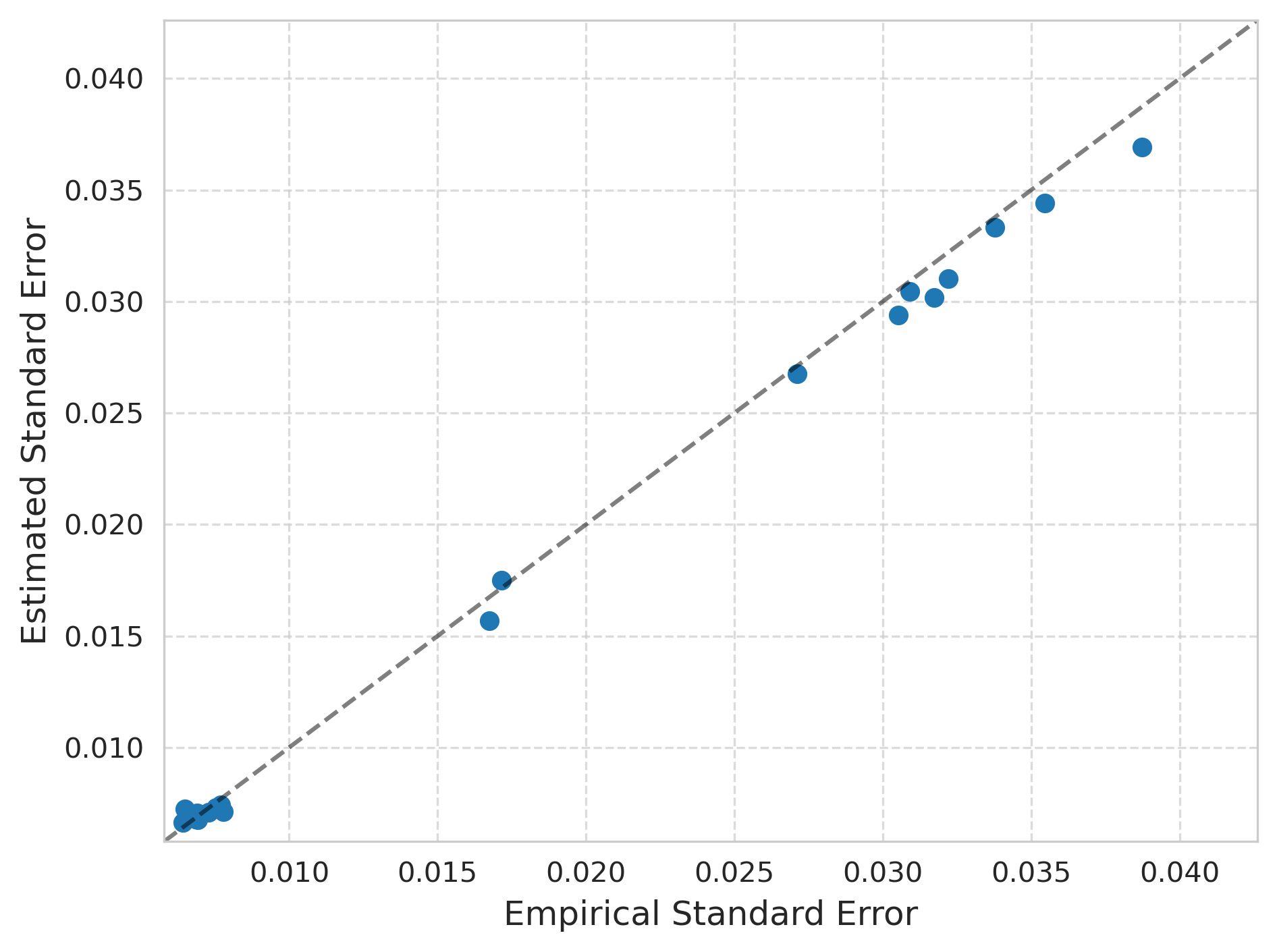}
        \caption{Empirical vs. means of estimated standard errors}
    \end{subfigure}%
    \begin{subfigure}{.55\textwidth}
        \centering
        \includegraphics[width=.9\linewidth]{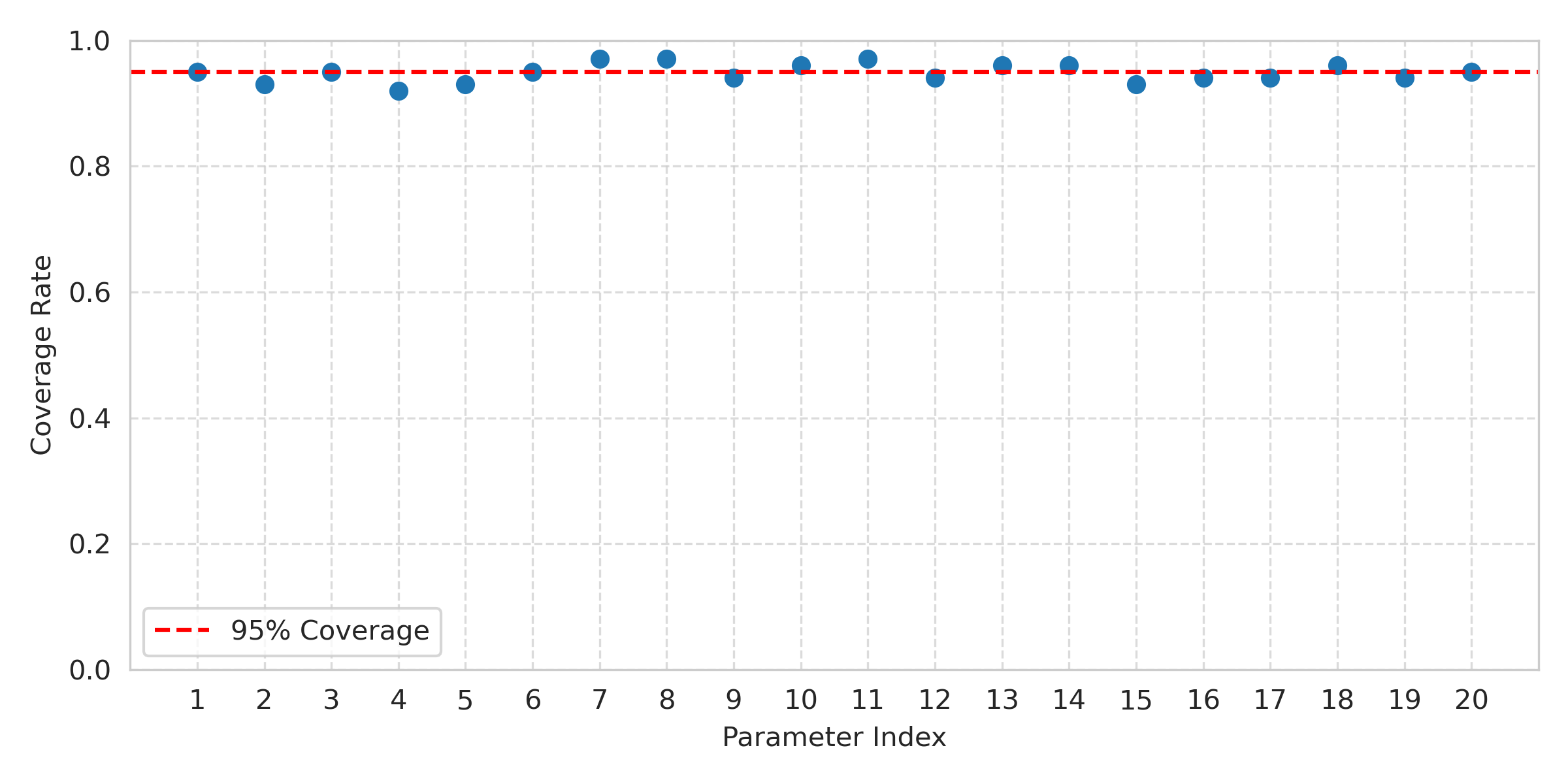}
        \caption{Coverage rates of 95\% confidence intervals}
    \end{subfigure}
    \caption{\small
    Simulation results in Study I-2 where the missingness is ignorable in the true model but non-ignorable in the estimated imputation model.
    }
    \label{f_sim_I_2}
\end{figure}

In Study I-2 (Figure \ref{f_sim_I_2}) the estimated model is overspecified by allowing for non-ignorability which is actually absent. The results are essentially the same as for the correctly specified Study I-1. This demonstrates that there is no cost to the safety-first approach of allowing for non-ignorable missingness even when this may be unnecessary.

\subsection{Parameter setting for Simulation Study III}
\label{ss_a_simIII}

\begin{table}[htbp]
\centering
\caption{Precision matrix $\Omega$ of the sparse Gaussian model for
the continuous variables.}
\label{tab:simu3_cont}
\begin{tabular}{
    c
    S[table-format=-1.2]
    S[table-format=-1.2]
    S[table-format=-1.2]
    S[table-format=-1.2]
    S[table-format=-1.2]
  }
  \toprule
  & {$Y_1$} & {$Y_2$} & {$Y_3$} & {$Y_4$} & {$Y_5$} \\
  \midrule
  {$Y_1$} & 2.92    & 0.00    & 0.00    & -0.46   & 2.36    \\
  {$Y_2$} & 0.00    & 0.79    & -0.69   & 0.00    & 0.00    \\
  {$Y_3$} & 0.00    & -0.69   & 2.01    & -1.22   & 0.00    \\
  {$Y_4$} & -0.46   & 0.00    & -1.22   & 1.78    & 0.00    \\
  {$Y_5$} & 2.36    & 0.00    & 0.00    & 0.00    & 2.46    \\
  \bottomrule
\end{tabular}
\end{table}

\begin{table}[htbp]
\centering
\caption{Precision matrix $\mathbf{S}$ for the Ising model for the
binary variables.}
\label{tab:simu3_bin}
\begin{tabular}{
    l
    S[table-format=-1.2]
    S[table-format=-1.2]
    S[table-format=-1.2]
    S[table-format=-1.2]
    S[table-format=-1.2]
  }
  \toprule
  & {$Y_6$} & {$Y_7$} & {$Y_8$} & {$Y_9$} & {$Y_{10}$} \\
  \midrule
  $Y_6$    & 0.00    & 0.51    & -0.76   & 0.61    & -0.47      \\
  $Y_7$    & 0.51    & 0.00    & 0.00    & 0.00    & 0.72       \\
  $Y_8$    & -0.76   & 0.00    & 0.00    & 0.00    & 0.97       \\
  $Y_9$    & 0.61    & 0.00    & 0.00    & 0.00    & 0.00       \\
  $Y_{10}$ & -0.47   & 0.72    & 0.97    & 0.00    & 0.00       \\
  \bottomrule
\end{tabular}
\end{table}

\subsection{Simulation results for estimates of conditional means}
\label{ss_a_sim_condmean}

Here the last-numbered variable ($Y_{10}$ in Study III, $Y_{20}$ in
the other studies) is binary.
What is reported in the figures in this subsection are simulation
results for the estimated conditional means of each of the other
variables given value 0 of this variable (first half of the
parameters in plots (b) and (d) of the figures) and value 1 of this
variable (second half of the parameters in plots (b) and (d) of the figures).

\clearpage

\begin{figure}[p]
\centering
\begin{subfigure}{.4\textwidth}
  \centering
  \includegraphics[width=.75\linewidth]{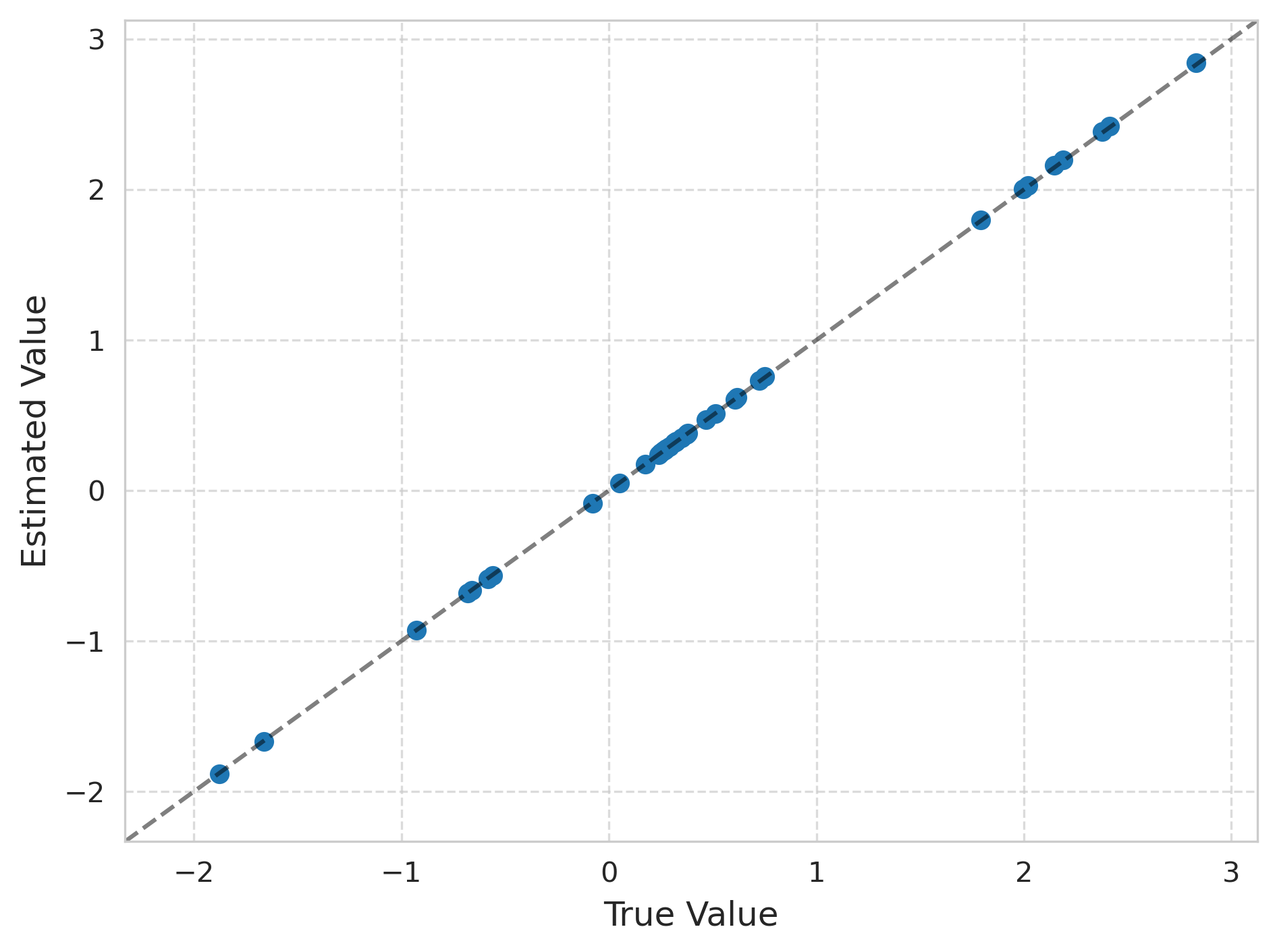}
  \caption{True parameter values vs.\ \\ \hspace*{2em} simulation
  means of point estimates}
\end{subfigure}%
\begin{subfigure}{.6\textwidth}
  \centering
  \includegraphics[width=.8\linewidth]{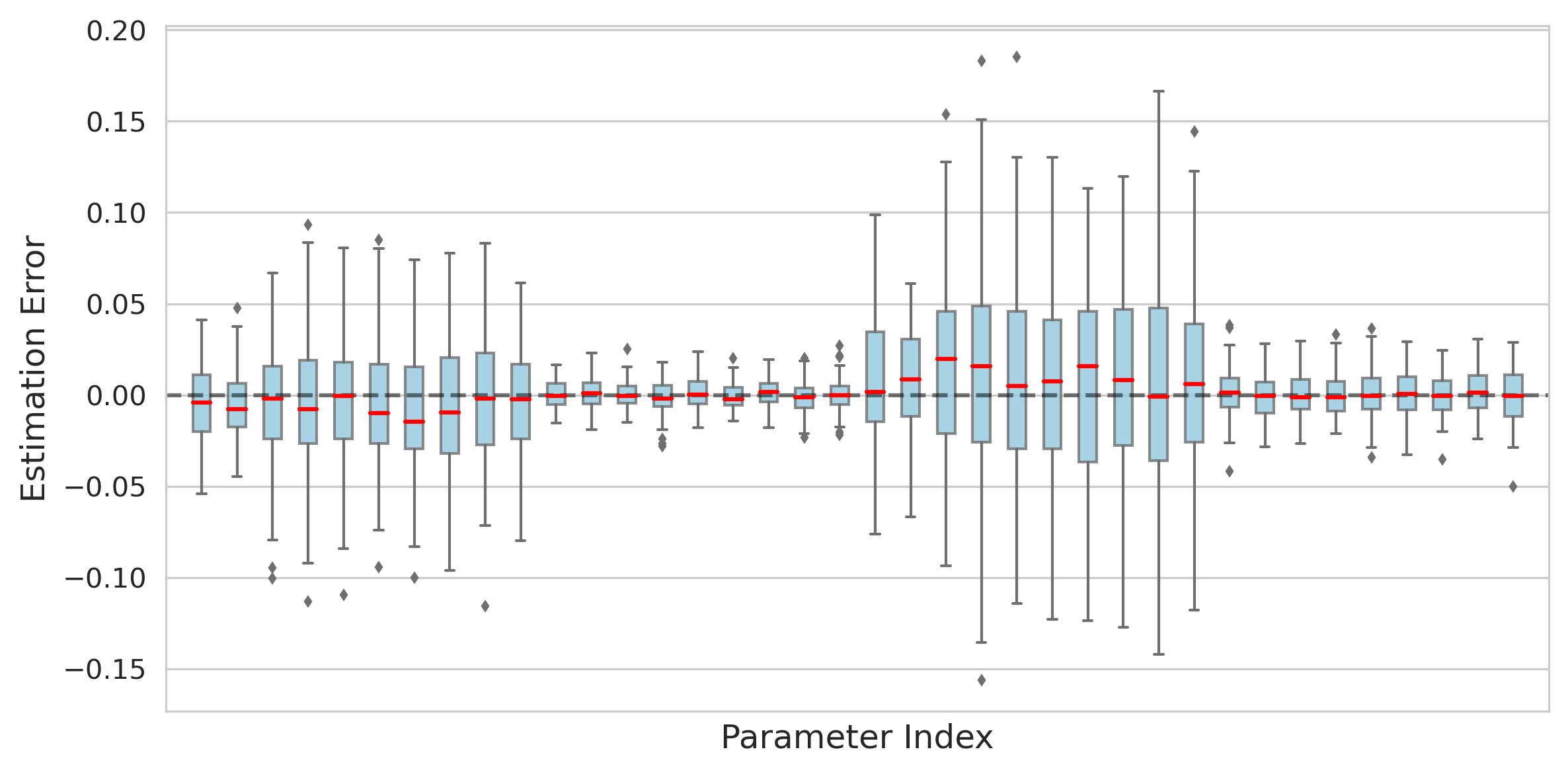}
  \caption{Distributions of point estimates $-$ true         values}
\end{subfigure}
\begin{subfigure}{.4\textwidth}
  \centering
  \includegraphics[width=.75\linewidth]{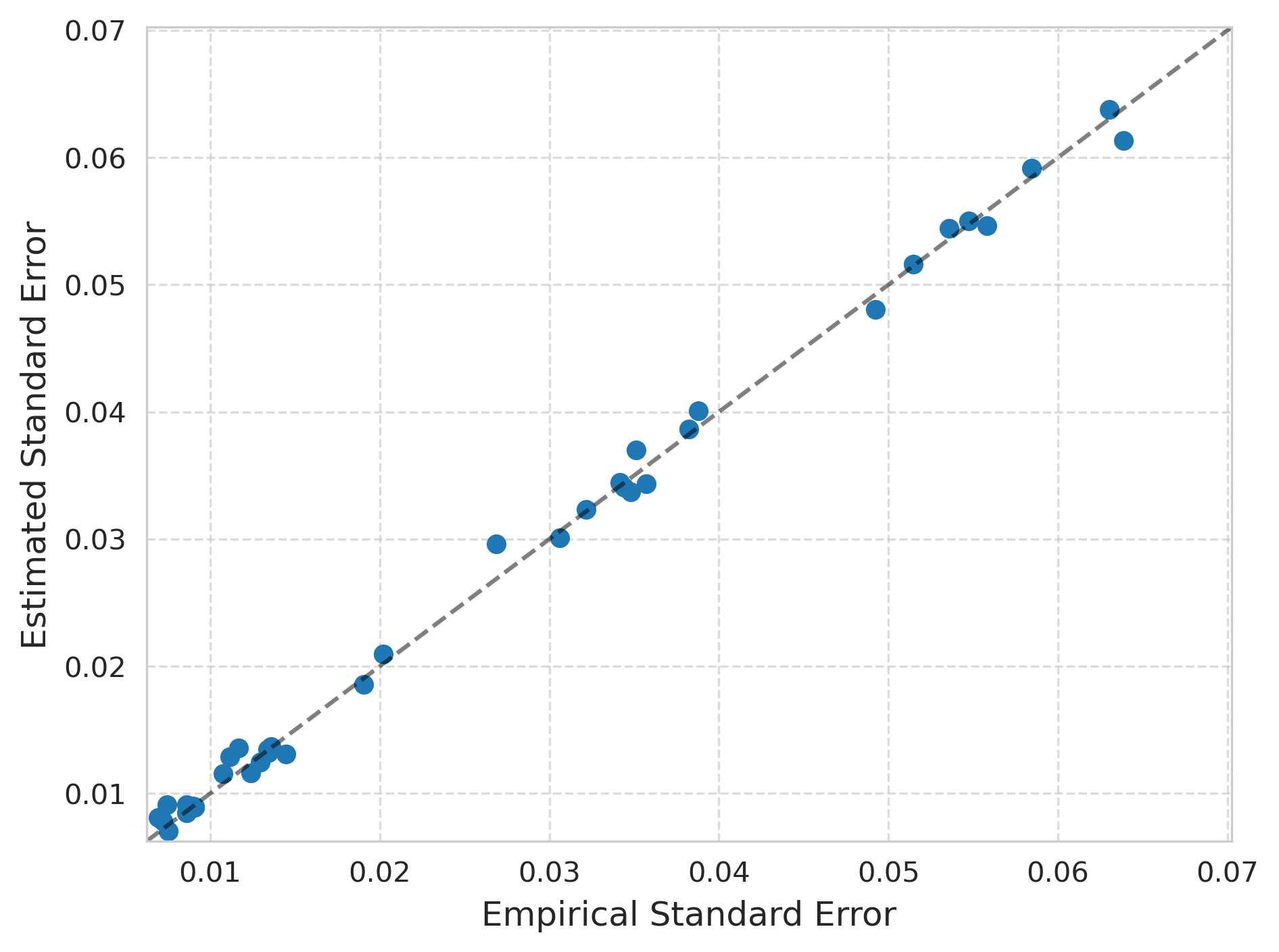}
  \caption{Simulation standard deviations vs.\ \\
  \hspace*{2em} means of estimated standard errors}
\end{subfigure}%
\begin{subfigure}{.6\textwidth}
  \centering
  \includegraphics[width=.8\linewidth]{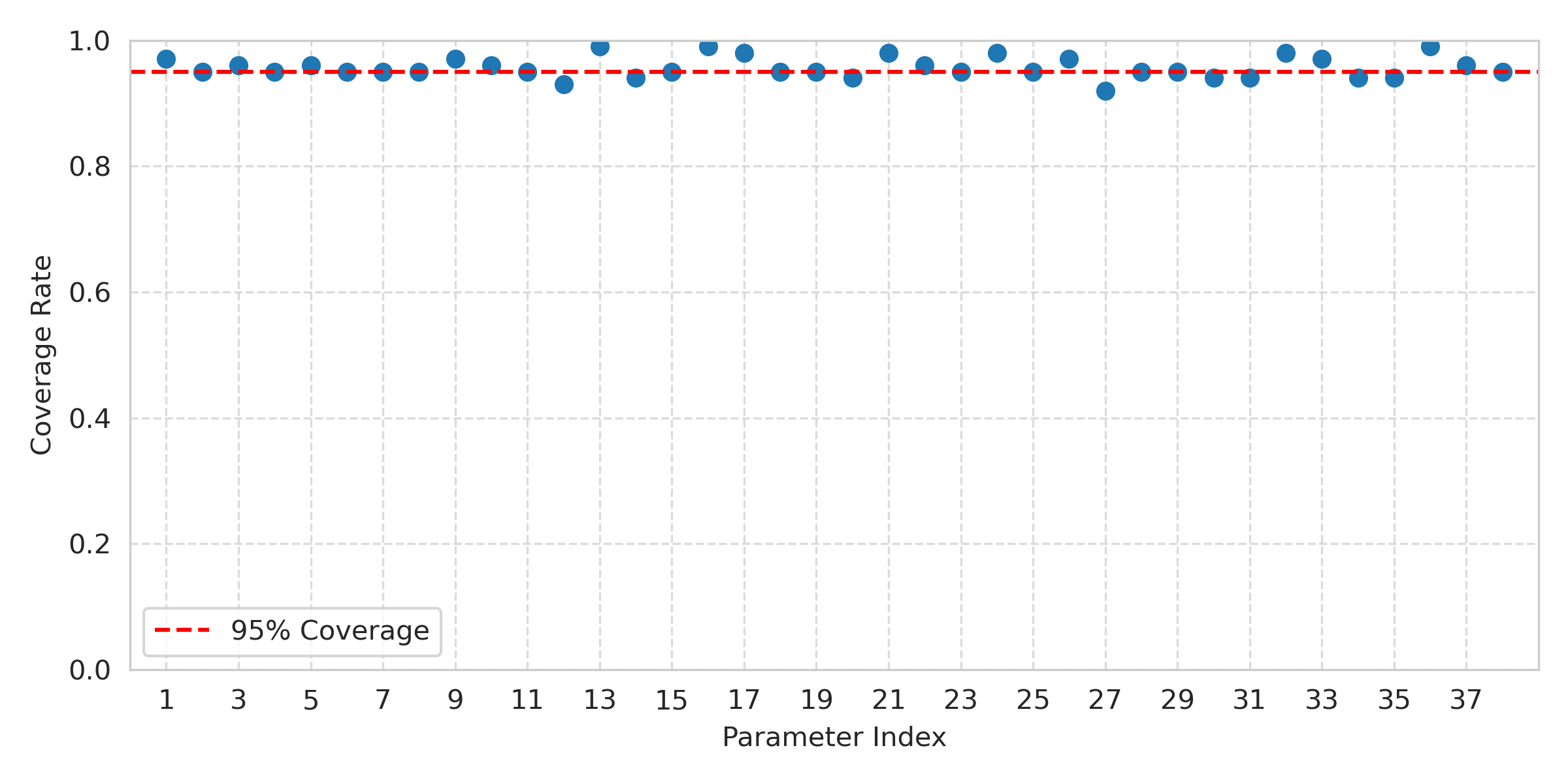}
  \caption{Coverage rates of 95\% confidence intervals.}
\end{subfigure}
\caption{
  Estimation results, over 100 simulation replicates, for conditional
  means of 10
  continuous variables (parameters 1--10 and 19--28) and 9
  binary variables (parameters 11--18 and 29--38) given the values of
  single binary variable in simulations (Study I-1)
  where both the true and assumed
  imputation model are of the latent variable form  (1) considered in
  the paper, both with ignorable missingness.
}
\label{f_a_sim_I_1}
\end{figure}

\begin{figure}[p]
\centering
\begin{subfigure}{.4\textwidth}
  \centering
  \includegraphics[width=.75\linewidth]{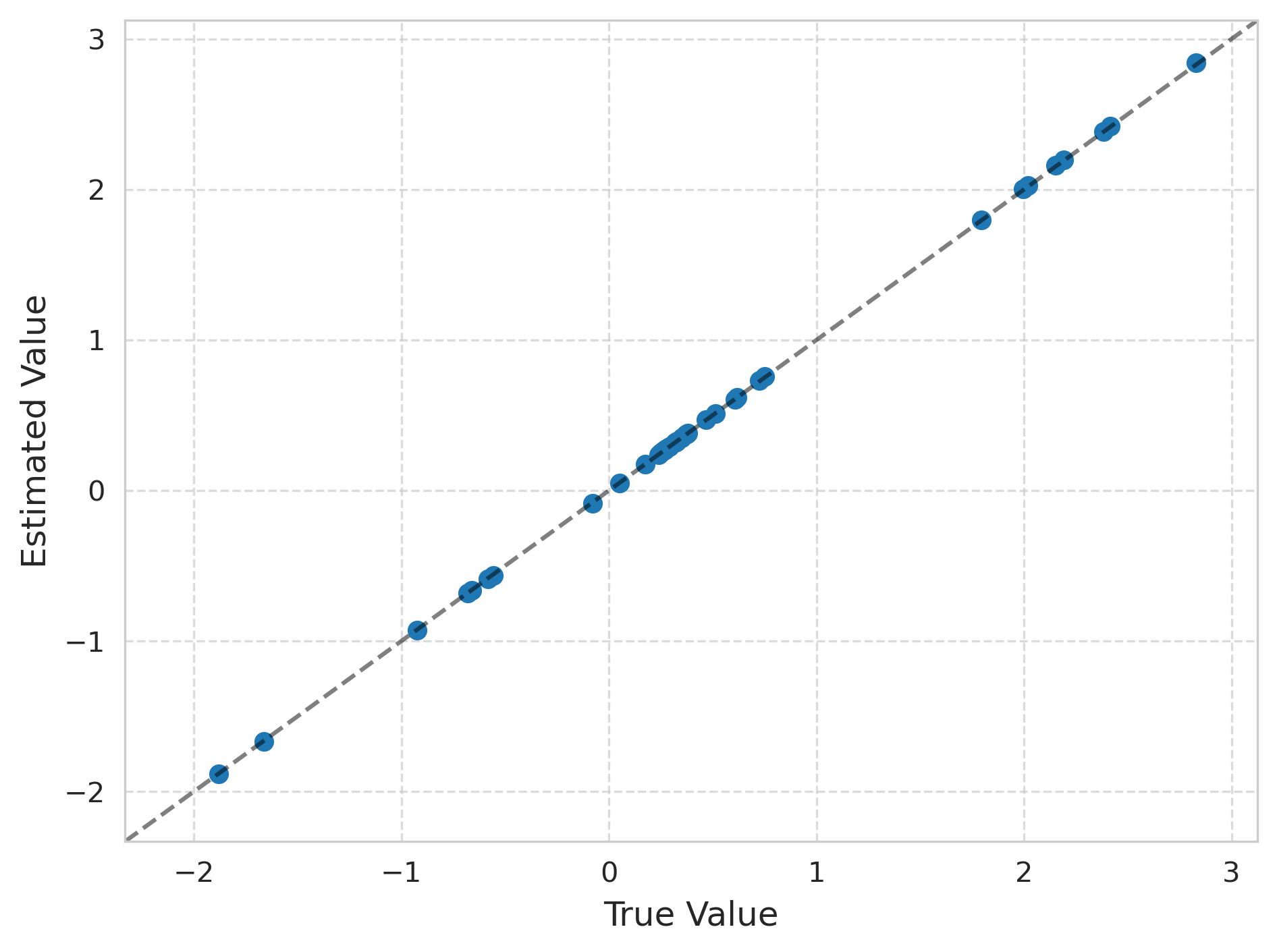}
  \caption{True parameter values vs.\ \\ \hspace*{2em} simulation
  means of point estimates}
\end{subfigure}%
\begin{subfigure}{.6\textwidth}
  \centering
  \includegraphics[width=.8\linewidth]{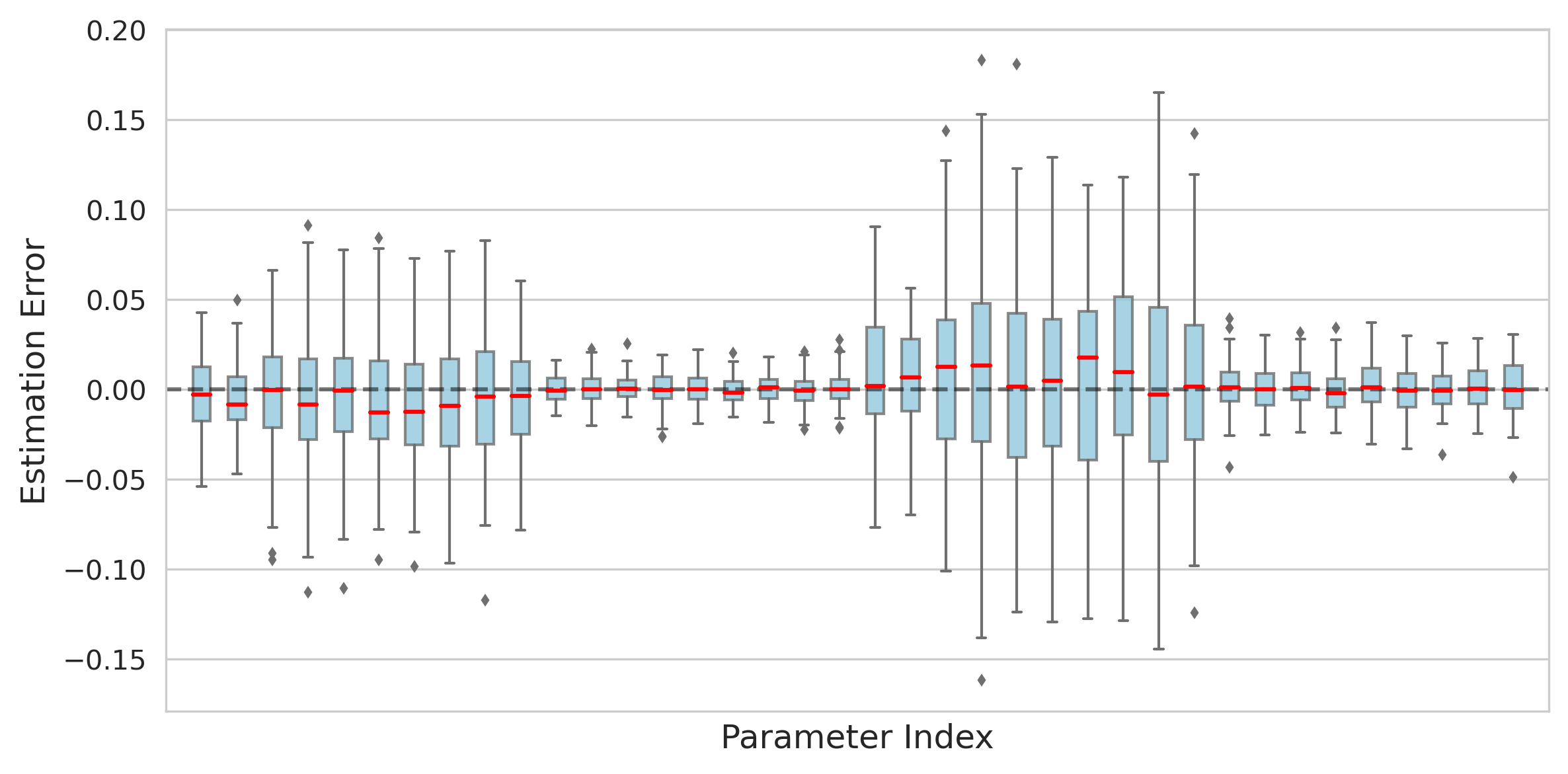}
  \caption{Distributions of point estimates $-$ true         values}
\end{subfigure}
\begin{subfigure}{.4\textwidth}
  \centering
  \includegraphics[width=.75\linewidth]{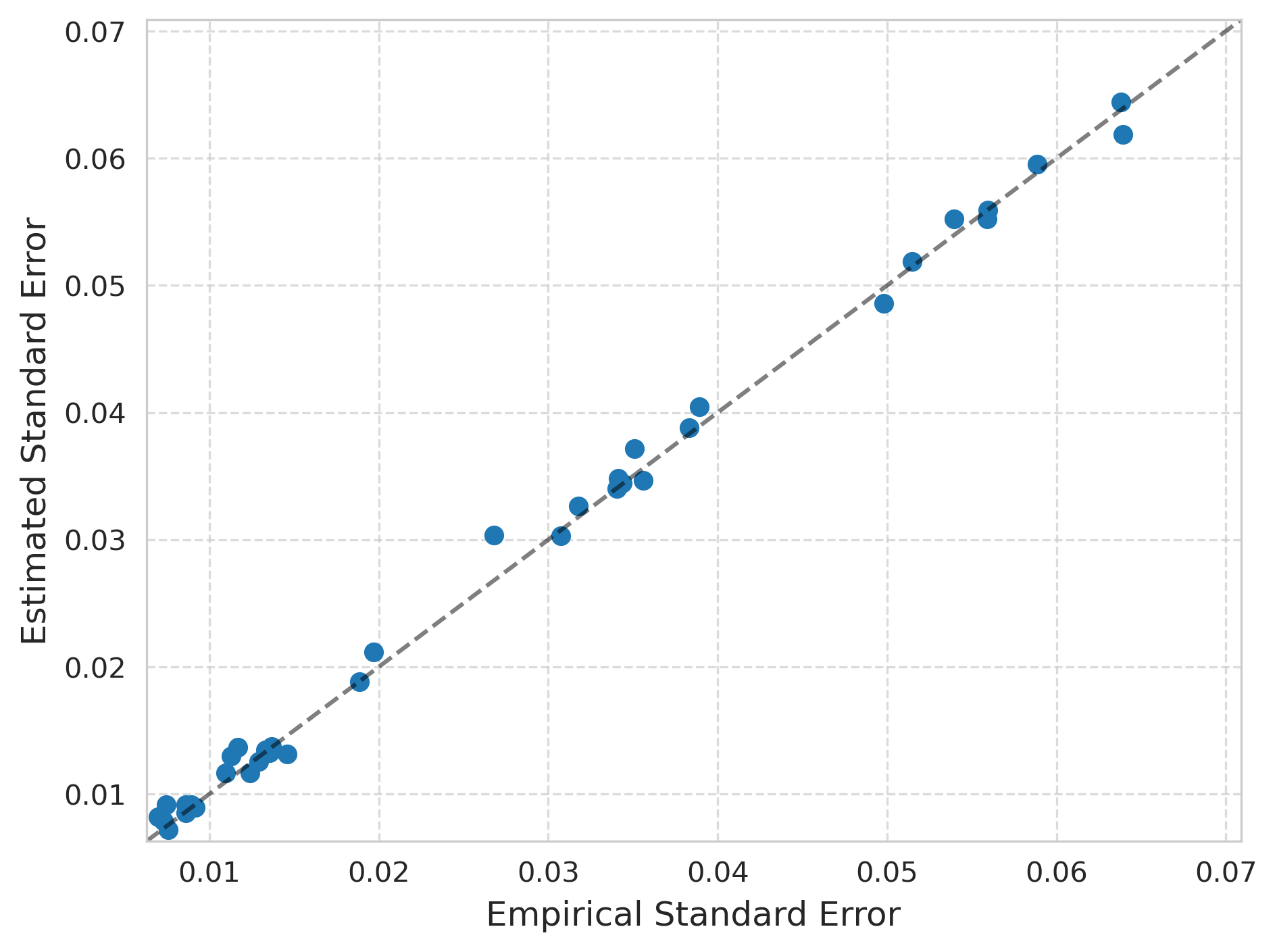}
  \caption{Simulation standard deviations vs.\ \\
  \hspace*{2em} means of estimated standard errors}
\end{subfigure}%
\begin{subfigure}{.6\textwidth}
  \centering
  \includegraphics[width=.8\linewidth]{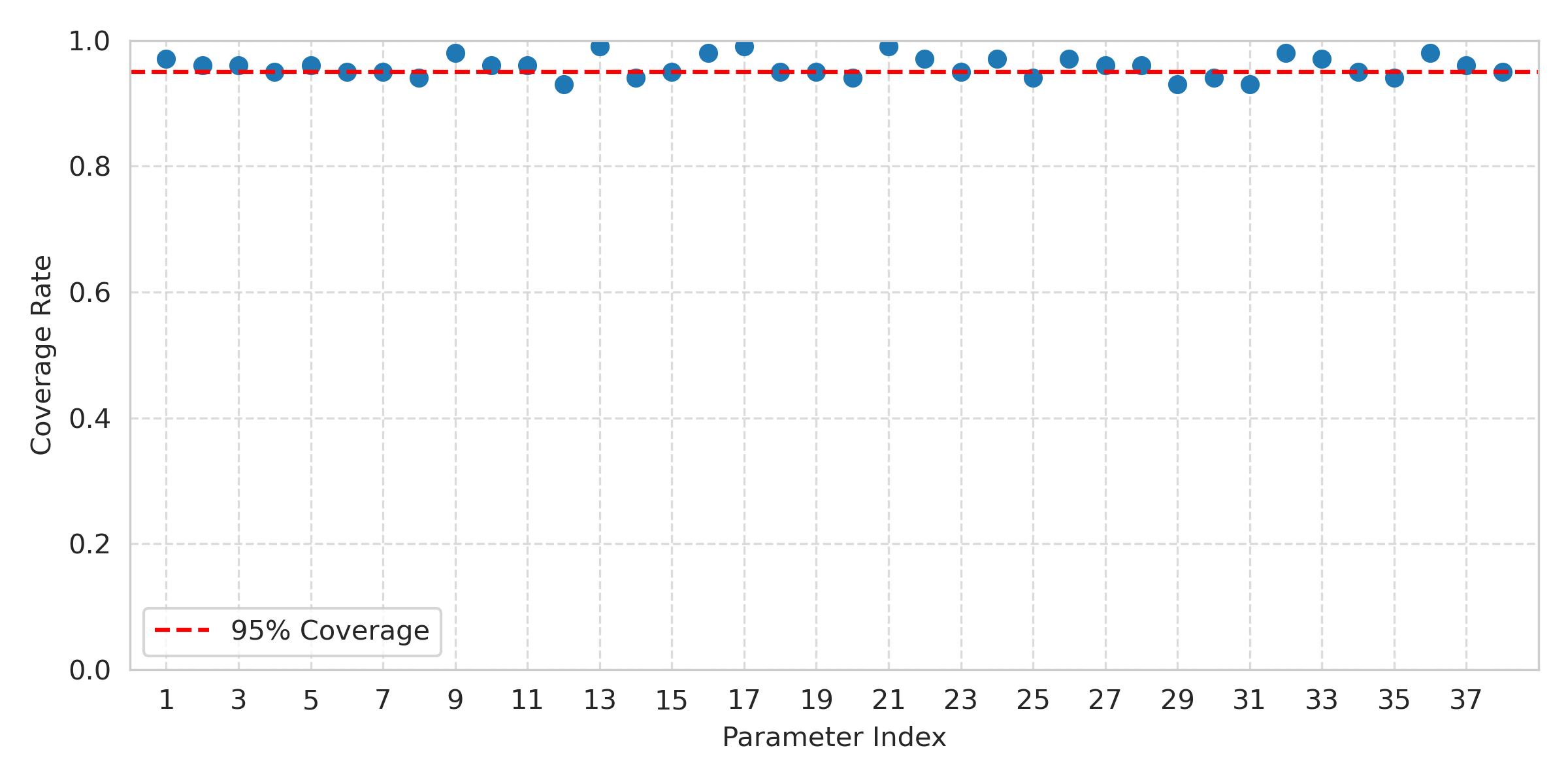}
  \caption{Coverage rates of 95\% confidence intervals}
\end{subfigure}
\caption{
  Simulation results in Study I-2 where
  the missingness is ignorable in the true model but non-ignorable in
  the estimated imputation model (see caption of Figure
  \ref{f_a_sim_I_1} for further settings).
}
\label{f_a_sim_I_2}
\end{figure}

\begin{figure}[p]
\centering
\begin{subfigure}{.4\textwidth}
  \centering
  \includegraphics[width=.75\linewidth]{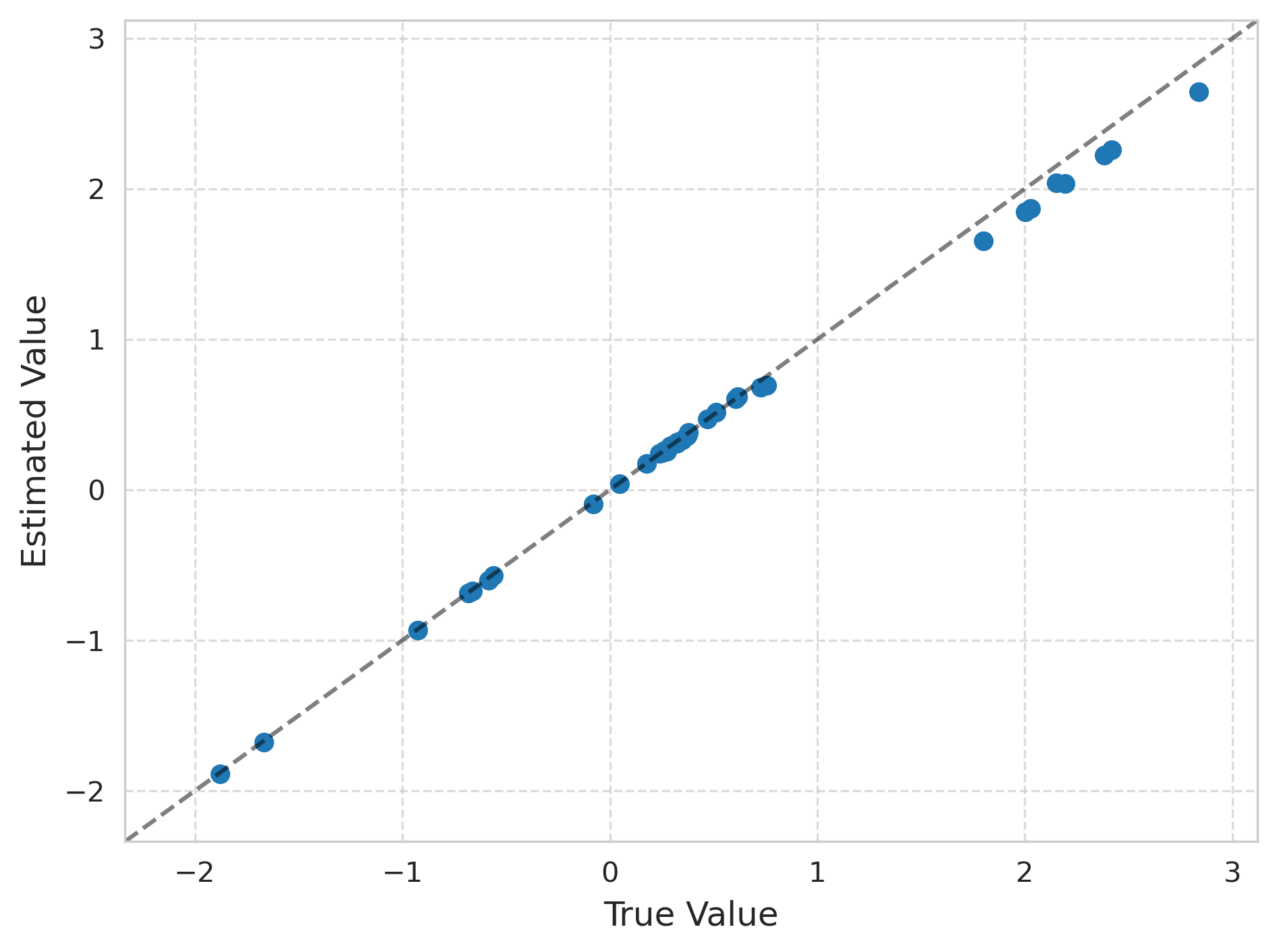}
  \caption{True parameter values vs.\ \\ \hspace*{2em} simulation
  means of point estimates}
\end{subfigure}%
\begin{subfigure}{.6\textwidth}
  \centering
  \includegraphics[width=.8\linewidth]{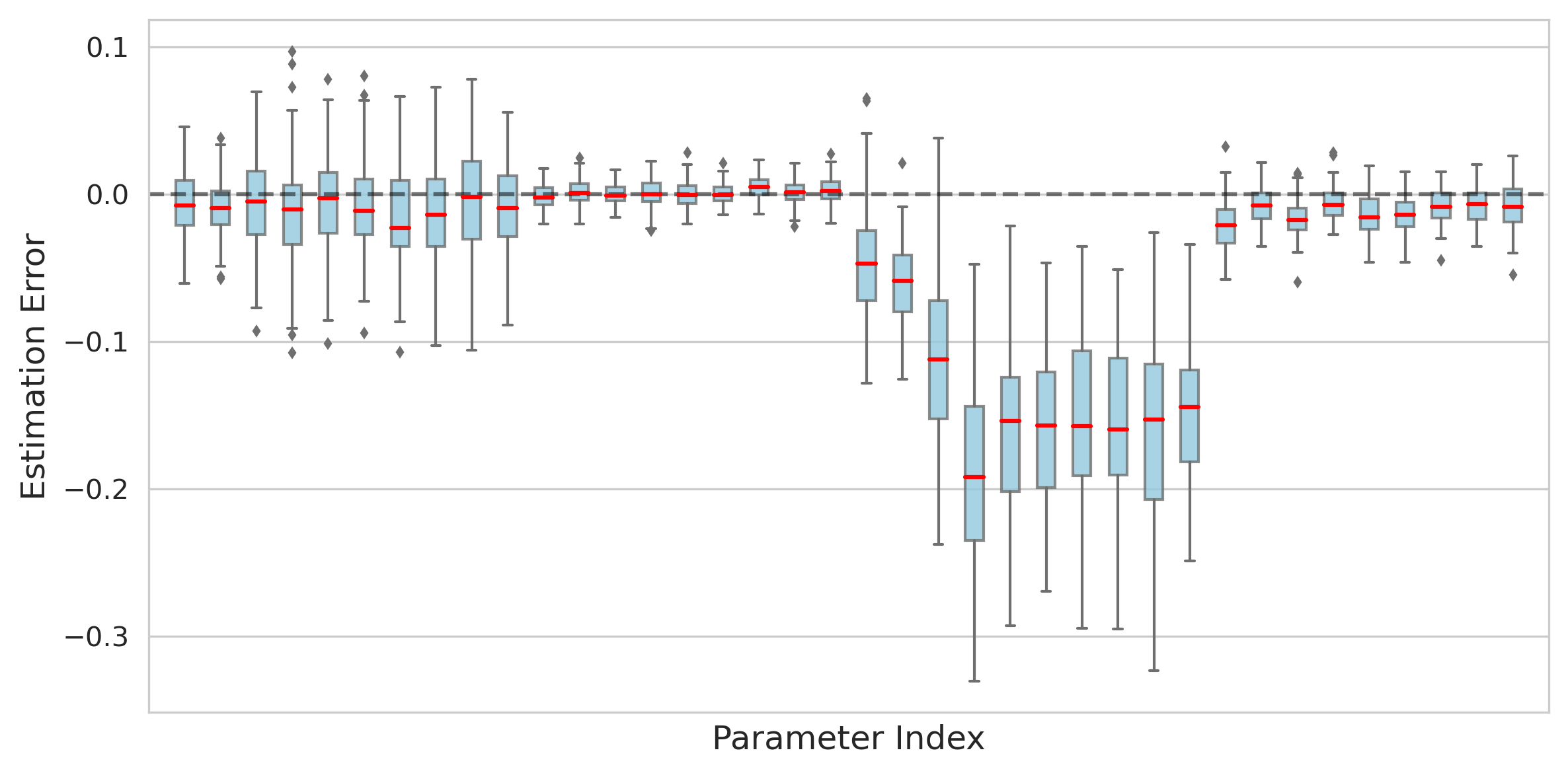}
  \caption{Distributions of point estimates $-$ true         values}
\end{subfigure}
\begin{subfigure}{.4\textwidth}
  \centering
  \includegraphics[width=.75\linewidth]{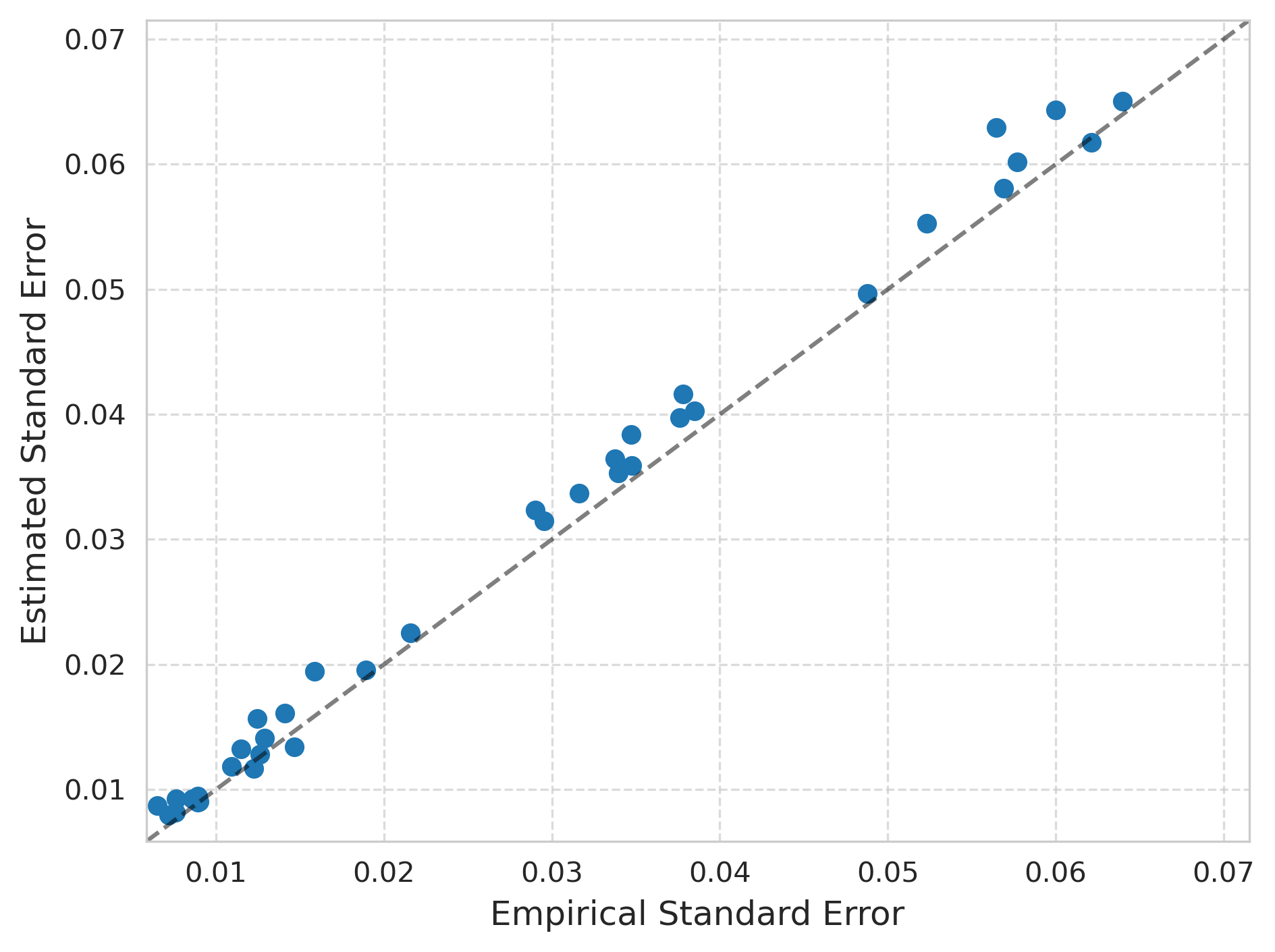}
  \caption{Simulation standard deviations vs.\ \\
  \hspace*{2em} means of estimated standard errors}
\end{subfigure}%
\begin{subfigure}{.6\textwidth}
  \centering
  \includegraphics[width=.8\linewidth]{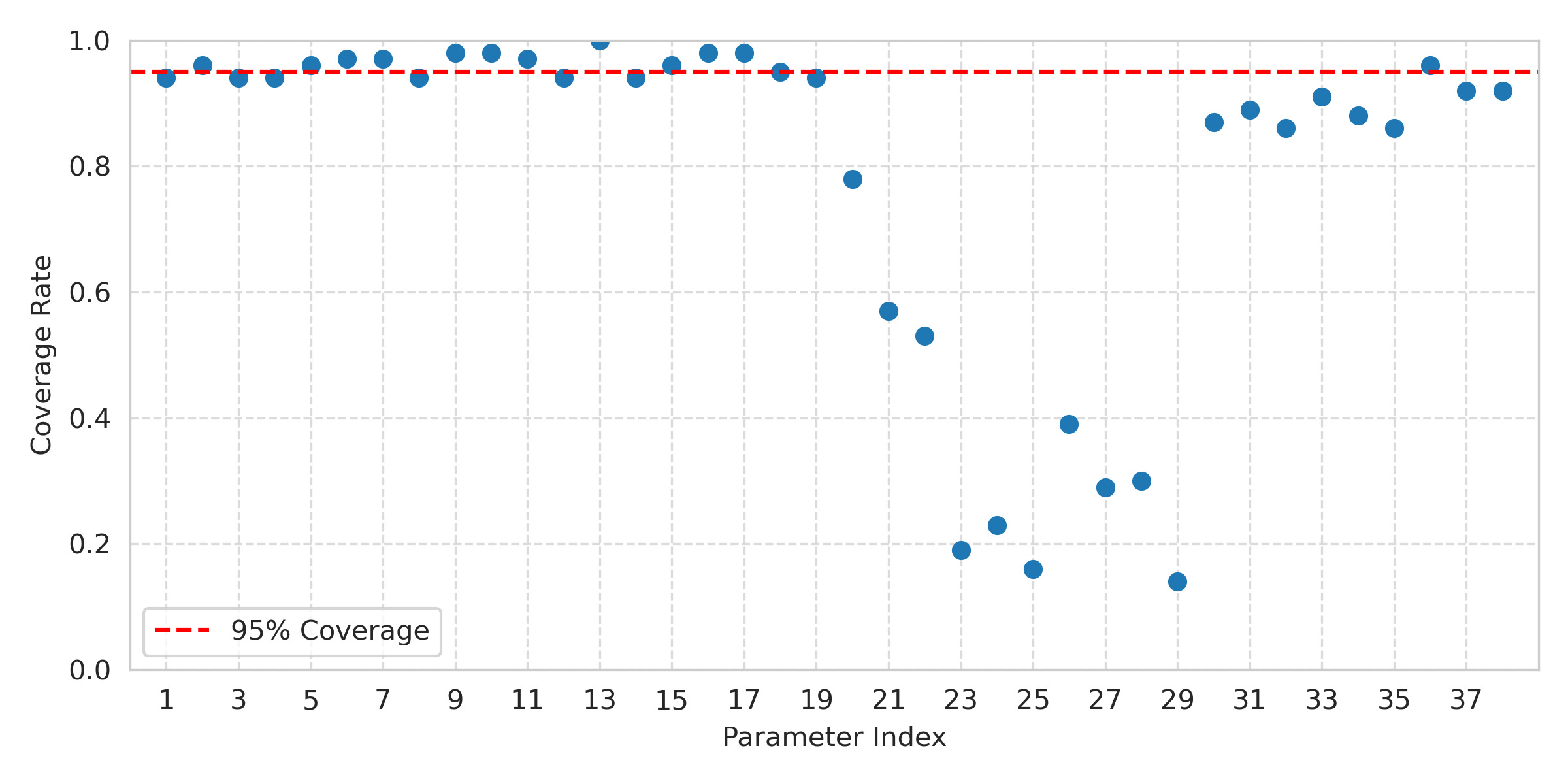}
  \caption{Coverage rates of 95\% confidence intervals}
\end{subfigure}
\caption{
  Simulation results in Study II-1 where
  the missingness is non-ignorable in the true model but ignorable in
  the estimated imputation model (see caption of Figure
  \ref{f_a_sim_I_1} for further settings).
}
\label{f_a_sim_II_1}
\end{figure}

\begin{figure}[p]
\centering
\begin{subfigure}{.4\textwidth}
  \centering
  \includegraphics[width=.75\linewidth]{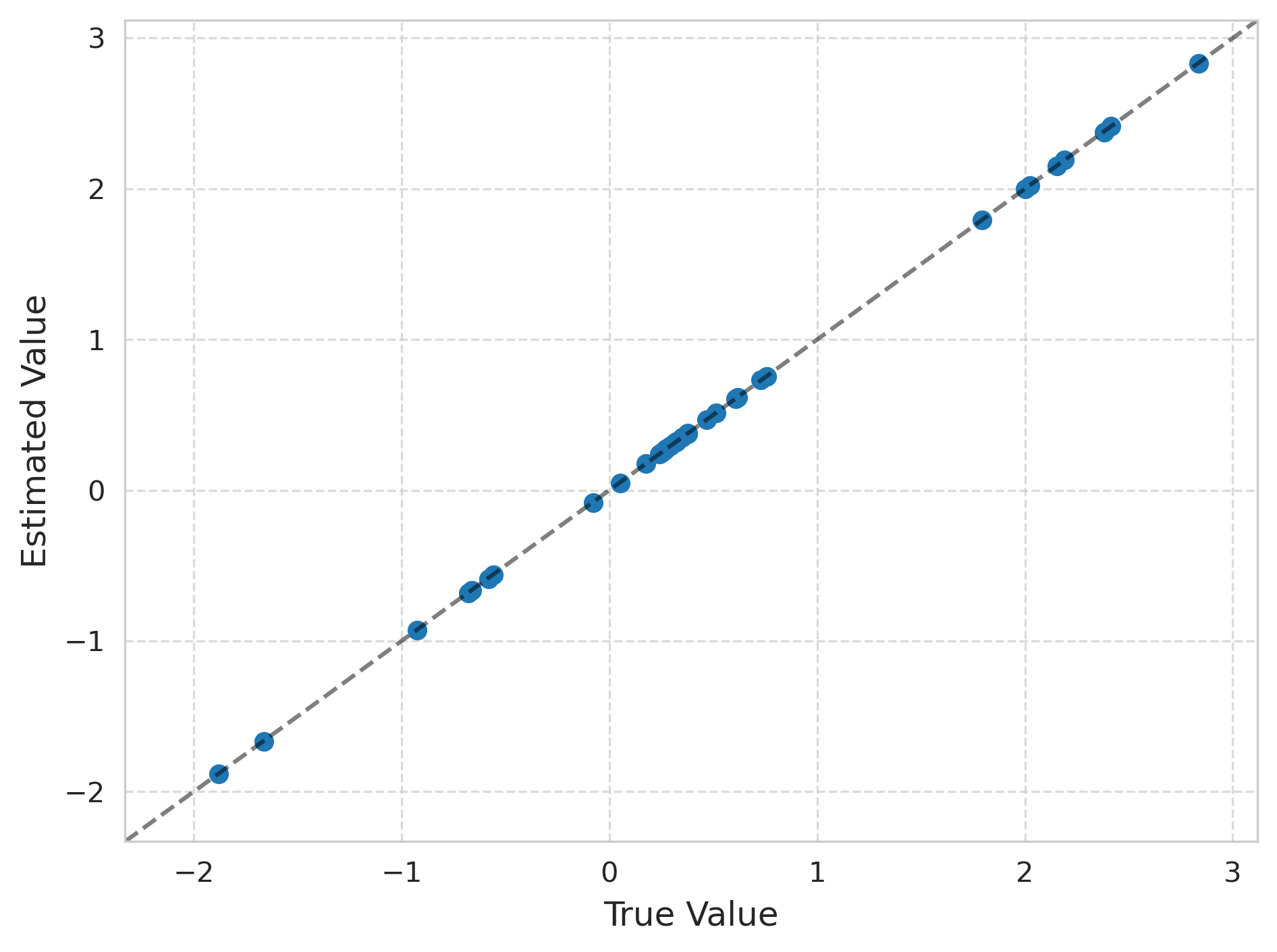}
  \caption{True parameter values vs.\ \\ \hspace*{2em} simulation
  means of point estimates}
  \caption{True value vs. estimated value.}
\end{subfigure}%
\begin{subfigure}{.6\textwidth}
  \centering
  \includegraphics[width=.8\linewidth]{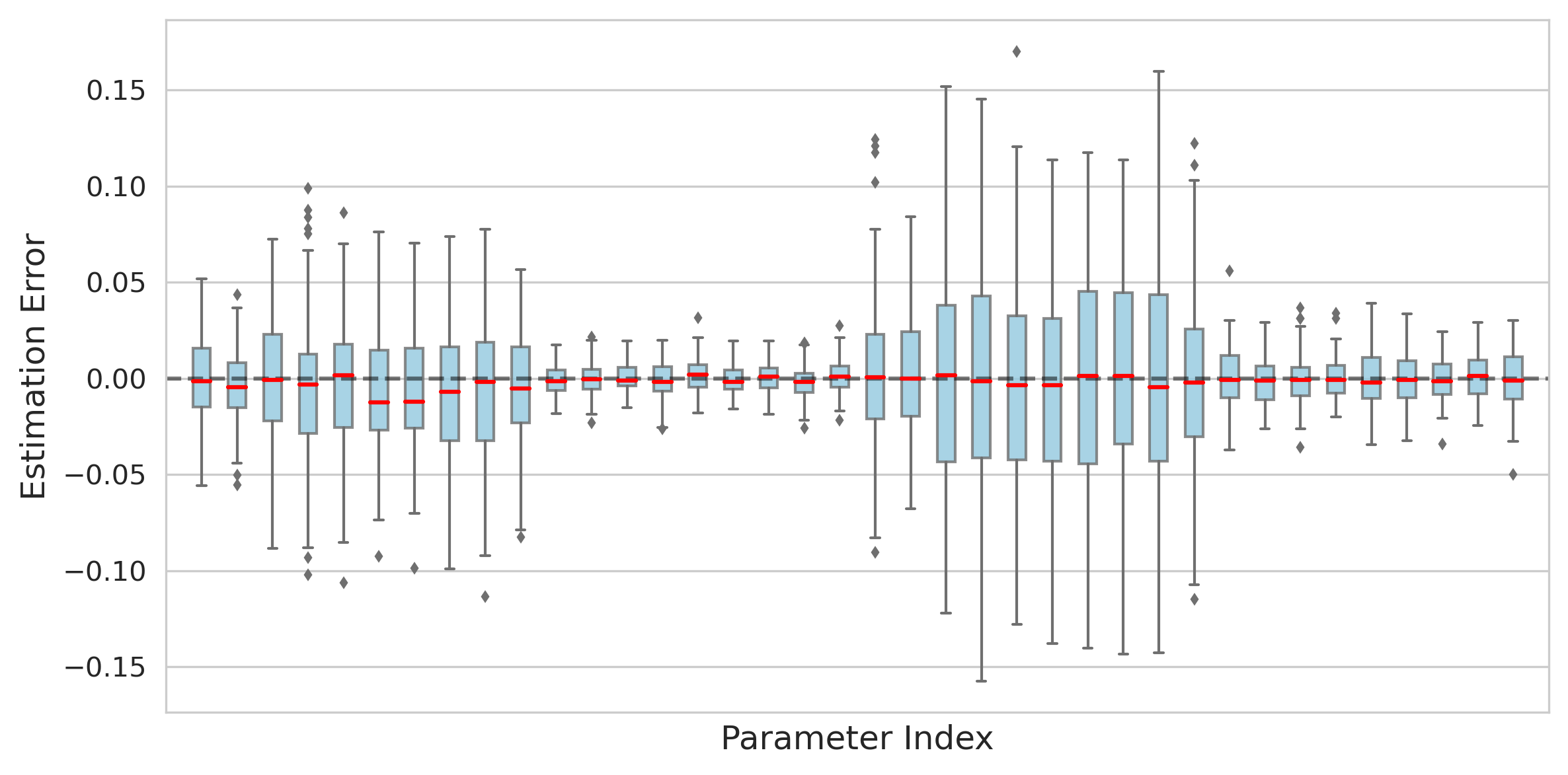}
  \caption{Distributions of point estimates $-$ true         values}
\end{subfigure}
\begin{subfigure}{.4\textwidth}
  \centering
  \includegraphics[width=.75\linewidth]{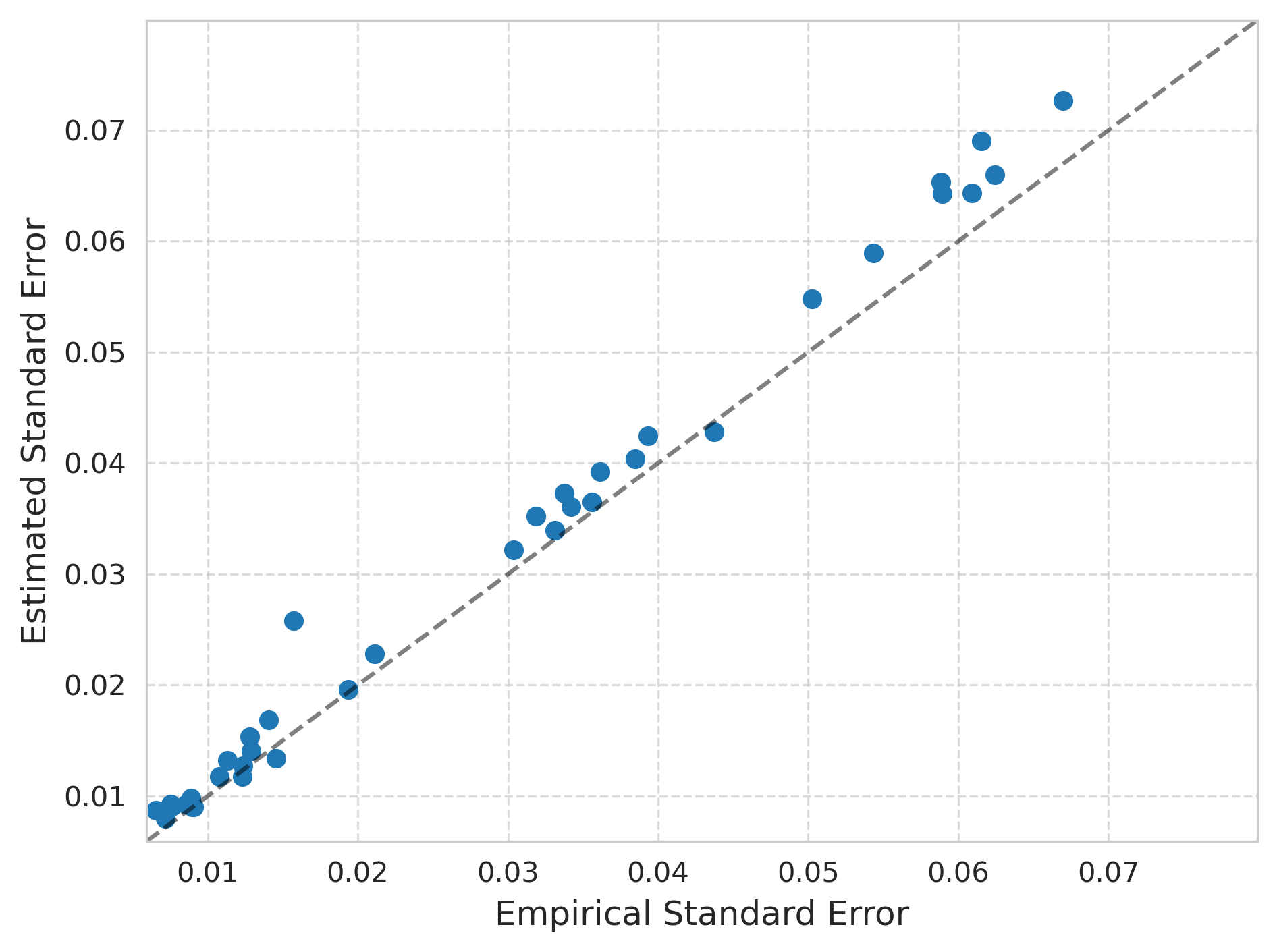}
  \caption{Simulation standard deviations vs.\ \\
  \hspace*{2em} means of estimated standard errors}
\end{subfigure}%
\begin{subfigure}{.6\textwidth}
  \centering
  \includegraphics[width=.8\linewidth]{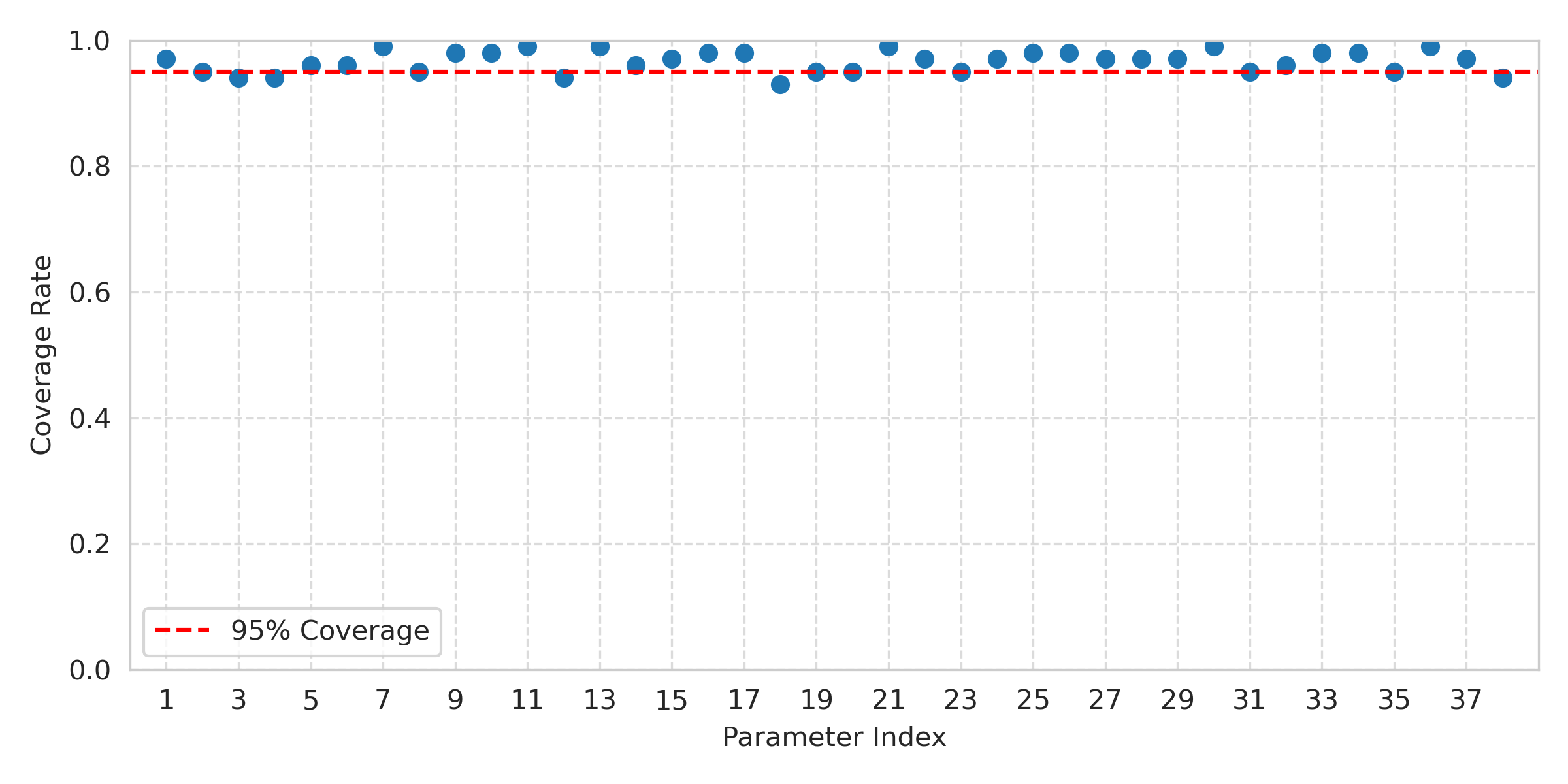}
  \caption{Coverage rates of 95\% confidence intervals}
\end{subfigure}
\caption{
  Simulation results in Study II-2 where
  the missingness is non-ignorable in both the true model and the
  estimated imputation model (see caption of Figure
  \ref{f_a_sim_I_1} for further settings).
}
\label{f_a_sim_II_2}
\end{figure}

\clearpage

\begin{figure}[htbp]
\begin{subfigure}{.5\textwidth}
  \centering
  \includegraphics[width=.8\linewidth]{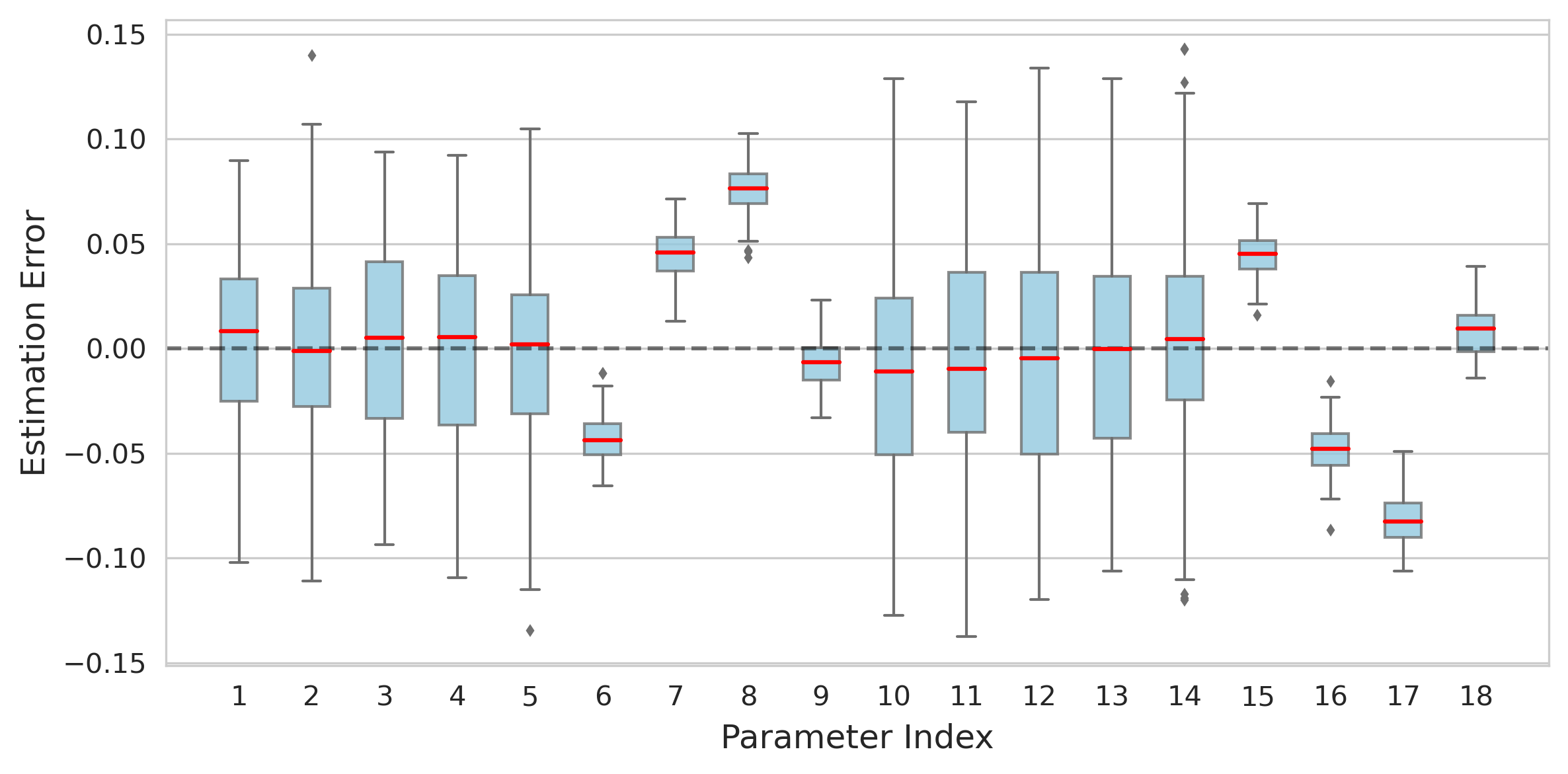}
  \caption{Distributions of point estimates $-$ true values
  ($K_{1}=1$)}
\end{subfigure}%
\begin{subfigure}{.5\textwidth}
  \centering
  \includegraphics[width=.8\linewidth]{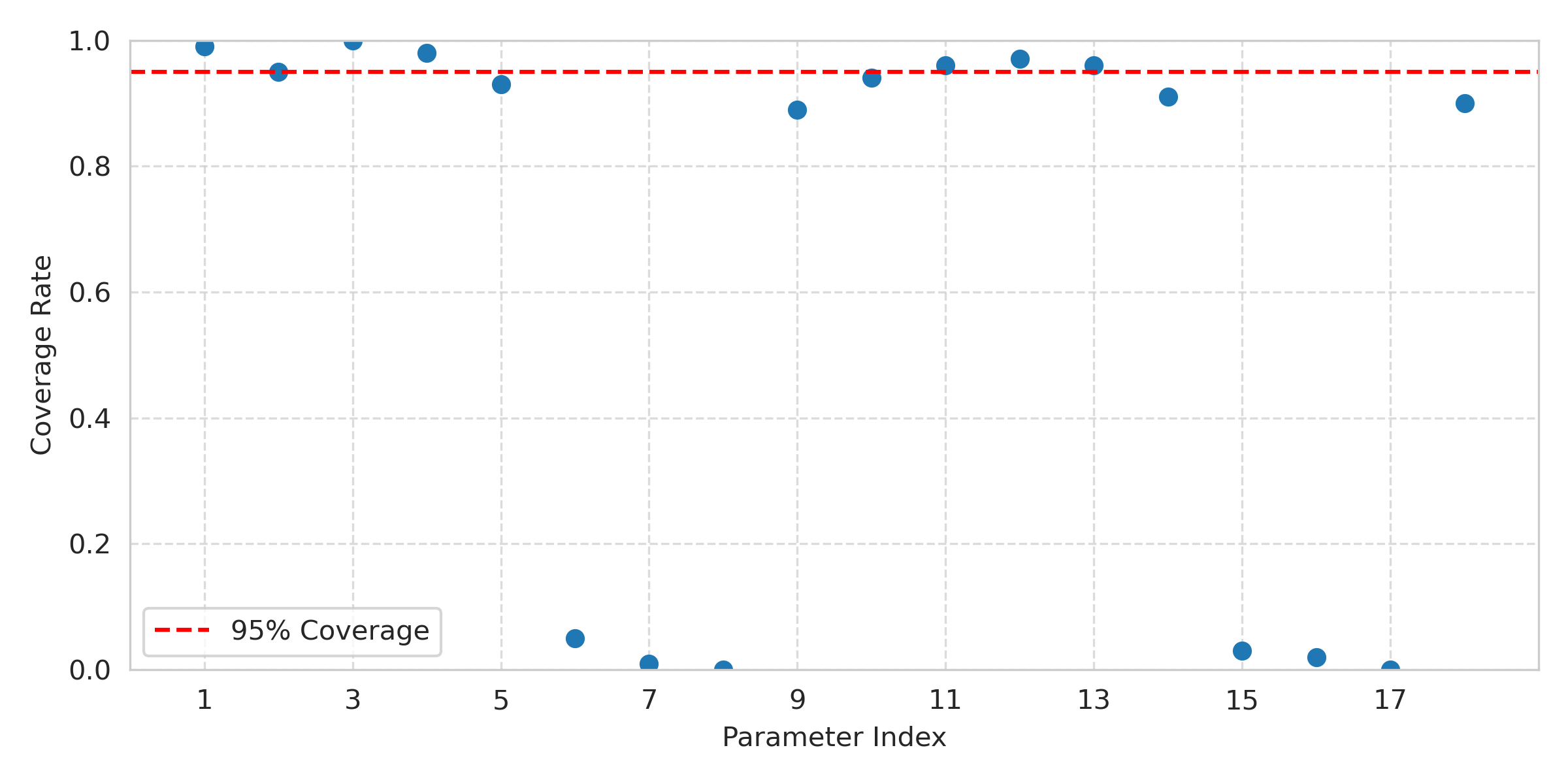}
  \caption{Coverage rates of 95\% confidence intervals ($K_1=1$).}
\end{subfigure}
\begin{subfigure}{.5\textwidth}
  \centering
  \includegraphics[width=.8\linewidth]{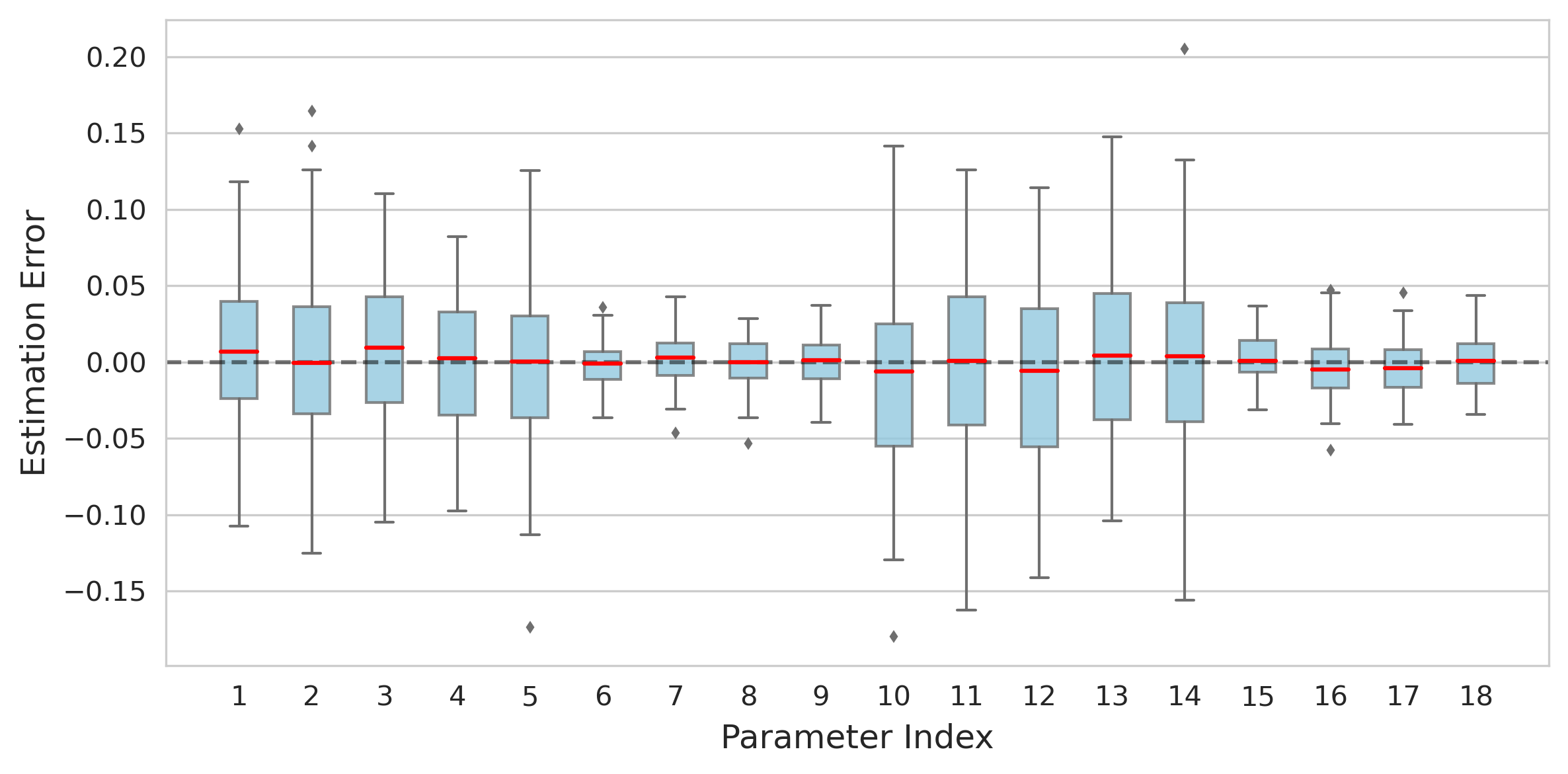}
  \caption{Distributions of point estimates $-$ true values
  ($K_{1}=4$)}
\end{subfigure}%
\begin{subfigure}{.5\textwidth}
  \centering
  \includegraphics[width=.8\linewidth]{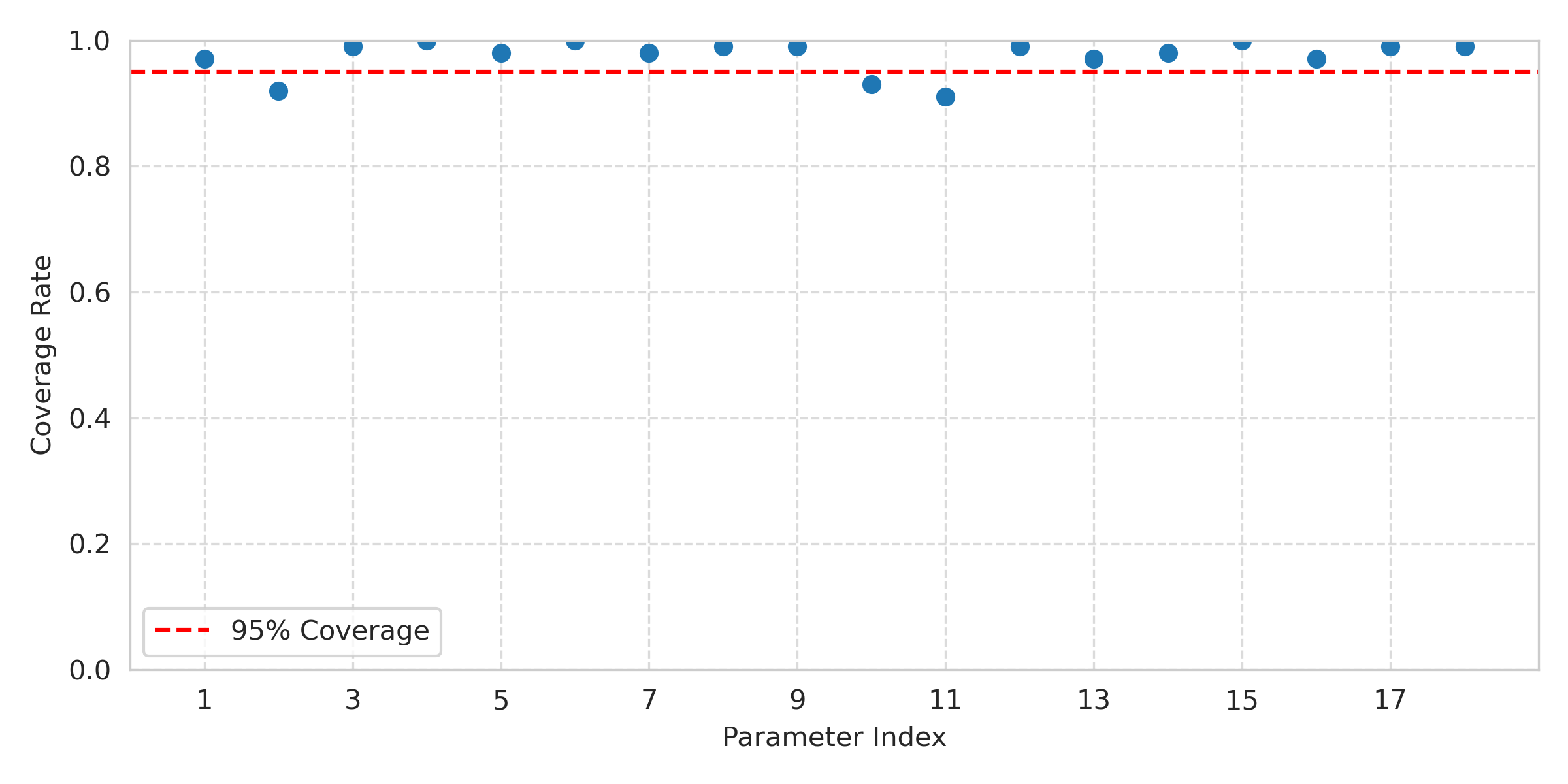}
  \caption{Coverage rates of 95\% confidence intervals ($K_1=4$).}
\end{subfigure}
\caption{Simulation Study III: Estimation results, over 100
  simulation replicates,
  for conditional
  means of 5
  continuous variables (parameters 1--5 and 10--14) and 4
  binary variables (parameters 6--9 and 15--18) given the values of
  single binary variable in simulations where
  the estimated imputation model is of the latent variable form
  (1) considered in the paper but the true data-genetating model is
  not of this form.
  Plots (a) and (b) are from simulations where the
  number of substantive factors $\boldsymbol{\eta}$ in the imputation
  model was $K_{1}=1$, and
  plots (c) and (d) from ones where it was $K_{1}=4$.
}
\label{f_a_sim_III}
\end{figure}

\subsection{Results for Study III with an ignorable estimated model}
\label{ss_a_IIIignorable}

\vspace*{-2ex}
\begin{figure}[p]
\begin{subfigure}{.5\textwidth}
  \centering
  \includegraphics[width=.8\linewidth]{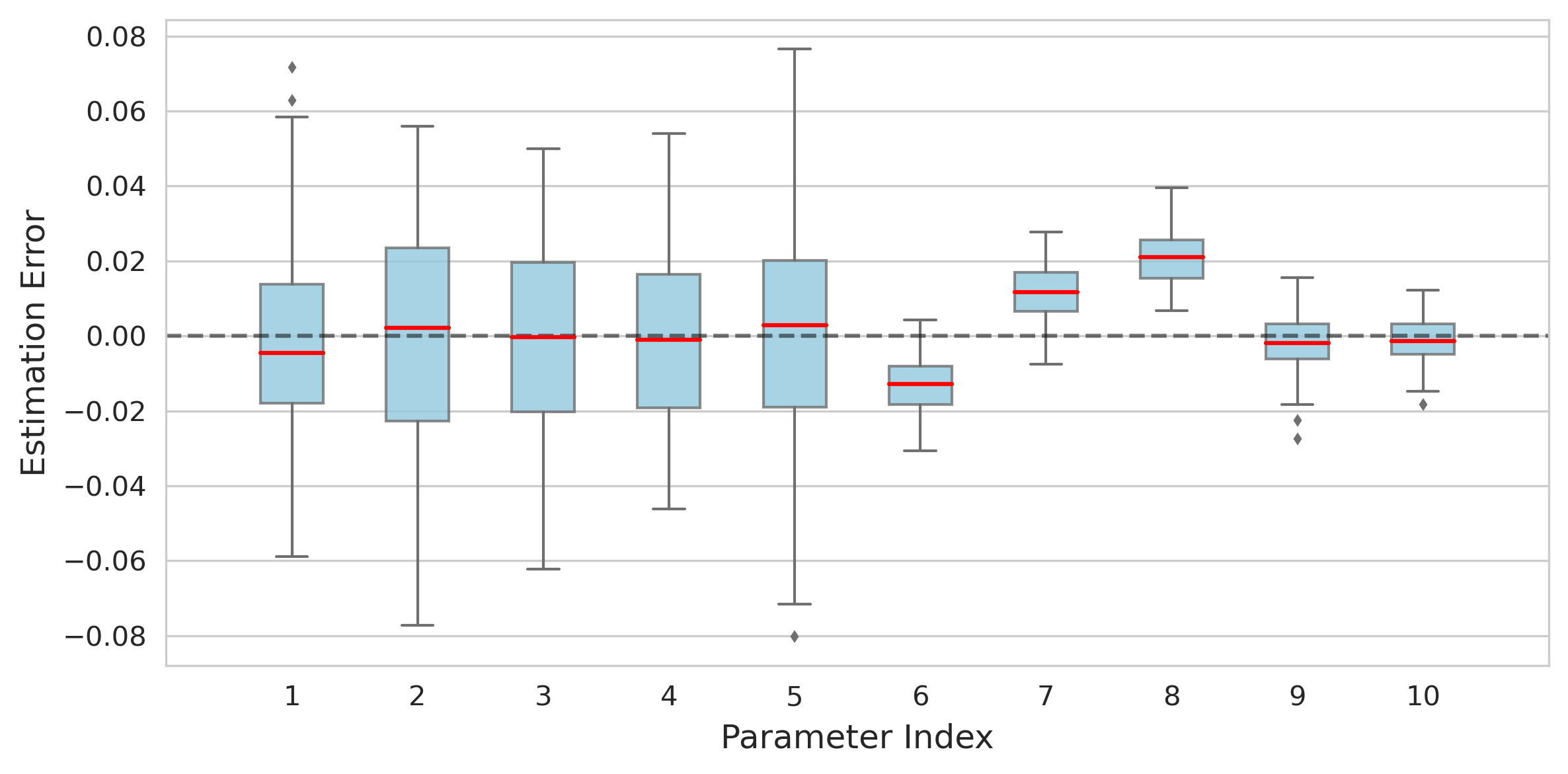}
  \caption{Distributions of point estimates $-$ true values
  ($K_{1}=1$)}
\end{subfigure}%
\begin{subfigure}{.5\textwidth}
  \centering
  \includegraphics[width=.8\linewidth]{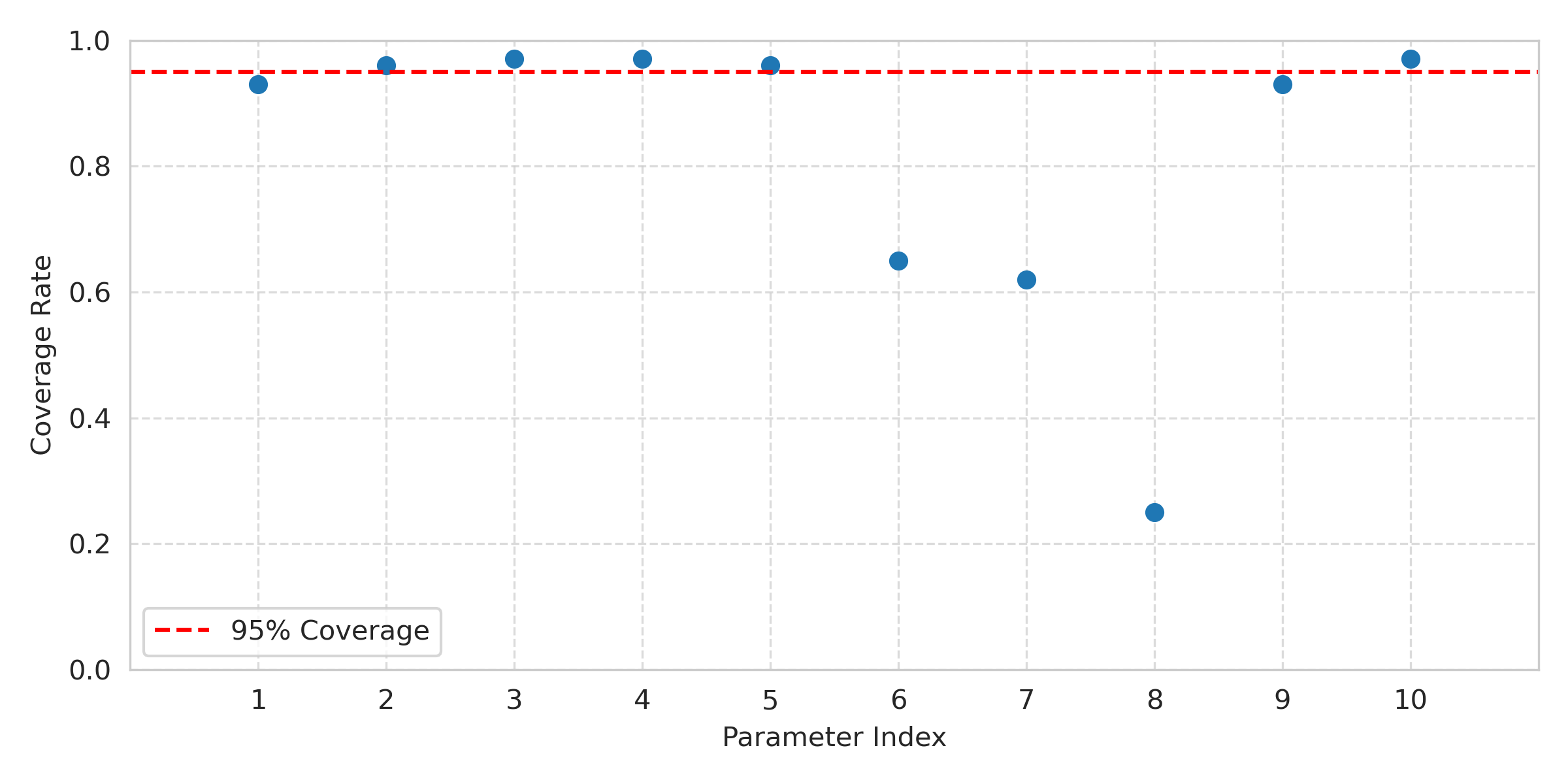}
  \caption{Coverage rates of 95\% confidence intervals ($K_1=1$).}
\end{subfigure}
\begin{subfigure}{.5\textwidth}
  \centering
  \includegraphics[width=.8\linewidth]{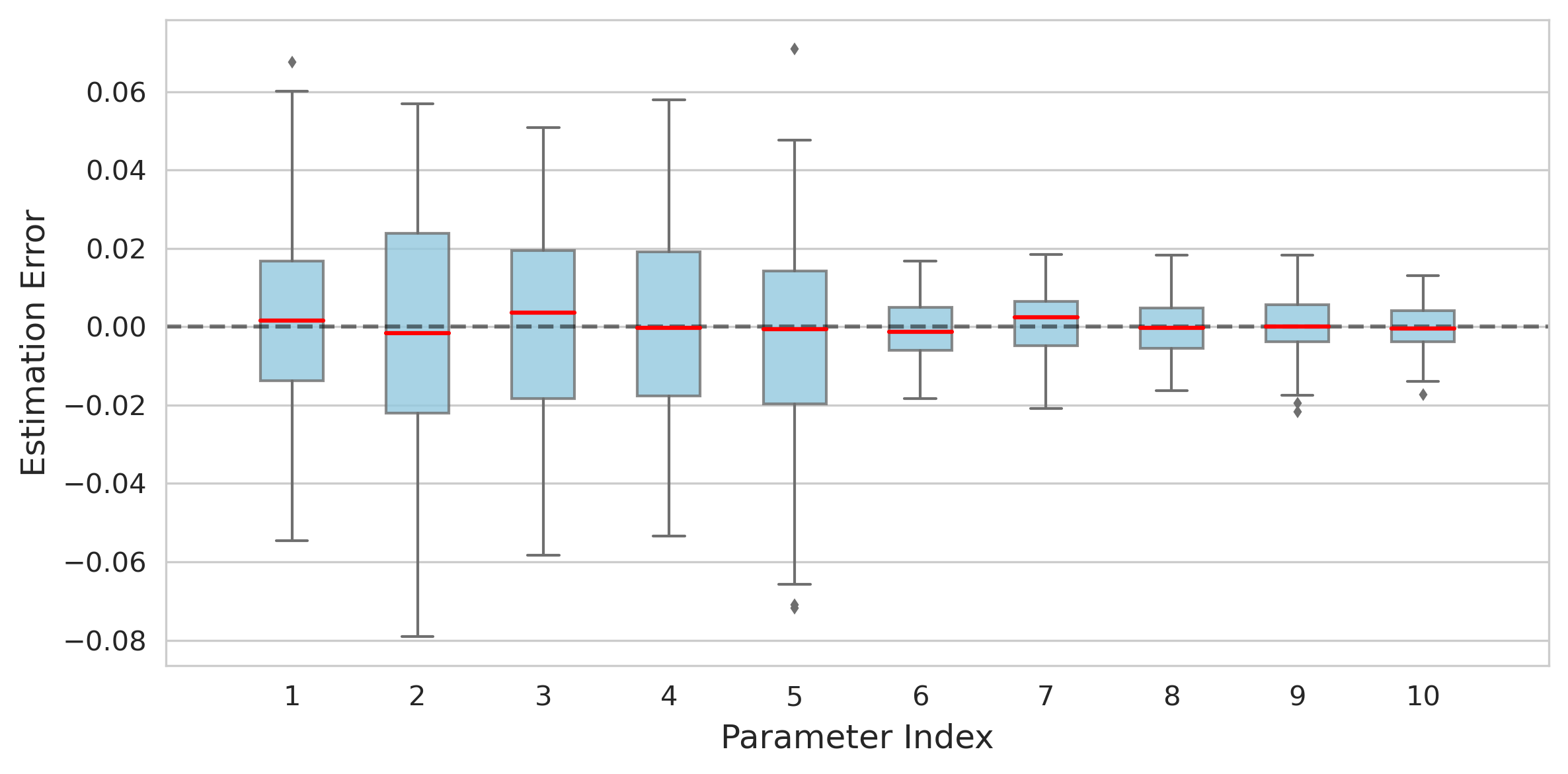}
  \caption{Distributions of point estimates $-$ true values
  ($K_{1}=4$)}
\end{subfigure}%
\begin{subfigure}{.5\textwidth}
  \centering
  \includegraphics[width=.8\linewidth]{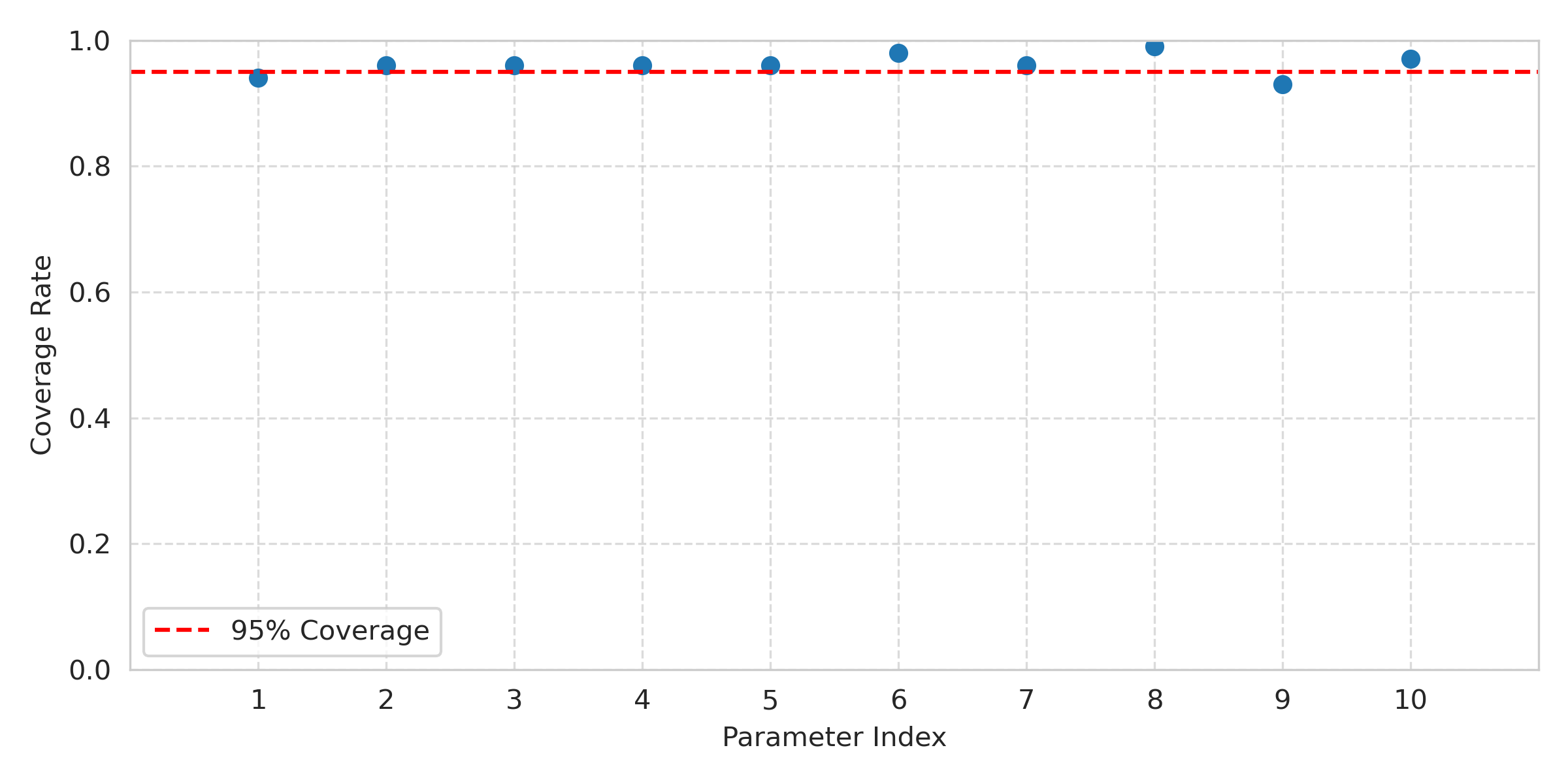}
  \caption{Coverage rates of 95\% confidence intervals ($K_1=4$).}
\end{subfigure}
\caption{
  Estimation results for marginal means in simulations where the data
  were generated as in Study III but the imputation model was
  estimated under assumption of ignorable missingness.
  The content and format of the plots is the same as in Figure
5 in the main text of the paper.}
\label{f_a_sim_III_ignorable}
\end{figure}

\vspace*{-2ex}
\begin{figure}[p]
\begin{subfigure}{.5\textwidth}
  \centering
  \includegraphics[width=.8\linewidth]{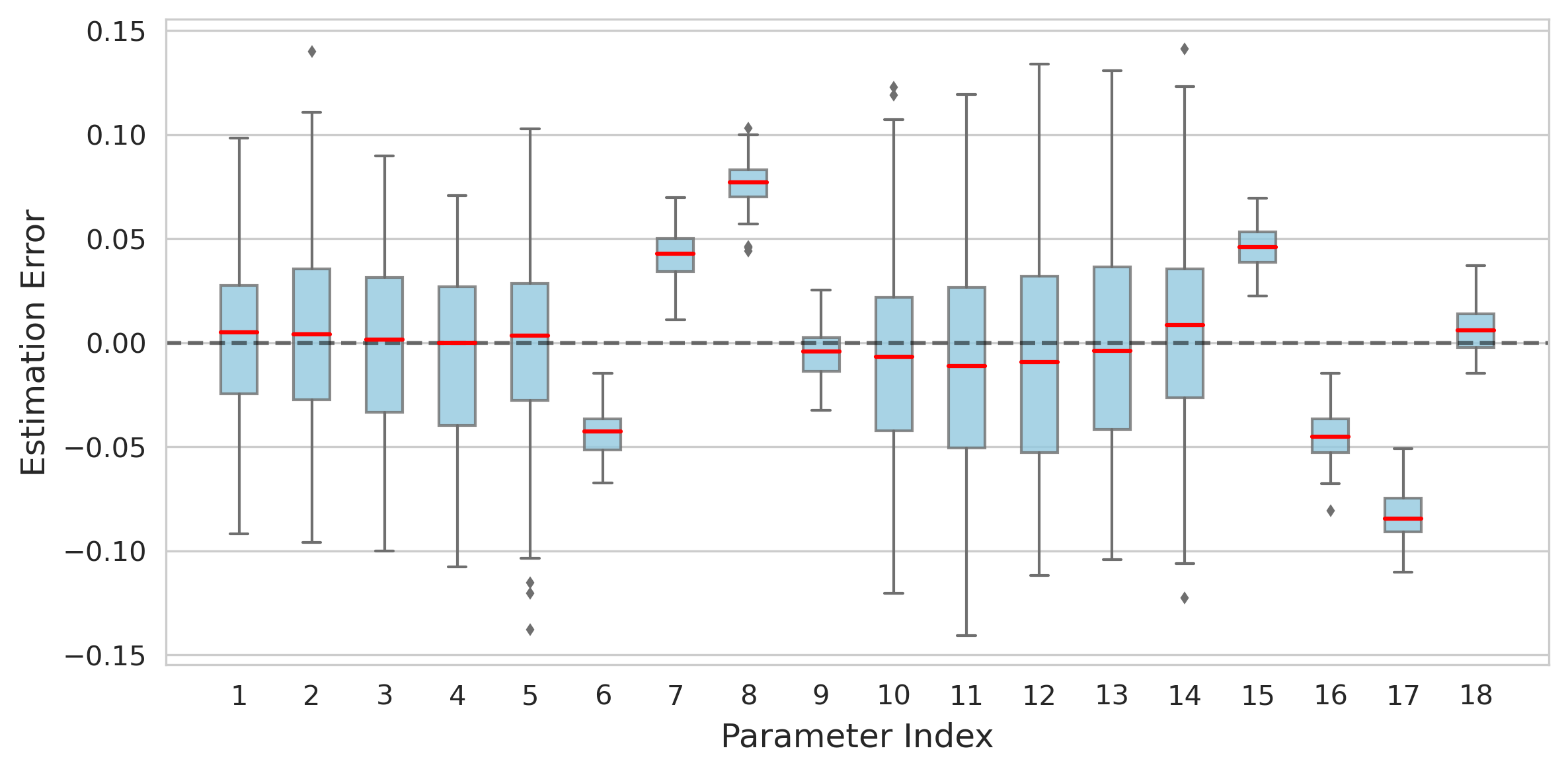}
  \caption{Distributions of point estimates $-$ true values
  ($K_{1}=1$)}
\end{subfigure}%
\begin{subfigure}{.5\textwidth}
  \centering
  \includegraphics[width=.8\linewidth]{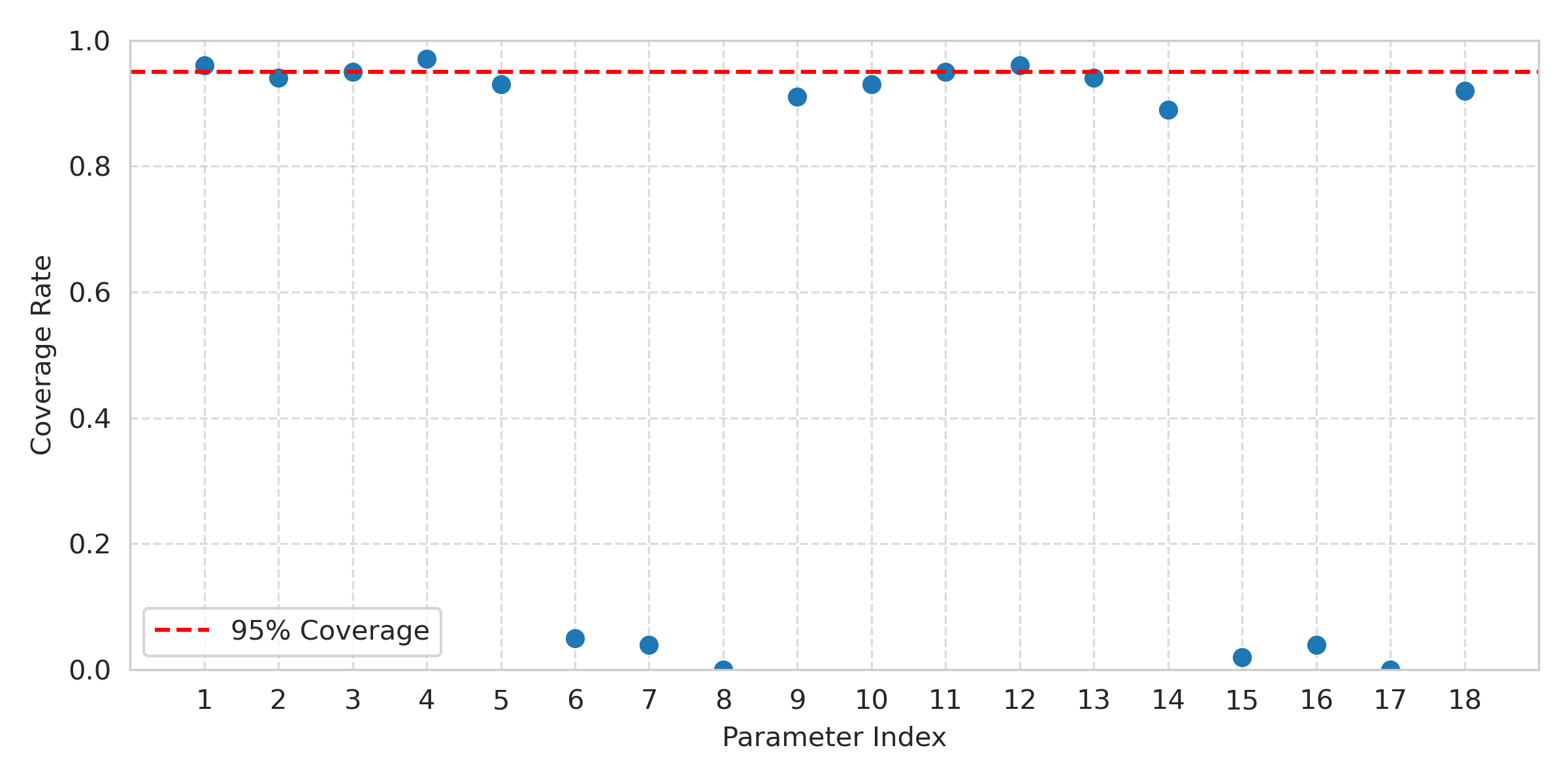}
  \caption{Coverage rates of 95\% confidence intervals ($K_1=1$).}
\end{subfigure}
\begin{subfigure}{.5\textwidth}
  \centering
  \includegraphics[width=.8\linewidth]{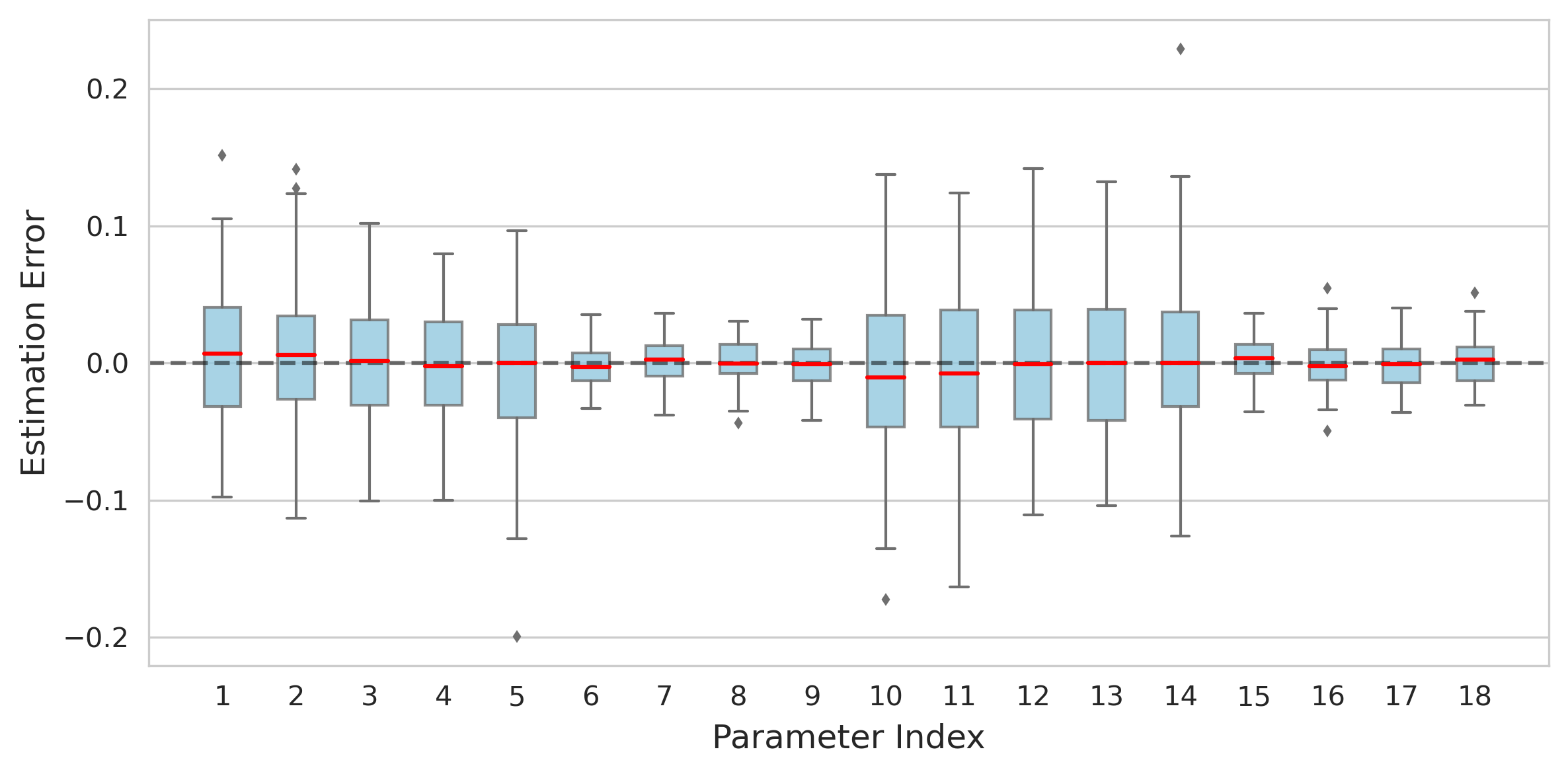}
  \caption{Distributions of point estimates $-$ true values
  ($K_{1}=4$)}
\end{subfigure}%
\begin{subfigure}{.5\textwidth}
  \centering
  \includegraphics[width=.8\linewidth]{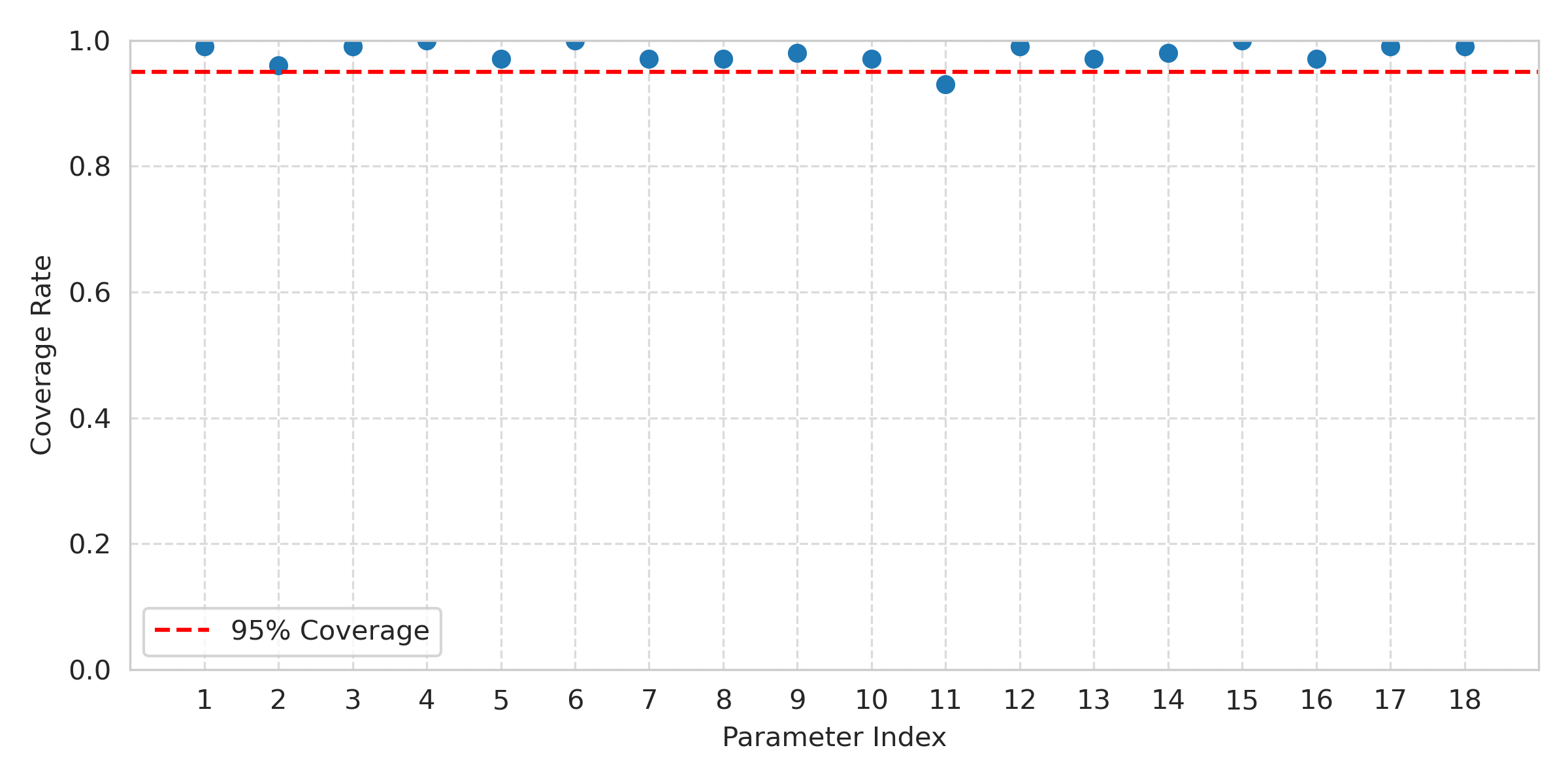}
  \caption{Coverage rates of 95\% confidence intervals ($K_1=4$).}
\end{subfigure}
\caption{
  Estimation results for conditional marginal means in simulations
  where the data were generated as in Study III but the imputation
  model was estimated under assumption of ignorable missingness.
  The content and format of the plots is the same as in Figure
\ref{f_a_sim_III} above.}
\label{f_a_sim_III_cond_ignorable}
\end{figure}

\clearpage

\section{Supplement to Section 6: Further information on the applied example}

\subsection{Details of dimension selection for the imputation model}
\label{ss_a_dimension_selection}

To select the dimensions of the substantive ($K_{1}$) and nonresponse ($K_{2}$) latent variables for the imputation model applied to the European Social Survey data, we used data from the two countries with the largest sample sizes, Italy and Austria. We compared seven settings for the latent dimensions.

The results are presented in Table~\ref{table:bic_supp}. The best models, as judged by the
BIC statistic, have dimensions $(K_{1},K_{2})=(1,1)$ for Italy and
(2,1) for Austria. However, BIC values for (1,1), (2,1), (3,1) are in
both cases all fairly similar and lower than those for larger numbers of
latent variables. To err on the side of caution in making the model more
flexible, as also suggested by the simulation study, we will use
latent dimensions of $(K_{1},K_{2})=(3,1)$.

For every choice of dimensions we also carried out a likelihood ratio test of the
hypothesis that $\boldsymbol{\kappa}=\mathbf{0}$. It gave a very small
$p$-value in every case, clearly indicating strong evidence of non-ignorable nonresponse.

\begin{table}[!htbp]
\centering
\caption{BIC statistics for estimated latent variable model
(1) with different numbers of substantive latent
variables ($\boldsymbol{\eta}$, dimension $K_{1}$) and nonresponse
latent variables ($\boldsymbol{\xi}$, dimension $K_{2}$), fitted to data
from Italy and Austria separately. In these
models nonresponse is non-ignorable with respect to
$\mathbf{\eta}$, i.e.\ parameter $\boldsymbol{\kappa}\ne \mathbf{0}$.
Likelihood ratio test of $\boldsymbol{\kappa}= \mathbf{0}$ gives
$p<0.001$ in all cases.
}
\label{table:bic_supp}
\begin{tabular}{@{}l  *{7}{r} @{}}
\toprule
& \multicolumn{7}{c}{Latent dimensions $(K_{1},K_{2})$} \\
Country  & (1,1) & (2,1) & (2,2) & (3,1) & (3,2) & (4,1) & (4,2) \\
\midrule
Italy    & \textbf{76465} & {76640} & {82245} & {77506} & {83124} & 81398 & {85843} \\
Austria    & {73379} & \textbf{73077} & {79265} & {75091} & 79783 & {75342} & {81533} \\
\bottomrule
\end{tabular}
\end{table}

\clearpage

\subsection{Covariates}

\begin{table}[htbp]
\centering
\vspace*{1ex}
\begin{tabular}{llrr}
  \hline
  Variable                                  & Coding
  & Mean                      & SD           \\
  \hline
  Number of household members               &
  & 2.56                      & 1.34         \\[.5ex]
  Age (years)                               &
  & 51.08                     & 18.62        \\[.5ex]
  Highest level of education                & ES-ISCED$^{1}$
  & 3.00                      & 1.80         \\[.5ex]
  Subjective general health
  & 0=Very good to 4=Very bad & 1.21  & 0.93 \\[.5ex]
  Main activity in last 7 days is paid work & 0=No, 1=Yes
  & 0.50                      &
  0.50
  \\[.5ex]
  Ever given birth to / fathered a child
  & 0=No, 1=Yes               & 0.30  & 0.46 \\[.5ex]
  Sex                                       & 0=Male, 1=Female
  & 0.54                      & 0.50         \\[.5ex]
  Ever lived with a spouse or partner       & 0=No, 1=Yes
  & 0.87                      & 0.37         \\[.5ex]
  Born in the country                       &
  0=No, 1=Yes                               & 0.90
  & 0.30                                     \\[.5ex]
  Own income is 0                           & 0=No, 1=Yes
  & 0.10                      & 0.30         \\[.5ex]
  \hline
  \multicolumn{4}{l}{\footnotesize{
      1 \hspace{.2em} 7-level harmonized coding used by the ESS, from
      0=Less than than lower secondary to
  }}
  \\
  \multicolumn{4}{l}{\footnotesize{
      \hspace{1em} 6=Higher tertiary education, here treated as a
      continuous scale.
      Responses coded as Other or
  }}
  \\
  \multicolumn{4}{l}{\footnotesize{
      \hspace*{1em}as not harmonizable to ES\_ISCED are treated as missing,
      and such respondents are omitted.
  }}
\end{tabular}
\caption{
  Description and summary statistics of the
  covariates ($\mathbf{x}$ in the notation of the paper) from
  round 9 (2018) of the European Social Survey (ESS) that are used in
  the analysis. The variables are all characteristics of the survey
  respondent. These variables are observed for all respondents in our
  analysis sample. The analysis also uses
  twenty variables of main interest ($\mathbf{y}$) that are subject
  to nonresponse; they are listed in Table 1 of the paper.
  The sample size for all variables is $N=48,577$.
}
\label{tab:covariates}
\end{table}

\subsection{Sample sizes and nonresponse rates}
\begin{table}[htbp]
\scriptsize
\centering
\begin{tabular}{lrrrrrrrrrrrr}
  \toprule
  &         & \multicolumn{10}{c}{Variable}
  \\
  &         &
  F41                 & G4      & G5                            & G6
  & G7      & G14a                          & G17a    &
  G18                 & G19     & G20
  \\
  Country             & $N$     & hinctnta                      &
  ifredu              & ifrjob  & evfredu                       & evfrjob &
  netifr              & occinfr & topinfr                       &
  btminfr             & wltdffr
  \\
  \midrule
  AT (Austria)        & 2465    & 17.12                         &
  4.42                & 4.91    & 2.23                          &
  2.35    & 5.92    &
  8.97                & 6.04    & 3.12                          &
  4.91
  \\
  BE (Belgium)        & 1750    & 7.66                          &
  2.51                & 2.29    & 0.74                          &
  0.57    & 0.74    &
  2.63                & 2.91    & 2.11                          &
  2.40
  \\
  BG (Bulgaria)       & 2165    & 13.63                         &
  8.78                & 9.70    & 8.22                          &
  8.55    & 6.84    &
  5.54                & 17.09   & 7.25                          &
  12.98
  \\
  CH (Switzerland)    & 1500    & 21.07                         &
  4.93                & 5.40    & 4.73                          & 4.20    & 3.87
  & 13.67   & 8.67                          & 6.00    & 9.33
  \\
  CY (Cyprus)         & 767     & 16.43                         &
  3.52                & 4.17    & 1.30                          &
  1.30    & 7.56    &
  22.69               & 7.17    & 1.83                          &
  3.91
  \\
  CZ (Czechia)        & 2374    & 32.10                         &
  6.02                & 4.04    & 1.98                          &
  1.52    & 13.69   &
  17.02               & 13.27   & 11.75                         &
  13.40
  \\
  DE (Germany)        & 2339    & 11.03                         &
  3.89                & 3.21    & 1.33                          &
  1.41    & 3.89    &
  9.75                & 9.41    & 6.63                          &
  5.64
  \\
  DK (Denmark)        & 1544    & 13.28                         &
  3.04                & 4.92    & 2.20                          &
  2.27    & 3.89    &
  6.74                & 1.30    & 1.10                          &
  6.35
  \\
  EE (Estonia)        & 1899    & 2.37                          &
  2.00                & 1.21    & 0.63                          &
  0.90    & 0.58    &
  6.69                & 2.90    & 1.16                          &
  3.21
  \\
  ES (Spain)          & 1635    & 26.97                         &
  5.44                & 6.97    & 3.91                          &
  4.16    & 4.59    &
  12.17               & 15.78   & 8.81                          &
  9.60
  \\
  FI (Finland)        & 1749    & 6.75                          &
  3.54                & 2.57    & 1.09                          &
  1.43    & 1.54    &
  4.69                & 4.06    & 3.09                          &
  4.12
  \\
  FR (France)         & 1999    & 10.61                         &
  6.10                & 6.00    & 1.75                          &
  1.90    & 2.15    &
  6.65                & 8.25    & 5.25                          &
  6.35
  \\
  GB (United Kingdom) & 2126    & 15.10                         &
  1.65                & 1.51    & 1.08                          & 1.13    &
  3.06                & 6.82    & 7.29                          &
  4.33                & 5.08
  \\
  HR (Croatia)        & 1758    & 24.80                         &
  2.22                & 3.64    & 2.33                          &
  2.05    & 4.49    &
  7.28                & 9.27    & 3.75                          &
  4.04
  \\
  HU (Hungary)        & 1638    & 39.87                         &
  3.91                & 4.88    & 2.93                          &
  2.50    & 7.69    &
  10.07               & 9.10    & 8.06                          &
  7.51
  \\
  IE (Ireland)        & 2133    & 26.77                         &
  2.16                & 1.64    & 1.41                          &
  1.41    & 3.56    &
  5.11                & 4.88    & 4.03                          &
  3.80
  \\
  IS (Iceland)        & 851     & 7.17                          &
  2.35                & 3.17    & 1.76                          &
  2.00    & 2.59    & 9.17
  & 8.34    & 3.88                          & 6.58
  \\
  IT (Italy)          & 2662    & 44.03                         &
  5.90                & 7.66    & 4.13                          &
  3.16    & 9.24    &
  11.65               & 11.38   & 6.69                          &
  9.47
  \\
  LT (Lithuania)      & 1812    & 13.47                         &
  3.92                & 6.51    & 4.47                          &
  4.30    & 5.13    &
  7.78                & 10.87   & 4.97                          &
  5.24
  \\
  LV (Latvia)         & 914     & 10.61                         &
  5.80                & 8.21    & 2.84                          &
  3.50    & 4.81    &
  10.83               & 9.74    & 4.49                          &
  6.89
  \\
  ME (Montenegro)     & 1169    & 20.62                         &
  1.63                & 1.80    & 0.94                          &
  1.37    & 8.55    &
  7.78                & 5.22    & 4.19                          &
  4.19
  \\
  NL (Netherlands)    & 1642    & 16.44                         &
  2.68                & 3.65    & 1.83                          & 1.34    & 3.41
  & 7.55    & 7.55                          & 6.03    & 5.24
  \\
  NO (Norway)         & 1355    & 7.31                          &
  2.29                & 2.14    & 0.74                          &
  0.96    & 1.40    & 3.39
  & 3.03    & 1.33                          & 4.35
  \\
  PL (Poland)         & 1452    & 37.47                         &
  5.17                & 5.92    & 3.99                          &
  3.93    & 5.72    &
  9.09                & 11.02   & 7.71                          &
  8.82
  \\
  PT (Portugal)       & 1032    & 18.90                         &
  3.68                & 5.14    & 2.71                          &
  2.91    & 16.18   &
  17.34               & 6.59    & 2.13                          &
  3.59
  \\
  RS (Serbia)         & 1976    & 23.58                         &
  4.25                & 6.93    & 3.49                          &
  3.49    & 7.89    &
  11.18               & 12.50   & 6.78                          &
  8.65
  \\
  SE (Sweden)         & 1502    & 5.99                          &
  1.86                & 3.20    & 1.73                          &
  1.73    & 1.73    & 6.32
  & 6.32    & 5.19                          & 7.32
  \\
  SI (Slovenia)       & 1304    & 12.42                         &
  3.60                & 3.30    & 2.15                          &
  1.23    & 2.99    &
  8.82                & 10.28   & 5.67                          &
  6.06
  \\
  SK (Slovakia)       & 1065    & 21.97                         &
  2.54                & 2.91    & 2.16                          &
  1.97    & 6.20    &
  7.42                & 6.85    & 4.88                          &
  5.35
  \\ \hline
  All                 & 48,577  & 18.92                         &
  3.94                & 4.48    & 2.52                          &
  2.45    & 5.18    & 8.85 &
  8.43                & 5.16    & 6.60
  \\
  \hline\hline
  &         &
  G2                  & G3      & G27                           & G28
  & G29     & G1                            & G26     & G30     & G31
  & G32                \\
  &         & gvintcz                       & poltran & sofrwrk &
  sofrpr              & sofrprv & frprtpl                       &
  sofrdst             & ppldsrv & jstprev                       &
  pcmpinj
  \\
  \midrule
  AT (Austria)        &         & 1.83                          &
  2.92                & 0.97    & 1.54                          &
  3.20    & 3.29    & 1.50 &
  1.58                & 1.87    & 4.26
  \\
  BE (Belgium)        &         & 1.54                          &
  2.06                & 0.57    & 0.80                          &
  1.03    & 2.51    & 0.69 &
  0.80                & 1.09    & 1.60
  \\
  BG (Bulgaria)       &         & 10.76                         &
  14.55               & 8.96    & 9.65                          &
  17.18   & 18.11   &
  11.09               & 11.82   & 12.89                         &
  17.23
  \\
  CH (Switzerland)    &         & 9.40                          &
  12.20               & 2.00    & 1.60                          &
  3.40    & 9.07    &
  2.27                & 3.40    & 2.87                          &
  4.07
  \\
  CY (Cyprus)         &         & 3.39                          &
  4.82                & 0.78    & 0.26                          &
  3.65    & 6.00    & 1.17 &
  1.17                & 1.04    & 4.82
  \\
  CZ (Czechia)        &         & 3.71                          &
  5.22                & 3.12    & 3.62                          &
  5.43    & 5.98    & 3.54 &
  2.61                & 2.74    & 4.76
  \\
  DE (Germany)        &         & 2.22                          &
  2.86                & 0.86    & 0.81                          &
  1.67    & 2.52    & 1.33 &
  1.62                & 0.94    & 1.97
  \\
  DK (Denmark)        &         & 4.40                          &
  4.79                & 1.04    & 0.97                          &
  3.30    & 2.72    & 1.23 &
  2.27                & 2.27    & 3.24
  \\
  EE (Estonia)        &         & 1.90                          &
  2.53                & 0.05    & 0.26                          &
  0.74    & 2.05    & 0.26 &
  0.37                & 0.16    & 0.68
  \\
  ES (Spain)          &         & 6.12                          &
  6.97                & 3.79    & 4.22                          &
  4.77    & 8.93    & 3.91 & 4.34
  & 4.89    & 7.89
  \\
  FI (Finland)        &         & 1.94                          &
  3.54                & 0.91    & 0.80                          &
  1.20    & 2.40    & 1.09 &
  1.60                & 1.09    & 1.43
  \\
  FR (France)         &         & 2.65                          &
  3.65                & 1.20    & 1.35                          &
  1.85    & 3.85    & 1.35 &
  1.65                & 1.90    & 2.05
  \\
  GB (United Kingdom) &         & 1.13                          &
  2.16                & 0.38    & 0.66                          &
  1.03    & 2.16    &
  0.75                & 0.52    & 0.85                          &
  1.41
  \\
  HR (Croatia)        &         & 1.99                          &
  3.24                & 1.88    & 1.08                          &
  3.53    & 3.41    & 1.93 &
  1.48                & 1.19    & 2.50
  \\
  HU (Hungary)        &         & 2.93                          &
  4.58                & 1.04    & 1.95                          &
  2.44    & 5.43    & 1.22 &
  1.34                & 2.01    & 3.85
  \\
  IE (Ireland)        &         & 2.53                          &
  4.50                & 0.89    & 0.75                          &
  2.30    & 4.08    & 1.36 &
  1.92                & 1.97    & 2.72
  \\
  IS (Iceland)        &         & 2.12                          &
  3.41                & 1.06    & 1.06                          &
  0.94    & 3.06    & 1.53 &
  2.59                & 1.88    & 2.47
  \\
  IT (Italy)          &         & 3.16                          &
  5.82                & 1.58    & 1.69                          &
  2.85    & 6.35    & 1.69 &
  2.07                & 1.92    & 2.67
  \\
  LT (Lithuania)      &         & 2.92                          &
  7.34                & 2.37    & 2.54                          &
  5.74    & 7.89    & 2.92 &
  3.48                & 3.09    & 6.51
  \\
  LV (Latvia)         &         & 5.14                          &
  10.18               & 3.06    & 5.69                          &
  6.02    & 8.64    & 6.78 &
  2.52                & 3.17    & 5.36
  \\
  ME (Montenegro)     &         & 4.02                          &
  6.16                & 1.45    & 1.20                          &
  1.54    & 6.76    & 1.37
  & 2.05    & 1.45                          & 2.40
  \\
  NL (Netherlands)    &         & 3.11                          &
  5.12                & 1.28    & 1.22                          &
  1.64    & 4.63    & 1.34
  & 1.71    & 1.77                          & 2.98
  \\
  NO (Norway)         &         & 0.81                          &
  3.62                & 0.59    & 0.66                          &
  1.03    & 1.70    & 0.15 &
  0.96                & 0.81    & 1.40
  \\
  PL (Poland)         &         & 3.58                          &
  5.03                & 1.45    & 3.31                          &
  3.65    & 6.75    & 2.34 &
  2.34                & 2.41    & 4.34
  \\
  PT (Portugal)       &         & 3.68                          &
  4.46                & 1.36    & 1.65                          &
  3.68    & 4.94    & 1.36 &
  2.33                & 2.42    & 2.42
  \\
  RS (Serbia)         &         & 5.67                          &
  8.76                & 1.32    & 1.87                          &
  5.77    & 8.35    & 2.28 &
  2.23                & 1.77    & 3.74
  \\
  SE (Sweden)         &         & 2.00                          &
  4.66                & 1.07    & 1.26                          &
  1.60    & 1.46    & 1.13 &
  3.26                & 1.86    & 2.20
  \\
  SI (Slovenia)       &         & 2.53                          &
  5.67                & 1.38    & 1.30                          &
  2.68    & 4.14    & 1.38 &
  1.53                & 1.61    & 2.76
  \\
  SK (Slovakia)       &         & 2.25                          &
  4.60                & 1.50    & 2.25                          &
  4.69    & 3.66    & 1.97 &
  2.25                & 1.88    & 3.29
  \\ \hline
  All                 &         & 3.43                          &
  5.31                & 1.71    & 1.98                          &
  3.51    & 5.25    & 2.15 & 2.40 & 2.36 & 3.78 \\
  \bottomrule
\end{tabular}
\caption{Sample sizes for the 29 countries in the analysis sample from the European
Social Survey (ESS), and nonresponse rates (\%)
for the twenty analysis variables listed in Table 1
of the paper. The variables are identified by their numbers in the
ESS questionnaire (as also shown in Table 1) and by
their names in the published ESS dataset.}
\label{t_a_ESS}
\end{table}

\clearpage
\subsection{Additional analyses}
\label{ss_a_sim_more}

\begin{figure}[!htbp]
\centering
\includegraphics[width=0.9\linewidth]{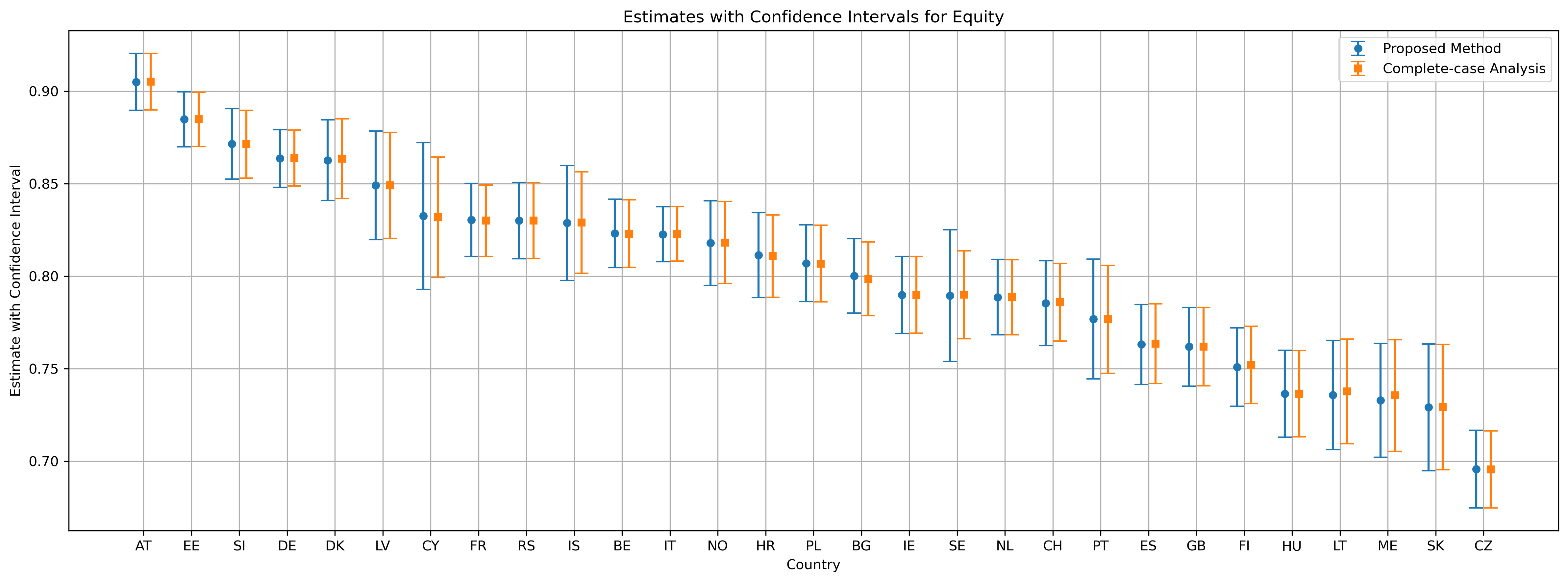}
\caption{
  Estimates of normative perceptions about equity of income
  distributions. The Figure shows proportions of people in adult
  populations 29 countries who would Agree Strongly or Agree with the
  statement that ``A society is fair when hard-working people earn
  more than others'', estimated from data from
  round 9 (2018) of the European Social Survey. Estimates are shown
  (with 95\% confidence intervals) based on only the observed
  responses to this question (in orange) and using the multiple
  imputation procedure described in the text (in blue). These results
  can be compared to ones for perceptions of Equality which are shown
  in Figure 6 in the main text of the paper.
}
\label{f_a_equity}
\end{figure}

\begin{figure}[!htbp]
\centering
\includegraphics[width=0.9\linewidth]{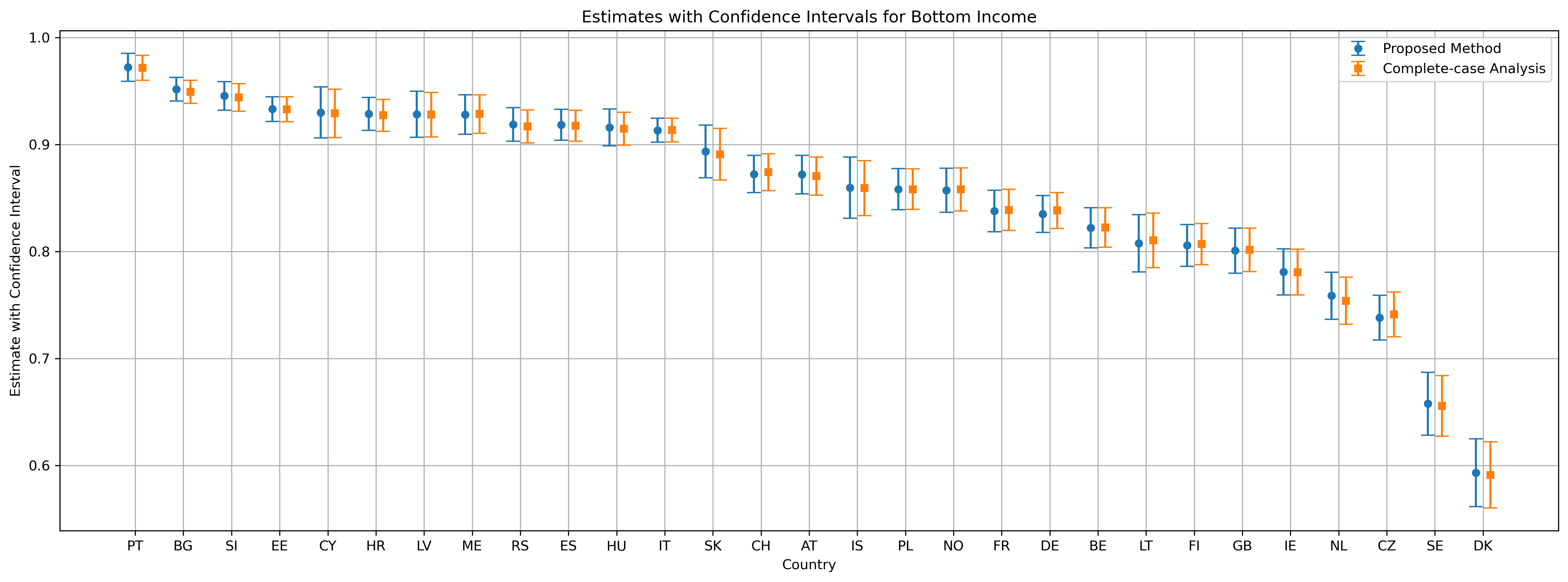}
\caption{
  Estimates of perceived unfairness of bottom incomes. The figure shows
  proportions of people in adult populations of 29 countries who
  would give a response to
  the statement
  ``Please think about the bottom 10\% of employees working full-time in
  [your country], earning less than [amount per month or per year]''
  that indicated that these incomes were unfairly low,
  estimated from data from
  round 9 (2018) of the European Social Survey. Estimates are shown
  (with 95\% confidence intervals) based on only the observed
  responses to this question (in orange) and using the multiple
  imputation procedure described in the text (in blue).
  These results can be compared to ones for perceptions about top
  10\% of incomes which are shown in Figure 7
  in the main text of the paper.
}
\label{f_a_bottom10}
\end{figure}

\clearpage
\bibliographystyle{apacite}
\bibliography{ref}